\documentclass[aps,pra,reprint,showkeys,floatfix]{revtex4-2}

\usepackage{amsmath}
\usepackage{amsfonts}
\usepackage{mathrsfs}
\usepackage[dvipsnames]{xcolor}
\usepackage{hyperref}
\hypersetup{
	colorlinks,
	linkcolor={red!50!black},
	citecolor={red!70!black},
	urlcolor={red!80!black}
}
\usepackage[capitalize]{cleveref}
\usepackage{multirow}
\usepackage[bb=px,oldbold]{mathalpha}
\usepackage{physics}
\usepackage{mhchem}
\usepackage{graphicx}
\usepackage{subfig}
\usepackage{colortbl}
\usepackage{cancel}
\usepackage{threeparttable}
\usepackage{booktabs}
\usepackage{mathtools}
\usepackage{tikz}
\usetikzlibrary {arrows.meta,decorations.text}

\DeclareMathOperator{\Span}{span}

\newcommand{\I}{\mathrm{i}}
\newcommand{\sm}{Supplemental Material}
\newcommand{\foreign}[1]{\textit{#1}}

\definecolor{OIblue}{HTML}{0072B2}
\definecolor{OIgreen}{HTML}{009E73}
\definecolor{OIverm}{HTML}{D55E00}
\definecolor{OIpurple}{HTML}{CC79A7}

\begin{document}

\title{Clifford Transformations for Fermionic Quantum Systems:\linebreak
	From Paulis to Majoranas to Fermions}

\author{I. Magoulas}
\email{ilias.magoulas@emory.edu}
\author{F.A. Evangelista}
\affiliation{Department of Chemistry and Cherry Emerson Center for Scientific Computation, Emory University, Atlanta, Georgia 30322, USA}

\begin{abstract}
Clifford gates and transformations, which map products of elementary Pauli or Majorana operators to other such products, are foundational in quantum computing, underpinning the stabilizer formalism, error-correcting codes, magic state distillation, quantum communication and cryptography, and qubit tapering. 
Moreover, circuits composed entirely of Clifford gates are classically simulatable, highlighting their computational significance.
In this work, we extend the concept of Clifford transformations to fermionic systems.
We demonstrate that fermionic Clifford transformations are generated by half-body and pair operators, providing a systematic framework for their characterization.
Additionally, we establish connections with fermionic mean-field theories and applications in qubit tapering, offering insights into their broader implications in quantum computing.
\end{abstract}


\maketitle

\section{Introduction}

Clifford gates, which constitute a subgroup of the unitary group, map one Pauli string (a product of elementary Pauli gates) to another under similarity transformations \cite{Gottesman.1997.quant-ph/9705052,Gottesman.1998.quant-ph/9807006,Gottesman.2024.QEC.draft}.
Although they do not form a universal gate set \cite{Gottesman.1998.10.1103/PhysRevA.57.127}, they are indispensable to quantum computing and quantum information science.
They form the backbone of the stabilizer formalism \cite{Gottesman.1996.10.1103/PhysRevA.54.1862,Calderbank.1997.10.1103/PhysRevLett.78.405,Calderbank.1998.10.1109/18.681315,Gottesman.1997.quant-ph/9705052}, utilized in quantum error correction \cite{Shor.1995.10.1103/PhysRevA.52.R2493,Steane.1996.10.1103/PhysRevLett.77.793}.
In turn, quantum error-correcting codes are employed to construct logical (fault-tolerant) qubits \cite{Shor.1995.10.1103/PhysRevA.52.R2493,Google.2023.10.1038/s41586-022-05434-1,Bravyi.2024.10.1038/s41586-024-07107-7,Reichardt.2024.2409.04628} and Clifford gates \cite{Gottesman.1998.10.1103/PhysRevA.57.127,Rengaswamy.2020.10.1109/TQE.2020.3023419}.
On the same note, Clifford gates play a crucial role in magic state distillation \cite{Bravyi.2005.10.1103/PhysRevA.71.022316,Rodriguez.2024.2412.15165,Daguerre.2025.10.1103/PhysRevResearch.7.023080,Itogawa.2025.10.1103/thxx-njr6,Daguerre.2025.2506.14169}, a technique used to implement gates beyond Clifford, such as the $T$ gate, in a fault-tolerant setting.

Among the Clifford unitaries one finds the basic gates to create superpositions and entanglement, namely, the Hadamard ($H$) and controlled-NOT (CNOT) gates.
Indeed, when the $H$ gate acts on one of the two single-qubit basis states, it creates an equal superposition of them.
The CNOT gate can be used to entangle the states of two qubits, producing a two-qubit state that cannot be expressed as the tensor product of the states of the individual qubits.
Thus, Clifford gates are needed to generate maximally entangled states, such as the Bell states \cite{Einstein.1935.10.1103/PhysRev.47.777,Bell.1964.10.1103/PhysicsPhysiqueFizika.1.195} and their multi-qubit generalizations \cite{Greenberger.1989.10.1007/978-94-017-0849-4_10,Greenberger.1990.10.1119/1.16243,Gisin.1998.10.1016/S0375-9601(98)00516-7,Sych.2009.10.1088/1367-2630/11/1/013006}.
Consequently, Clifford gates are useful in areas such as quantum communication \cite{Bennett.1992.10.1103/PhysRevLett.69.2881,Bennett.1993.10.1103/PhysRevLett.70.1895} and quantum cryptography \cite{Ekert.1991.10.1103/PhysRevLett.67.661,Hillery.1999.10.1103/PhysRevA.59.1829}.

Beyond quantum error correction, communication, and cryptography, Clifford gates and transformations are valuable for quantum simulations as well.
According to the Gottesman--Knill theorem \cite{Gottesman.1998.10.1103/PhysRevA.57.127,Gottesman.1998.quant-ph/9807006}, quantum circuits comprised exclusively of Clifford gates can be efficiently simulated classically.
Additionally, Clifford transformations lie at the heart of the qubit reduction technique known as qubit tapering \cite{Bravyi.2017.1701.08213,Setia.2020.10.1021/acs.jctc.0c00113}.
In this scheme, one exploits the $\mathbb{Z}_2$ symmetries of a given Hamiltonian to eliminate redundant qubits from quantum simulations.

Recent progress \cite{Microsoft.2023.10.1103/PhysRevB.107.245423,Microsoft.2025.10.1038/s41586-024-08445-2} in Majorana-based topological quantum computing \cite{Kitaev.2001.10.1070/1063-7869/44/10S/S29,Bravyi.2002.10.1006/aphy.2002.6254,Kitaev.2003.10.1016/S0003-4916(02)00018-0,Freedman.2003.10.1090/S0273-0979-02-00964-3,Nayak.2008.10.1103/RevModPhys.80.1083}, in which Majorana zero modes serve as qubits, has led to an extension of the Clifford group from Pauli to Majorana operators \cite{Mudassar.2024.10.1103/PhysRevA.110.032430,Bettaque.2025.2407.11319}.
In this context, the Majorana Clifford group contains unitary operations that map under similarity transformations one Majorana string (product of elementary Majorana operators) to another.
Although Majorana-based quantum devices are inherently more robust to errors due to fermionic superselection rules and nonlocal qubit encoding, Majorana Clifford gates are nevertheless expected to be essential for fault tolerance \cite{Mudassar.2024.10.1103/PhysRevA.110.032430,Bettaque.2025.2407.11319}.

In this work, motivated by the significance of Clifford gates in quantum computing and by the recent developments in Majorana Clifford gates, we explore Clifford transformations involving general products of fermionic creation and annihilation operators.
In particular, we investigate how these transformations can be systematically generated, demonstrating that half-body and pair operators act as the fundamental building blocks.
By doing so, we draw connections between fermionic Clifford transformations and established concepts in quantum many-body physics, such as fermionic mean-field theories \cite{Hartree.1928.10.1017/S0305004100011919,Hartree.1928.10.1017/S0305004100011920,Slater.1930.10.1103/PhysRev.35.210.2,Fock.1930.10.1007/BF01340294,Fock.1930.10.1007/BF01330439,Fukutome.1977.10.1143/PTP.57.1554,Fukutome.1981.10.1002/qua.560200502,Fukutome.1981.10.1143/PTP.65.809,Echenique.2007.10.1080/00268970701757875,Moussa.2012.1208.1086,Henderson.2024.10.1063/5.0188155} and Bogoliubov transformations \cite{Bogoljubov.1958.10.1007/BF02745585,Valatin.1958.10.1007/BF02745589}.
Additionally, we show that fermionic Clifford transformations naturally arise in the qubit tapering procedure \cite{Bravyi.2017.1701.08213,Setia.2020.10.1021/acs.jctc.0c00113}.
Through this exploration, we aim to bridge the gap between fermionic systems and the broader context of quantum information science, shedding light on the role of Clifford transformations in understanding and optimizing fermionic quantum systems.

The paper is structured as follows.
In \cref{sec:pauli,sec:majorana}, we review the salient features of the Pauli and Majorana algebras, respectively, and their relationship with Clifford transformations.
Subsequently, in \cref{sec:fermion}, we discuss the main finding of this study, namely, the realization of Clifford transformations for fermions.
In \cref{sec:tapering}, we use the minimum-basis-set hydrogen molecule as an example to illustrate the connection of fermionic Clifford transformations to qubit tapering.
Finally, in \cref{sec:mean-field}, we demonstrate the connection of fermionic Clifford unitaries to mean-field theories.

\section{Background Information}
\subsection{Pauli Algebra and Clifford Transformations}\label{sec:pauli}

In this section, we review the essential elements of the Pauli algebra and Clifford transformations pertinent to this work.
The interested reader is referred to the textbooks by Nielsen and Chuang \cite{Nielsen.2010.10.1017/CBO9780511976667} and Gottesman \cite{Gottesman.2024.QEC.draft} for additional information.

The Pauli matrices $\sigma_x$, $\sigma_y$, and $\sigma_z$, introduced by Pauli in 1927 to describe spin-$\frac{1}{2}$ systems \cite{Pauli.1927.10.1007/BF01397326}, represent fundamental single-qubit gates in quantum computing.
As shown in Section S1 in the \sm, although the set $\{\sigma_0, \sigma_x, \sigma_y, \sigma_z\}$, where $\sigma_0 \equiv I$ denotes the identity matrix, is not closed under matrix multiplication due to phase factors, closure can be achieved by extending it to
\begin{equation}\label{pgroup_1}
	\mathcal{P}_1 = \left\{\I^\alpha \sigma_j \mid \alpha\in\mathbb{Z}_4, j\in\{0,x,y,z\}\right\}.
\end{equation}
This extended set forms the single-qubit Pauli group, $\mathcal{P}_1$.
In the case of $M$ qubits, the $M$-qubit Pauli group is defined as
\begin{equation}\label{pgroup_m}
	\mathcal{P}_M = \left\{\I^\alpha P \mid \alpha\in\mathbb{Z}_4, P \in \left\{I,\sigma_x,\sigma_y,\sigma_z\right\}^{\otimes M}\right\},
\end{equation}
with $P$ typically referred to as a Pauli string.

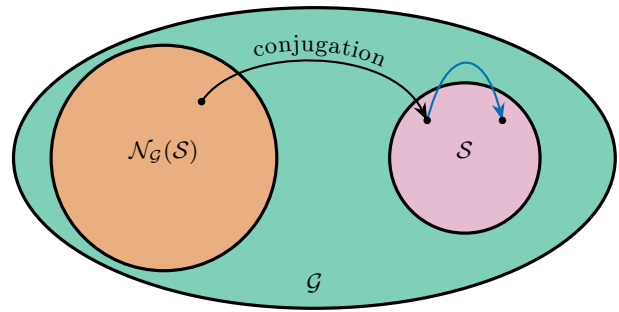
\begin{figure}[h!]
	\centering
	\begin{tikzpicture}
		\filldraw[color=black!100, fill=OIgreen!50, very thick] (0,0) ellipse (4 and 2);
		\filldraw[color=black!100, fill=OIverm!50, very thick] (-2,0) circle (1.5);
		\filldraw[color=black!100, fill=OIpurple!50, very thick] (2,0) circle (1);
		\draw[-Stealth, thick,
		postaction={decorate, decoration={text along path, text={conjugation}, text align=center, raise=4pt}}] (-1.5,0.75) .. controls (-1,1.5) and (1,1.5) .. (1.5,0.5);
		\draw[-Stealth, thick, color=OIblue] (1.5,0.5) .. controls (1.75,1.5) and (2.25,1.5) .. (2.5,0.5);
		\fill[color=black!100] (-1.5,0.75) circle (0.05);
		\fill[color=black!100] (1.5,0.5) circle (0.05);
		\fill[color=black!100] (2.5,0.5) circle (0.05);
		\node[anchor=base] at (0,-1.75) {$\mathcal{G}$};
		\node[anchor=base] at (-2,0) {$\mathcal{N}_{\mathcal{G}}(\mathcal{S})$};
		\node[anchor=base] at (2,0) {$\mathcal{S}$};
	\end{tikzpicture}
	\caption{
			Schematic illustration of the normalizer $\mathcal{N}_\mathcal{G}(\mathcal{S})$ of a subset $\mathcal{S}$ of a group $\mathcal{G}$.
			Elements of $\mathcal{N}_\mathcal{G}(\mathcal{S})$ map under conjugation an element of $\mathcal{S}$ to another element of $\mathcal{S}$.
			Note that if $\mathcal{S}$ is a subgroup of $\mathcal{G}$, then it is a subgroup of $\mathcal{N}_\mathcal{G}(\mathcal{S})$ as well.
		}
	\label{fig:normalizer}
\end{figure}

In group theory, the normalizer of a subset $\mathcal{S}$ of a group $\mathcal{G}$ is the subgroup of $\mathcal{G}$ that preserves $\mathcal{S}$ under similarity transformations, \foreign{i.e.}, $\mathcal{N_G}(\mathcal{S}) = \{g\in \mathcal{G} \mid g^{-1} s g \in \mathcal{S}, \forall s \in \mathcal{S} \}$ (see \cref{fig:normalizer}).
The normalizer of the $M$-qubit Pauli group in the unitary group $\mathcal{U}(2^M)$ is called the Clifford group \cite{Gottesman.1997.quant-ph/9705052,Gottesman.1998.quant-ph/9807006}, denoted as $\mathcal{C}_{\mathcal{P}_M}$.
In other words, the elements of the Clifford group, called Clifford gates, map any $M$-qubit Pauli string to another via similarity transformation, called Clifford transformation. The fact that $\mathcal{P}_M$ is a subgroup of $\mathcal{U}(2^M)$ implies that the Pauli group is also a subgroup of the Clifford group.

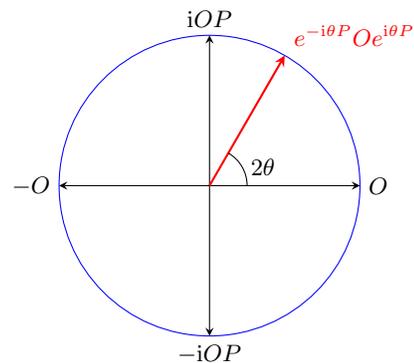
\begin{figure}[tbh!]
	\centering
	\begin{tikzpicture}[>=stealth, scale=1]
		\draw[->] (0,0) -- (2,0) node[right] {$O$};
		\draw[->] (0,0) -- (0,2) node[above] {$\I OP$};
		\draw[->] (0,0) -- (-2,0) node[left] {$-O$};
		\draw[->] (0,0) -- (0,-2) node[below] {$-\I OP$};
		\draw[->, thick, red] (0,0) -- (60:2cm) node[above right] {\textcolor{red}{$e^{-\I \theta P} O e^{\I \theta P}$}};
		\draw (0.5cm,0) arc (0:60:0.5cm) node[midway, right] {$2\theta$};
		\draw[blue] (0,0) circle [radius=2cm];
	\end{tikzpicture}
	\caption{
		Illustration of the continuous rotation of the Pauli string $O$ into $\I OP$ under the unitary transformation $\exp(-\I \theta P) O \exp(\I \theta P)$, with $\{O,P\}=0$.	
	}
	\label{fig:pauli_rotation}
\end{figure}

The Clifford group can be generated by three types of gates, namely, phase ($S$), $H$, and CNOT gates \cite{Gottesman.1997.quant-ph/9705052}, $\mathcal{C}_{\mathcal{P}_M} = \langle \exp(\I\theta) I, S_i, H_i, \text{CNOT}_{ij} \rangle$, where $i,j = 0,1,\ldots, M-1$ designate qubit indices and $\theta = \left[ 0, 2\pi\right)$.
Note that $\exp(\I\theta) I$ is included in the generating set since if $U\in \mathcal{C}_{\mathcal{P}_M}$ then $\exp(\I\theta)U \in \mathcal{C}_{\mathcal{P}_M}$ as well.
For the purposes of this study, we focus on an alternative representation of Clifford gates based on complex exponentials of Pauli strings, $\exp(\I \theta P)$.
To determine the conditions under which such an operator belongs to the Clifford group, we proceed as follows.
Using the fact that Pauli strings, similarly to the Pauli matrices, are Hermitian involutions, a Taylor expansion of $\exp(\I\theta P)$ yields
\begin{equation}
	\label{pauli_exp}
	e^{\I \theta P} = \cos(\theta)I+\I \sin(\theta)P.
\end{equation}
Using \cref{pauli_exp} and the fact that two Pauli strings either commute or anticommute, we obtain that the similarity transformation of one Pauli string by the exponential of another is given by
\begin{equation}
	\label{pauli_st}
	e^{-\I \theta P} O e^{\I \theta P} =
	\begin{cases}
		O, & [O,P] = 0 \\
		\cos(2\theta) O + \I \sin(2\theta)OP, & \{O,P\} = 0
	\end{cases}.
\end{equation}
In the non-trivial, anticommuting case, the above transformation performs a continuous rotation between the Pauli strings $O$ and $\I OP$ (see \cref{fig:pauli_rotation}).
A simple inspection of \cref{pauli_st} reveals that for $\theta = k\frac{\pi}{4}, k \in \mathbb{Z}$, the transformation maps $O$ onto one of the Pauli strings $\pm O$ or $\pm \I OP$.
Consequently, any exponentiated Pauli string of the form $\exp(\I\theta P)$ with $\theta$ as above is a Clifford gate.
To ensure that the Clifford transformation of $O$ will generate the new  Pauli string $\pm \I OP$, the restriction $\theta = (2k+1)\frac{\pi}{4}, k \in \mathbb{Z}$, may be imposed.

At this point, it is worth mentioning that complex exponentials of Pauli strings form another generating set for the Pauli Clifford group.
Indeed, it is straightforward to verify that \cite{Barenco.1995.10.1103/PhysRevA.52.3457,Nielsen.2010.10.1017/CBO9780511976667}
	\begin{equation}\label{eq:clifford_s}
		S_i = e^{\I \frac{\pi}{4}} e^{-\I \frac{\pi}{4}\sigma_z^{(i)}},
	\end{equation}
	\begin{equation}\label{eq:clifford_h}
		H_i = e^{\I \frac{3\pi}{2}} e^{\I \frac{\pi}{4} \sigma_z^{(i)}} e^{\I \frac{\pi}{4} \sigma_x^{(i)}} e^{\I \frac{\pi}{4} \sigma_z^{(i)}},
	\end{equation}
	and
	\begin{equation}\label{eq:clifford_cnot}
		\text{CNOT}_{ij} = e^{\I \frac{\pi}{4}} e^{-\I \frac{\pi}{4} \sigma_z^{(i)}} e^{-\I \frac{\pi}{4} \sigma_x^{(j)}} e^{\I \frac{\pi}{4} \sigma_z^{(i)} \sigma_x^{(j)}},
	\end{equation}
where the superscript inside parentheses denotes the qubit on which the corresponding Pauli gate is acting on.
As a result, an alternative generating set for the Pauli Clifford group is $\mathcal{C}_{\mathcal{P}_M} = \langle \exp(\I \theta)I,\allowbreak \exp(\I \frac{\pi}{4} \sigma_x^{(i)}),\allowbreak \exp(\I \frac{\pi}{4} \sigma_z^{(i)}),\allowbreak \exp(\I \frac{\pi}{4} \sigma_z^{(i)} \sigma_x^{(j)}) \rangle$.

\subsection{Majorana Algebra and Clifford Transformations}\label{sec:majorana}

Moving on to the next step in the complexity of the underlying algebra, we proceed to the case of Majorana fermions. 
Proposed in 1937 by Ettore Majorana \cite{Majorana.1937.10.1007/BF02961314}, Majorana fermions are particles that are their own antiparticle.
Although no elementary Majorana fermions have been conclusively observed (with neutrinos being the only possible candidates within the Standard Model \cite{Wilczek.2009.10.1038/nphys1380}), the formation of Majorana quasi-particles is theoretically possible in condensed matter systems \cite{Ma.2017.10.1016/j.physb.2017.01.028}.
However, their experimental detection has yet to be conclusively confirmed \cite{Frolov.2021.10.1038/d41586-021-00954-8,Legg.2025.2502.19560}.
In the era of quantum computing, Majorana quasi-particles have enjoyed a renewed interest, as they are anticipated to enable more stable qubits suitable for fault-tolerant fermionic computing \cite{Kitaev.2001.10.1070/1063-7869/44/10S/S29,Bravyi.2002.10.1006/aphy.2002.6254,Kitaev.2003.10.1016/S0003-4916(02)00018-0}.
Indeed, significant efforts toward their experimental realization are currently underway \cite{Microsoft.2023.10.1103/PhysRevB.107.245423,Microsoft.2025.10.1038/s41586-024-08445-2}.

Let $a$ and $a^\dagger$ denote the standard fermionic annihilation and creation operators, respectively, defining a single complex fermionic mode associated with a given single-particle state.
From these, one can generate two Majorana operators, \foreign{i.e.}, two real Majorana modes, given by
\begin{equation}\label{gamma1}
	\gamma_1 = a^\dagger + a
\end{equation}
and
\begin{equation}
	\gamma_2 = \I(a^\dagger - a).
\end{equation}
Note that $\gamma_1^\dagger = \gamma_1$ and $\gamma_2^\dagger = \gamma_2$, as anticipated for Majorana fermions.
Furthermore, $\gamma_1^2 = \gamma_2^2 = I$, implying that $\gamma_1$ and $\gamma_2$ are Hermitian involutions.
As shown in Section S2 of the \sm, the set $\{\gamma_1, \gamma_2\}$ does not form a group since it is not closed under multiplication.
However, the group structure can be attained by the augmented set \cite{Bettaque.2025.2407.11319}
\begin{equation}\label{mgroup_2}
	\mathcal{M}_2 = \left\{\I^\alpha \gamma_i \mid \alpha,i\in\mathbb{Z}_4\right\}, 
\end{equation}
where $\gamma_0 \equiv I$ and $\gamma_3 \equiv \I \gamma_2 \gamma_1 = I-2a^\dagger a$.
Note that the auxiliary operator $\gamma_3$ was introduced to facilitate a direct comparison with the Pauli group.
The set $\mathcal{M}_2$ is known as the Majorana group for two Majorana modes.
As shown in Section S2 of the \sm, the single-qubit Pauli and two-Majorana--mode Majorana groups are isomorphic, $\mathcal{M}_2 \cong \mathcal{P}_1$, with $\gamma_1 \mapsto \sigma_x$, $\gamma_2 \mapsto \sigma_y$, and $\gamma_3 \mapsto \sigma_z$ under the isomorphism.

In the case of $M$ fermionic modes, the corresponding Majorana group is defined as \cite{Bettaque.2025.2407.11319}
\begin{equation}\label{mgroup_2m}
	\begin{split}
	\mathcal{M}_{2M} ={}& \left\{ \I^\alpha \Gamma \mid \alpha\in\mathbb{Z}_4,\vphantom{\prod_{k=1}^M}\right.\\
	&\left.\Gamma = \prod_{k=1}^M \gamma_p^{(k)}, \gamma_p^{(k)} \in \{\gamma_0^{(k)}, \gamma_1^{(k)}, \gamma_2^{(k)}, \gamma_3^{(k)}\} \right\},
	\end{split}
\end{equation}
where the $k$ superscript denotes that the given operator acts on the $k$th single-particle state. The operator product $\Gamma$ is typically referred to as a Majorana string.
The fermionic anticommutation relations obeyed by Majorana operators, which read
\begin{equation}\label{acomm_m}
	\{\gamma_p^{(k)},\gamma_q^{(l)}\} = 2\delta_{pq} \delta_{kl}, \quad p,q \in \{1,2\},
\end{equation}
introduce additional sign factors when multiplying two Majorana strings.
Consequently, in higher dimensions, the isomorphism between the Pauli and Majorana groups requires mappings that are non-local.
For example, in the Jordan--Wigner transformation \cite{Jordan.1928.10.1007/BF01331938}, one has
\begin{equation}
	\gamma_1^{(k)} \mapsto \sigma_x^{(k)} \prod_{p = 1}^{k-1} \sigma_z^{(p)},
\end{equation}
\begin{equation}
	\gamma_2^{(k)} \mapsto \sigma_y^{(k)} \prod_{p = 1}^{k-1} \sigma_z^{(p)},
\end{equation}
and
\begin{equation}
	\gamma_3^{(k)} \mapsto \sigma_z^{(k)}.
\end{equation}

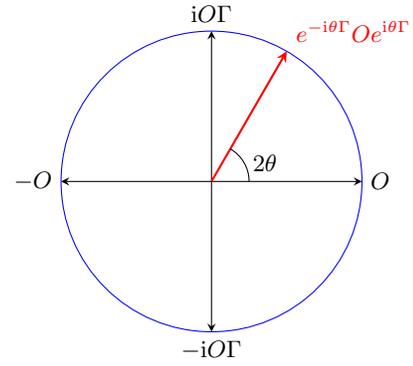
\begin{figure}[tbh!]
	\centering
	\begin{tikzpicture}[>=stealth, scale=1]
		\draw[->] (0,0) -- (2,0) node[right] {$O$};
		\draw[->] (0,0) -- (0,2) node[above] {$\I O\Gamma$};
		\draw[->] (0,0) -- (-2,0) node[left] {$-O$};
		\draw[->] (0,0) -- (0,-2) node[below] {$-\I O\Gamma$};
		\draw[->, thick, red] (0,0) -- (60:2cm) node[above right] {\textcolor{red}{$e^{-\I\theta\Gamma}Oe^{\I\theta\Gamma}$}};
		\draw (0.5cm,0) arc (0:60:0.5cm) node[midway, right] {$2\theta$};
		\draw[blue] (0,0) circle [radius=2cm];
	\end{tikzpicture}
	\caption{
		Illustration of the continuous rotation of the Majorana string $O$ into $\I O\Gamma$ under the unitary transformation $\exp(-\I \theta \Gamma) O \exp(\I \theta \Gamma)$, with $\Gamma^\dagger = \Gamma$ and $\{O,\Gamma\}$.
		For an anti-Hermitian $\Gamma$, the imaginary unit is dropped.
			}
	\label{fig:majorana_rotation}
\end{figure}

Similar to the Pauli case, the Majorana Clifford group on 2$M$ Majorana modes, $\mathcal{C}_{\mathcal{M}_{2M}}$, is defined as the set of unitaries $U\in\mathcal{U}(2^M)$ that map one Majorana string to another via similarity transformation \cite{Mudassar.2024.10.1103/PhysRevA.110.032430,Bettaque.2025.2407.11319}.
As was the case with the Pauli group, the Majorana group is a subgroup of the Majorana Clifford group.
For the purposes of this work, we focus on Majorana Clifford operators generated by exponentiating a Majorana string.
However, the anticommutation relations, \cref{acomm_m}, introduce a minor complication.
In contrast to Pauli strings, which are always Hermitian involutions, Majorana strings can be either Hermitian involutions,
\begin{equation}\label{mstring_hermitian}
	\Gamma^\dagger = \Gamma, \quad \Gamma^2 = I, \quad L=1,4,5,8,9,\ldots,
\end{equation}
or anti-Hermitian skew-involutions,
\begin{equation}\label{mstring_antihermitian}
	\Gamma^\dagger = -\Gamma, \quad \Gamma^2=-I, \quad L=2,3,6,7,10,11,\ldots.
\end{equation}
This depends on the length $L$ of a given Majorana string, which is defined as the number of $\gamma_1$ and $\gamma_2$ operators in the string (recall that $\gamma_3 \equiv \I \gamma_2 \gamma_1$).
Therefore, there are two families of Majorana unitaries, namely, those generated by a Hermitian Majorana string,
\begin{equation}\label{mclifford_hermitian}
		U = e^{\I\theta \Gamma}, \quad L=1,4,5,8,9,\ldots,
\end{equation}
and those based on an anti-Hermitian one,
\begin{equation}\label{mclifford_antihermitian}
	U = e^{\theta \Gamma}, \quad L=2,3,6,7,10,11,\ldots.
\end{equation}
To determine the conditions under which unitaries of the form of \cref{mclifford_hermitian,mclifford_antihermitian} belong to the Majorana Clifford group, we follow a procedure similar to the Pauli case.
We first express the above unitaries in closed-form by Taylor expanding the exponentials and using \cref{mstring_hermitian,mstring_antihermitian}, obtaining
\begin{equation}\label{majorana_hermitian_exp}
	e^{\I\theta \Gamma} = \cos(\theta)I+\I\sin(\theta) \Gamma, \quad L=1,4,5,8,9,\ldots
\end{equation}
and
\begin{equation}\label{majorana_antihermitian_exp}
	e^{\theta \Gamma} = \cos(\theta) I + \sin(\theta)\Gamma, \quad L=2,3,6,7,10,11,\ldots.
\end{equation}
Using \cref{majorana_hermitian_exp,majorana_antihermitian_exp} and the fact that two Majorana strings either commute or anticommute, we arrive at the following closed-form expressions for the similarity transformation of one Majorana string by the unitary exponential of another:
\begin{equation}\label{mST_hermitian}
e^{-\I\theta\Gamma}Oe^{\I\theta\Gamma} =
	\begin{cases}
		O, &[O,\Gamma] = 0\\
		\cos(2\theta)O + \I\sin(2\theta)O\Gamma, &\{ O,\Gamma \}  = 0
	\end{cases} 
\end{equation}
for $L = 1,4,5,8,9, \ldots$
and
\begin{equation}\label{mST_antihermitian}
	e^{-\theta\Gamma}Oe^{\theta\Gamma} =
	\begin{cases}
		O, &[O,\Gamma] = 0\\
		\cos(2\theta)O + \sin(2\theta)O\Gamma, &\{ O,\Gamma \}  = 0
	\end{cases}
\end{equation}
when $L=2,3,6,7,10,11,\ldots$.
In the non-trivial, anticommuting case, the above transformations perform continuous rotations between the Majorana strings $O$ and $\I O\Gamma$ [\cref{mST_hermitian}] or $O\Gamma$ [\cref{mST_antihermitian}] (see \cref{fig:majorana_rotation}).
As in the Pauli case, these transformations constitute Majorana Clifford operations when $\theta = k\frac{\pi}{4},k\in\mathbb{Z}$, while the restriction $\theta = (2k+1)\frac{\pi}{4}$ ensures that a new Majorana string will be generated.

The isomorphism between the Pauli and Majorana groups directly implies the isomorphism of the corresponding Clifford groups, $\mathcal{C}_{\mathcal{M}_{2M}} \cong \mathcal{C}_{\mathcal{P}_M}$.
Thus, a generating set of the Majorana Clifford group can be readily obtained by the inverse JW transformation of the generators of the Pauli Clifford group, \foreign{i.e.}, $\mathcal{C}_{\mathcal{M}_{2M}} = \langle \exp(\I \theta)I,\allowbreak \exp(\I^p \frac{\pi}{4} \gamma_1^{(p)} \prod_{q=1}^{p-1} \gamma_2^{(q)} \gamma_1^{(q)}),\allowbreak \exp(-\frac{\pi}{4} \gamma_2^{(p)} \gamma_1^{(p)}),\allowbreak \exp(\I^{q+1} \frac{\pi}{4} \gamma_2^{(p)} \gamma_1^{(p)} \gamma_1^{(q)} \prod_{r=1}^{q-1} \gamma_2^{(r)} \gamma_1^{(r)}) \rangle$.
At this point, it is worth mentioning that Majorana fermions obey the fermionic parity superselection rule, which states that it is impossible to create a coherent superposition of states with even and odd numbers of fermions.
As a result, the only physically realizable Majorana strings are those that commute with the fermionic parity operator, which in the language of Majorana operators takes the form
\begin{equation}
	\Pi = \I^{M} \prod_{p=1}^{M} \gamma_2^{(p)} \gamma_1^{(p)}.
\end{equation}
One can readily observe that parity-preserving Majorana strings must have an even length.
Furthermore, since Clifford transformations at most convert the Majorana string $O$ to $O\Gamma$ [see \cref{mstring_hermitian,mstring_antihermitian}], the transformation will conserve fermionic parity as long as the generator $\Gamma$ has an even length.
Consequently, parity-preserving Majorana Clifford unitaries are generated by even-length Majorana strings, and form a subgroup of the Majorana Clifford group \cite{Mudassar.2024.10.1103/PhysRevA.110.032430,Bettaque.2025.2407.11319}.
It has been shown that the generators of the parity-preserving Majorana Clifford group have the form $\exp(\I \frac{\pi}{4} \gamma_a^{(p_a)} \gamma_b^{(p_b)} \gamma_c^{(p_c)} \gamma_d^{(p_d)})$ \cite{McLauchlan.2022.10.1103/PhysRevLett.128.180504}, known as the $\text{BRAID}_4$ gate \cite{Mudassar.2024.10.1103/PhysRevA.110.032430,Bettaque.2025.2407.11319}.

\section{Fermionic Algebra and Clifford Transformations}\label{sec:fermion}

Having reviewed the algebra and Clifford transformations associated with Pauli and Majorana strings, we now focus on the study's main subject, namely, (Dirac) fermions.

We start with a single fermionic mode, defined by the pair of $a$ and $a^\dagger$ second-quantized operators, associated with a given single-particle state.
Here, the involutory property of elementary Pauli and Majorana operators is replaced by the $a^2=(a^\dagger)^2=0$ nilpotency conditions.
This indicates immediately that the fermionic algebra differs fundamentally from both the Pauli and Majorana algebras.
Although the $\{a, a^\dagger\}$ set is not closed under operator multiplication, closure can be attained by extending it to
\begin{equation}
	\mathcal{F}_1 = \left\{0, I, a, a^\dagger, n, h\right\},
\end{equation}
where $n\equiv a^\dagger a$ and $h \equiv a a^\dagger$ denote the particle and hole number operators, respectively.
Note that despite the linear dependence of $n$ and $h$ ($h = I-n$), both operators naturally emerge as distinct elements during operator multiplication.
With the exception of the identity element $I$, the elements of $\mathcal{F}_1$ are not invertible.
Thus, in contrast to the Pauli $\mathcal{P}_1$ and Majorana $\mathcal{M}_2$ groups, $\mathcal{F}_1$ is not a group but a monoid.

In the case of $M$ fermionic modes, the corresponding monoid is given by
\begin{equation}
	\begin{split}
	\mathcal{F}_{M} ={}& \left\{\I^\alpha F \mid \alpha\in\mathbb{Z}_4,\vphantom{\prod_{k=1}^M}\right.\\
	&\left. F = \prod_{k=1}^M f_k, f_k \in \left\{ 0, I, a_k, a_k^\dagger, n_k, h_k \right\} \right\},
	\end{split}
\end{equation}
where the subscript $k$ indicates that the operator is acting on the $k$th single-particle state.
The second-quantized operator product $F$ is typically called a fermionic string.
Similar to the Majorana case, the usual fermionic anticommutation relations, namely,
\begin{equation}\label{acomm_f}
	\{a_p, a_q\} = \{a_p^\dagger, a_q^\dagger\} = 0, \{a_p, a_q^\dagger\} = \delta_{pq},
\end{equation}
imply that the ordering of operators in a string is important. In this work, we order the operators in the fermionic string as follows:
\begin{equation}\label{fstring}
	\begin{split}
	F &= a_{p_1}^\dagger \cdots a_{p_k}^\dagger a_{q_l} \cdots a_{q_1} h_{r_1} \cdots h_{r_m} n_{s_1} \cdots n_{s_n}\\
	&= a_{q_1 \ldots q_l}^{p_1 \ldots p_k} h_{r_1 \ldots r_m} n_{s_1 \ldots s_n},
	\end{split}
\end{equation}
where all indices are assumed to be distinct and $p_1 < \cdots < p_k$, $q_1 < \cdots < q_l$, $r_1 < \cdots < r_m$, and $s_1 < \cdots < s_n$.
Since we are working with the full algebra for $M$ fermionic modes, the expression for $F$ is generic, encompassing all possible scenarios regarding its nature, including, for example, particle-nonconserving and spin-flip operators.
Note that if $F$ is simply a product of number operators, then it is Hermitian and idempotent.
Otherwise, it is nilpotent.

In light of all the above, it becomes apparent that the fermionic algebra is fundamentally distinct from those of Pauli and Majorana strings.
It is thus interesting to see how the concept of Clifford operators and transformations discussed for the Pauli and Majorana groups can be extended to the case of fermions.
For example, in contrast to Pauli and Majorana strings, it is impossible for fermionic strings to act as Clifford operators as they are not invertible.
Driven by our interest in quantum computing, we seek unitary operators that map one fermionic string to another via similarity transformation.
Formally, since fermionic strings are not, in general, unitary, the fermionic Clifford group will not be a normalizer, but rather the setwise stabilizer \cite{Rotman.2012.10.1007/978-1-4612-4176-8} of the fermionic monoid in the unitary group.

In this work, we explore the conditions under which a fermionic Clifford transformation can be generated by the exponential of a fermionic string.
Similar to the case of Majorana fermions, the unitarity condition can be satisfied by either an anti-Hermitian,
\begin{equation}\label{fClifford_antihermitian}
	U = e^{\theta A}, \quad A = F - F^\dagger,
\end{equation}
or Hermitian,
\begin{equation}\label{fClifford_hermitian}
	U = e^{\I\theta H}, \quad H = F+F^\dagger,
\end{equation}
generator, with $F$ being a generic fermionic string [see \cref{fstring}].
We have recently shown that the similarity transformation of a single fermionic string $O$ by the above unitaries can be exactly expressed by the following closed forms \cite{Evangelista.2025.10.1103/PhysRevA.111.042825} (see, also, \cite{Jayakumar.2025.2510.10957} for an alternative derivation):
\begin{equation}\label{fST_antihermitian}
	\begin{split}
	e^{-\theta A} O e^{\theta A} ={}& O + \frac{\sin\left( \sqrt{\alpha} \theta \right)}{\sqrt{\alpha}} [O,A]\\
	&+ \frac{1-\cos\left( \sqrt{\alpha} \theta \right)}{\alpha} [[O,A],A]
	\end{split}
\end{equation}
and
\begin{equation}\label{fST_hermitian}
	\begin{split}
	e^{-\I \theta H} O e^{\I \theta H} ={}& O + \I \frac{\sin\left(\sqrt{\beta} \theta\right)}{\sqrt{\beta}} [O,H]\\
	&+ \frac{\cos\left( \sqrt{\beta} \theta \right) - 1}{\beta}[[O,H],H].
	\end{split}
\end{equation}
The values of the parameters $\alpha$ and $\beta$ depend on the structure of the two fermionic strings:
\begin{itemize}
	\item $\alpha,\beta = 1$, if $A[O,A]A = 0$ and $H[O,H]H = 0$,
	\item $\alpha,\beta = 4$, if $A[O,A]A = [O,A]$ and $H[O,H]H = [O,H]$,
\end{itemize}
with these two cases encompassing all possible combinations of fermionic operators.
The above expressions serve as the starting point for finding the conditions under which the unitaries shown in \cref{fClifford_antihermitian,fClifford_hermitian} act as Clifford.

After expressing the $A$ and $H$ generators in terms of $F$ and $F^\dagger$, one readily obtains that \cref{fST_antihermitian,fST_hermitian} perform a continuous rotation between the fermionic string $O$ and twelve other fermionic strings, namely, $FO$, $OF$, $F^\dagger O$, $O F^\dagger$, $FOF$, $F^\dagger O F^\dagger$, $FOF^\dagger$, $F^\dagger OF$, $OFF^\dagger$, $OF^\dagger F$, $FF^\dagger O$, and $F^\dagger FO$.
As such, these similarity transformations are far more complicated than the corresponding ones for the Pauli, \cref{pauli_st}, and Majorana, \cref{mclifford_hermitian,mclifford_antihermitian}, algebras.
To be able to map one fermionic string to another, the combined effect of the single and double commutators appearing in \cref{fST_antihermitian,fST_hermitian} must be, at most, to cancel the original fermionic string $O$ and introduce another.
This type of action places, in general, various constraints on the form of $F$ and the parameter $\theta$.
Indeed, as demonstrated in Section S3 of the \sm, fermionic Clifford transformations can only be generated by Hermitian and anti-Hermitian linear combinations of half-body ($a_p^\dagger \pm a_p$) and pair ($a_p^\dagger a_q^\dagger \pm a_q a_p$, $a_p^\dagger a_q \pm a_q^\dagger a_p$) operators with $\theta = k\frac{\pi}{2}$, $k\in\mathbb{Z}$.
Similar to the Pauli and Majorana cases, to map one fermionic string to another, the restriction $\theta = (2k+1)\frac{\pi}{2}$, $k\in\mathbb{Z}$, may be enforced.
In the Hermitian case, Clifford transformations can also be generated by number operators, $n_p$, for arbitrary values of $\theta$.

The physical significance of fermionic Clifford transformations becomes apparent once we examine their action.
To that end, in \cref{table_antihermitian,table_hermitian}, we have tabulated the analytic expressions for the fermionic Clifford transformations reported in this work.
We begin the discussion with the simpler case in which the string $O$ to be transformed shares no index with the generator of the Clifford transformation.
For the pair generators, which have an even length and necessarily commute with $O$, the Clifford transformation is trivial.
Depending on the length $L_O$ of the fermionic string $O$, the Clifford transformation generated by half-body operators will either be trivial (even $L_O$) or introduce a minus sign (odd $L_O$).
\begin{table*}
	\caption{
		Analytic expressions for the fermionic Clifford transformations based on anti-Hermitian generators.\protect\footnotemark[1]
	}
	{\resizebox{\textwidth}{!}{
			\begin{ruledtabular}
				\begin{tabular}{ccc}
					$A$ & $O$ & $e^{-(2k+1)\frac{\pi}{2}A}Oe^{(2k+1)\frac{\pi}{2}A}$\\
					\colrule
					\multirow{5}{*}[-20pt]{$a_{\textcolor{blue}{\boldsymbol{c}}}^\dagger - a_{\textcolor{blue}{\boldsymbol{c}}}$} & $a_{u_1 \ldots u_f}^{t_1 \ldots t_e} h_{v_1 \ldots v_g} n_{w_1 \ldots w_h}$ & $(-1)^{L_O} a_{u_1 \ldots u_f}^{t_1 \ldots t_e} h_{v_1 \ldots v_g} n_{w_1 \ldots w_h}$\\[10pt]
					& $a_{u_1 \ldots u_f}^{t_1 \ldots t_e} h_{v_1 \ldots v_g} n_{w_1 \ldots w_{i-1} \,\textcolor{blue}{\boldsymbol{c}}\, w_{i+1} \ldots w_h}$ & $(-1)^{L_O} a_{u_1 \ldots u_f}^{t_1 \ldots t_e} h_{v_1 \ldots v_g} n_{w_1 \ldots w_{i-1} w_{i+1} \ldots w_h} h_{\textcolor{blue}{\boldsymbol{c}}}$\\[10pt]
					& $a_{u_1 \ldots u_f}^{t_1 \ldots t_e} h_{v_1 \ldots v_{i-1} \,\textcolor{blue}{\boldsymbol{c}}\, v_{i+1} \ldots v_g} n_{w_1 \ldots w_h}$ & $(-1)^{L_O} a_{u_1 \ldots u_f}^{t_1 \ldots t_e} h_{v_1 \ldots v_{i-1} v_{i+1} \ldots v_g} n_{w_1 \ldots w_h} n_{\textcolor{blue}{\boldsymbol{c}}}$ \\[10pt]
					& $a_{u_1 \ldots u_f}^{t_1 \ldots t_{i-1} \,\textcolor{blue}{\boldsymbol{c}}\, t_{i+1} \ldots t_e} h_{v_1 \ldots v_g} n_{w_1 \ldots w_h}$ & $(-1)^i a_{u_1 \ldots u_f}^{t_1 \ldots t_{i-1} t_{i+1} \ldots t_e} h_{v_1 \ldots v_g} n_{w_1 \ldots w_h} a_{\textcolor{blue}{\boldsymbol{c}}}$\\[10pt]
					& $a_{u_1 \ldots u_{i-1} \,\textcolor{blue}{\boldsymbol{c}}\, u_{i+1} \ldots u_f}^{t_1 \ldots t_e} h_{v_1 \ldots v_g} n_{w_1 \ldots w_h}$ & $(-1)^{L_O + i + 1} a_{u_1 \ldots u_{i-1} u_{i+1} \ldots u_f}^{t_1 \ldots t_e} h_{v_1 \ldots v_g} n_{w_1 \ldots w_h} a_{\textcolor{blue}{\boldsymbol{c}}}^\dagger$ \\[20pt]
					\multirow{5}{*}[-20pt]{$a_{\textcolor{blue}{\boldsymbol{c}}}^\dagger a_{\textcolor{red}{\boldsymbol{d}}}^\dagger - a_{\textcolor{red}{\boldsymbol{d}}} a_{\textcolor{blue}{\boldsymbol{c}}}$\footnotemark[2]} & $a_{u_1 \ldots u_f}^{t_1 \ldots t_e} h_{v_1 \ldots v_g} n_{w_1 \ldots w_h}$ & $a_{u_1 \ldots u_f}^{t_1 \ldots t_e} h_{v_1 \ldots v_g} n_{w_1 \ldots w_h}$\\[10pt]
					& $a_{u_1 \ldots u_f}^{t_1 \ldots t_e} h_{v_1 \ldots v_g} n_{w_1 \ldots w_{i-1} \,\textcolor{blue}{\boldsymbol{c}}\, w_{i+1} \ldots w_h}$ & $a_{u_1 \ldots u_f}^{t_1 \ldots t_e} h_{v_1 \ldots v_g} n_{w_1 \ldots w_{i-1} w_{i+1} \ldots w_h} h_{\textcolor{red}{\boldsymbol{d}}}$\\[10pt]
					& $a_{u_1 \ldots u_f}^{t_1 \ldots t_e} h_{v_1 \ldots v_{i-1} \,\textcolor{blue}{\boldsymbol{c}}\, v_{i+1} \ldots v_g} n_{w_1 \ldots w_h}$ & $a_{u_1 \ldots u_f}^{t_1 \ldots t_e} h_{v_1 \ldots v_{i-1} v_{i+1} \ldots v_g} n_{w_1 \ldots w_h} n_{\textcolor{red}{\boldsymbol{d}}}$ \\[10pt]
					& $a_{u_1 \ldots u_f}^{t_1 \ldots t_{i-1} \,\textcolor{blue}{\boldsymbol{c}}\, t_{i+1} \ldots t_e} h_{v_1 \ldots v_g} n_{w_1 \ldots w_h}$ & $(-1)^{L_O+i+k} a_{u_1 \ldots u_f}^{t_1 \ldots t_{i-1} t_{i+1} \ldots t_e} h_{v_1 \ldots v_g} n_{w_1 \ldots w_h} a_{\textcolor{red}{\boldsymbol{d}}}$\\[10pt]
					& $a_{u_1 \ldots u_{i-1} \,\textcolor{blue}{\boldsymbol{c}}\, u_{i+1} \ldots u_f}^{t_1 \ldots t_e} h_{v_1 \ldots v_g} n_{w_1 \ldots w_h}$ & $(-1)^{k + i + 1} a_{u_1 \ldots u_{i-1} u_{i+1} \ldots u_f}^{t_1 \ldots t_e} h_{v_1 \ldots v_g} n_{w_1 \ldots w_h} a_{\textcolor{red}{\boldsymbol{d}}}^\dagger$ \\[20pt]
					\multirow{5}{*}[-20pt]{$a_{\textcolor{blue}{\boldsymbol{c}}}^\dagger a_{\textcolor{red}{\boldsymbol{d}}} - a_{\textcolor{red}{\boldsymbol{d}}}^\dagger a_{\textcolor{blue}{\boldsymbol{c}}}$\footnotemark[2]} & $a_{u_1 \ldots u_f}^{t_1 \ldots t_e} h_{v_1 \ldots v_g} n_{w_1 \ldots w_h}$ & $a_{u_1 \ldots u_f}^{t_1 \ldots t_e} h_{v_1 \ldots v_g} n_{w_1 \ldots w_h}$\\[10pt]
					& $a_{u_1 \ldots u_f}^{t_1 \ldots t_e} h_{v_1 \ldots v_g} n_{w_1 \ldots w_{i-1} \,\textcolor{blue}{\boldsymbol{c}}\, w_{i+1} \ldots w_h}$ & $a_{u_1 \ldots u_f}^{t_1 \ldots t_e} h_{v_1 \ldots v_g} n_{w_1 \ldots w_{i-1} w_{i+1} \ldots w_h} n_{\textcolor{red}{\boldsymbol{d}}}$\\[10pt]
					& $a_{u_1 \ldots u_f}^{t_1 \ldots t_e} h_{v_1 \ldots v_{i-1} \,\textcolor{blue}{\boldsymbol{c}}\, v_{i+1} \ldots v_g} n_{w_1 \ldots w_h}$ & $a_{u_1 \ldots u_f}^{t_1 \ldots t_e} h_{v_1 \ldots v_{i-1} v_{i+1} \ldots v_g} n_{w_1 \ldots w_h} h_{\textcolor{red}{\boldsymbol{d}}}$ \\[10pt]
					& $a_{u_1 \ldots u_f}^{t_1 \ldots t_{i-1} \,\textcolor{blue}{\boldsymbol{c}}\, t_{i+1} \ldots t_e} h_{v_1 \ldots v_g} n_{w_1 \ldots w_h}$ & $(-1)^{L_O+i+k} a_{u_1 \ldots u_f}^{t_1 \ldots t_{i-1} t_{i+1} \ldots t_e} h_{v_1 \ldots v_g} n_{w_1 \ldots w_h} a_{\textcolor{red}{\boldsymbol{d}}}^\dagger$\\[10pt]
					& $a_{u_1 \ldots u_{i-1} \,\textcolor{blue}{\boldsymbol{c}}\, u_{i+1} \ldots u_f}^{t_1 \ldots t_e} h_{v_1 \ldots v_g} n_{w_1 \ldots w_h}$ & $(-1)^{k + i + 1} a_{u_1 \ldots u_{i-1} u_{i+1} \ldots u_f}^{t_1 \ldots t_e} h_{v_1 \ldots v_g} n_{w_1 \ldots w_h} a_{\textcolor{red}{\boldsymbol{d}}}$ \\
				\end{tabular}
			\end{ruledtabular}
	}}
	\footnotetext[1]{The analytic expressions for cases in which $a_c^\dagger a_d^\dagger - a_d a_c$ or $a_c^\dagger a_d - a_d^\dagger a_c$ share two common indices with the string $O$ can be obtained by performing the transformation in two steps, each step involving one common index.}
	\footnotetext[2]{If the common index is $d$ instead of $c$, the sign of any expression whose sign factor depends on $k$ must be inverted.}
	\label{table_antihermitian}
\end{table*}

\begin{table*}
	\caption{
		Analytic expressions for the fermionic Clifford transformations based on Hermitian generators.\protect\footnotemark[1]
	}
	{\resizebox{\textwidth}{!}{
			\begin{ruledtabular}
				\begin{tabular}{ccc}
					$H$ & $O$ & $e^{-(2k+1)\frac{\pi}{2}\I H}Oe^{(2k+1)\frac{\pi}{2}\I H}$\\
					\colrule
					\multirow{5}{*}[-20pt]{$a_{\textcolor{blue}{\boldsymbol{c}}}^\dagger + a_{\textcolor{blue}{\boldsymbol{c}}}$} & $a_{u_1 \ldots u_f}^{t_1 \ldots t_e} h_{v_1 \ldots v_g} n_{w_1 \ldots w_h}$ & $(-1)^{L_O} a_{u_1 \ldots u_f}^{t_1 \ldots t_e} h_{v_1 \ldots v_g} n_{w_1 \ldots w_h}$\\[10pt]
					& $a_{u_1 \ldots u_f}^{t_1 \ldots t_e} h_{v_1 \ldots v_g} n_{w_1 \ldots w_{i-1} \,\textcolor{blue}{\boldsymbol{c}}\, w_{i+1} \ldots w_h}$ & $(-1)^{L_O} a_{u_1 \ldots u_f}^{t_1 \ldots t_e} h_{v_1 \ldots v_g} n_{w_1 \ldots w_{i-1} w_{i+1} \ldots w_h} h_{\textcolor{blue}{\boldsymbol{c}}}$\\[10pt]
					& $a_{u_1 \ldots u_f}^{t_1 \ldots t_e} h_{v_1 \ldots v_{i-1} \,\textcolor{blue}{\boldsymbol{c}}\, v_{i+1} \ldots v_g} n_{w_1 \ldots w_h}$ & $(-1)^{L_O} a_{u_1 \ldots u_f}^{t_1 \ldots t_e} h_{v_1 \ldots v_{i-1} v_{i+1} \ldots v_g} n_{w_1 \ldots w_h} n_{\textcolor{blue}{\boldsymbol{c}}}$ \\[10pt]
					& $a_{u_1 \ldots u_f}^{t_1 \ldots t_{i-1} \,\textcolor{blue}{\boldsymbol{c}}\, t_{i+1} \ldots t_e} h_{v_1 \ldots v_g} n_{w_1 \ldots w_h}$ & $(-1)^{i+1} a_{u_1 \ldots u_f}^{t_1 \ldots t_{i-1} t_{i+1} \ldots t_e} h_{v_1 \ldots v_g} n_{w_1 \ldots w_h} a_{\textcolor{blue}{\boldsymbol{c}}}$\\[10pt]
					& $a_{u_1 \ldots u_{i-1} \,\textcolor{blue}{\boldsymbol{c}}\, u_{i+1} \ldots u_f}^{t_1 \ldots t_e} h_{v_1 \ldots v_g} n_{w_1 \ldots w_h}$ & $(-1)^{L_O + i} a_{u_1 \ldots u_{i-1} u_{i+1} \ldots u_f}^{t_1 \ldots t_e} h_{v_1 \ldots v_g} n_{w_1 \ldots w_h} a_{\textcolor{blue}{\boldsymbol{c}}}^\dagger$ \\[20pt]
					\multirow{5}{*}[-20pt]{$a_{\textcolor{blue}{\boldsymbol{c}}}^\dagger a_{\textcolor{red}{\boldsymbol{d}}}^\dagger + a_{\textcolor{red}{\boldsymbol{d}}} a_{\textcolor{blue}{\boldsymbol{c}}}$\footnotemark[2]} & $a_{u_1 \ldots u_f}^{t_1 \ldots t_e} h_{v_1 \ldots v_g} n_{w_1 \ldots w_h}$ & $a_{u_1 \ldots u_f}^{t_1 \ldots t_e} h_{v_1 \ldots v_g} n_{w_1 \ldots w_h}$\\[10pt]
					& $a_{u_1 \ldots u_f}^{t_1 \ldots t_e} h_{v_1 \ldots v_g} n_{w_1 \ldots w_{i-1} \,\textcolor{blue}{\boldsymbol{c}}\, w_{i+1} \ldots w_h}$ & $a_{u_1 \ldots u_f}^{t_1 \ldots t_e} h_{v_1 \ldots v_g} n_{w_1 \ldots w_{i-1} w_{i+1} \ldots w_h} h_{\textcolor{red}{\boldsymbol{d}}}$\\[10pt]
					& $a_{u_1 \ldots u_f}^{t_1 \ldots t_e} h_{v_1 \ldots v_{i-1} \,\textcolor{blue}{\boldsymbol{c}}\, v_{i+1} \ldots v_g} n_{w_1 \ldots w_h}$ & $a_{u_1 \ldots u_f}^{t_1 \ldots t_e} h_{v_1 \ldots v_{i-1} v_{i+1} \ldots v_g} n_{w_1 \ldots w_h} n_{\textcolor{red}{\boldsymbol{d}}}$ \\[10pt]
					& $a_{u_1 \ldots u_f}^{t_1 \ldots t_{i-1} \,\textcolor{blue}{\boldsymbol{c}}\, t_{i+1} \ldots t_e} h_{v_1 \ldots v_g} n_{w_1 \ldots w_h}$ & $(-1)^{L_O+i+k+1} \I a_{u_1 \ldots u_f}^{t_1 \ldots t_{i-1} t_{i+1} \ldots t_e} h_{v_1 \ldots v_g} n_{w_1 \ldots w_h} a_{\textcolor{red}{\boldsymbol{d}}}$\\[10pt]
					& $a_{u_1 \ldots u_{i-1} \,\textcolor{blue}{\boldsymbol{c}}\, u_{i+1} \ldots u_f}^{t_1 \ldots t_e} h_{v_1 \ldots v_g} n_{w_1 \ldots w_h}$ & $(-1)^{k + i + 1} \I a_{u_1 \ldots u_{i-1} u_{i+1} \ldots u_f}^{t_1 \ldots t_e} h_{v_1 \ldots v_g} n_{w_1 \ldots w_h} a_{\textcolor{red}{\boldsymbol{d}}}^\dagger$ \\[20pt]
					\multirow{5}{*}[-20pt]{$a_{\textcolor{blue}{\boldsymbol{c}}}^\dagger a_{\textcolor{red}{\boldsymbol{d}}} + a_{\textcolor{red}{\boldsymbol{d}}}^\dagger a_{\textcolor{blue}{\boldsymbol{c}}}$\footnotemark[2]} & $a_{u_1 \ldots u_f}^{t_1 \ldots t_e} h_{v_1 \ldots v_g} n_{w_1 \ldots w_h}$ & $a_{u_1 \ldots u_f}^{t_1 \ldots t_e} h_{v_1 \ldots v_g} n_{w_1 \ldots w_h}$\\[10pt]
					& $a_{u_1 \ldots u_f}^{t_1 \ldots t_e} h_{v_1 \ldots v_g} n_{w_1 \ldots w_{i-1} \,\textcolor{blue}{\boldsymbol{c}}\, w_{i+1} \ldots w_h}$ & $a_{u_1 \ldots u_f}^{t_1 \ldots t_e} h_{v_1 \ldots v_g} n_{w_1 \ldots w_{i-1} w_{i+1} \ldots w_h} n_{\textcolor{red}{\boldsymbol{d}}}$\\[10pt]
					& $a_{u_1 \ldots u_f}^{t_1 \ldots t_e} h_{v_1 \ldots v_{i-1} \,\textcolor{blue}{\boldsymbol{c}}\, v_{i+1} \ldots v_g} n_{w_1 \ldots w_h}$ & $a_{u_1 \ldots u_f}^{t_1 \ldots t_e} h_{v_1 \ldots v_{i-1} v_{i+1} \ldots v_g} n_{w_1 \ldots w_h} h_{\textcolor{red}{\boldsymbol{d}}}$ \\[10pt]
					& $a_{u_1 \ldots u_f}^{t_1 \ldots t_{i-1} \,\textcolor{blue}{\boldsymbol{c}}\, t_{i+1} \ldots t_e} h_{v_1 \ldots v_g} n_{w_1 \ldots w_h}$ & $(-1)^{L_O+i+k+1} \I a_{u_1 \ldots u_f}^{t_1 \ldots t_{i-1} t_{i+1} \ldots t_e} h_{v_1 \ldots v_g} n_{w_1 \ldots w_h} a_{\textcolor{red}{\boldsymbol{d}}}^\dagger$\\[10pt]
					& $a_{u_1 \ldots u_{i-1} \,\textcolor{blue}{\boldsymbol{c}}\, u_{i+1} \ldots u_f}^{t_1 \ldots t_e} h_{v_1 \ldots v_g} n_{w_1 \ldots w_h}$ & $(-1)^{k + i + 1} \I a_{u_1 \ldots u_{i-1} u_{i+1} \ldots u_f}^{t_1 \ldots t_e} h_{v_1 \ldots v_g} n_{w_1 \ldots w_h} a_{\textcolor{red}{\boldsymbol{d}}}$ \\
				\end{tabular}
			\end{ruledtabular}
	}}
	\footnotetext[1]{The analytic expressions for cases in which $a_c^\dagger a_d^\dagger + a_d a_c$ or $a_c^\dagger a_d + a_d^\dagger a_c$ share two common indices with the string $O$ can be obtained by performing the transformation in two steps, each step involving one common index.}
	\footnotetext[2]{If the common index is $d$ instead of $c$, the sign of any expression whose sign factor depends on $k$ must be inverted.}
	\label{table_hermitian}
\end{table*}

Next, we move on to the more interesting case where $O$ has a common index, \foreign{e.g.}, $c$, with the generator of the Clifford transformation. In this scenario, transformations generated by half-body operators, $a_c^\dagger \pm a_c$, perform particle--hole conjugation, essentially returning a string in which the operator with the common index is replaced by its particle--hole conjugate, namely, $a_c \rightarrow a_c^\dagger$, $n_c \rightarrow h_c$, and vice versa.
When the generator is a single excitation operator, \foreign{e.g.}, $a_c^\dagger a_d \pm a_d^\dagger a_c$, the Clifford transformation produces a string in which the common index $c$ is replaced by the unique index $d$ appearing in the generator without modifying the operator type, namely, $a_c \rightarrow a_d$, $n_c \rightarrow n_d$, \foreign{etc}.
For example, these Clifford unitaries can be employed to perform a spin-flip transformation, exchanging spin-up operators by their spin-down counterparts and vice versa.
Such transformations are generated by operators of the form $a_{C\uparrow}^\dagger a_{C\downarrow} \pm a_{C\downarrow}^\dagger a_{C\uparrow}$, where $C$ denotes the $C$th spatial orbital, and $\uparrow$ and $\downarrow$ denote the $s_z = \frac{1}{2}$ and $s_z = -\frac{1}{2}$ spin states, respectively.
As shown in \cref{table_antihermitian,table_hermitian}, a Clifford transformation based on the pair creation/annihilation operator $a_c^\dagger a_d^\dagger \pm a_d a_c$  is identical, up to a phase factor, to the combined effect of Clifford transformations generated by $a_c^\dagger \pm a_c$ and $a_c^\dagger a_d \pm a_d^\dagger a_c$.
To be precise, the Clifford transformation generated by $a_c^\dagger a_d^\dagger \pm a_d a_c$ replaces the common index $c$ by the unique index $d$ and also inverts the particle--hole character of the operator, namely, $a_c \rightarrow a_d^\dagger$, $n_c \rightarrow h_d$, \foreign{etc}.
For example, when the Clifford unitary takes the form $a_{C\uparrow}^\dagger a_{C\downarrow}^\dagger \pm a_{C\downarrow} a_{C\uparrow}$, the transformation simultaneously inverts both the spin and the particle--hole character of the operator with the common index.

Since the pair operators have even length, they commute with operators that have no common indices.
As a result, the analytic expressions of Clifford transformations in which the string $O$ and the pair generator share two indices can be readily obtained by expressing them as two separate transformations, each having only one common index.
Qualitatively, when the generator is a single-excitation operator, the Clifford transformation, up to a phase factor, simply exchanges the two common indices.
The corresponding Clifford transformation relying on the pair creation/annihilation generator exchanges the two common indices and also inverts their particle--hole character.

Further insights into the mechanism of fermionic Clifford transformations can be gained by translating them to the language of Majorana operators, using the relations
\begin{subequations}\label{eq:f2m}
	\begin{align}
	a_p^\dagger &= \frac{1}{2} (\gamma_1^{(p)} - \I\gamma_2^{(p)})\\
	a_p &= \frac{1}{2} (\gamma_1^{(p)} + \I\gamma_2^{(p)})\\
	n_p &= \frac{1}{2} (I + \I\gamma_1^{(p)}\gamma_2^{(p)})\\
	h_p &= \frac{1}{2} (I - \I\gamma_1^{(p)}\gamma_2^{(p)}).
	\end{align}
\end{subequations}
As a result, a fermionic string $O$ with length $L_O$ will give rise to a linear combination of $2L_O$ Majorana strings.
Without loss of generality, we focus on the case where the generators share one common index, namely, $c$, with the string to be transformed, becoming
\begin{subequations}\label{eq:f2m_half}
	\begin{align}
	a_c^\dagger - a_c &= -\I \gamma_2^{(c)}\\
	a_c^\dagger + a_c &= \gamma_1^{(c)}
	\end{align}
\end{subequations}
for the half-body operators,
\begin{subequations}\label{eq:f2m_single}
	\begin{align}
	a_c^\dagger a_d - a_d^\dagger a_c &= \frac{1}{2} (\gamma_1^{(c)}\gamma_1^{(d)} + \gamma_2^{(c)} \gamma_2^{(d)})\\
	a_c^\dagger a_d + a_d^\dagger a_c &= \frac{1}{2} \I (\gamma_1^{(c)}\gamma_2^{(d)} - \gamma_2^{(c)} \gamma_1^{(d)})
	\end{align}
\end{subequations}
for the single-excitation operators, and
\begin{subequations}\label{eq:f2m_pair}
	\begin{align}
	a_c^\dagger a_d^\dagger - a_d a_c &= \frac{1}{2} (\gamma_1^{(c)}\gamma_1^{(d)} - \gamma_2^{(c)} \gamma_2^{(d)})\\ a_c^\dagger a_d^\dagger + a_d a_c &= -\frac{1}{2} \I (\gamma_1^{(c)} \gamma_2^{(d)} + \gamma_2^{(c)} \gamma_1^{(d)})
	\end{align}
\end{subequations}
in the case of the pair creation/annihilation operators.
To ensure that the Clifford transformation will produce a new fermionic string, we focus on angles of the form $(2k+1)\frac{\pi}{2}$, $k\in\mathbb{Z}$.
From \cref{eq:f2m_half,eq:f2m_single,eq:f2m_pair} and the fact that $(2k+1)\frac{\pi}{2}\in l \frac{\pi}{4}$, with $k,l\in\mathbb{Z}$, one can readily obtain that fermionic Clifford unitaries will be mapped to Majorana Cliffords.

We begin by examining the Clifford transformations generated by half-body operators.
For the chosen values of $\theta$, the Clifford transformations in the Majorana space invert the sign of the Majorana string to be transformed (see \cref{fig:majorana_rotation}), unless they are trivial.
As shown in Section S2 of the \sm, whether two Majorana strings commute or anticommute depends on their lengths and the number of common operators that they have.
Since half-body operators have unit length and since there is only one common index, the transformation will be trivial for half of the $2L_O$ Majorana strings while inverting the sign of the other half.
The overall result is the replacement of $\gamma_1^{(c)}\pm\I\gamma_2^{(c)}$ by $\gamma_1^{(c)}\mp\I\gamma_2^{(c)}$ and $(I \pm \I \gamma_1 \gamma_2)$ by $(I \mp \I \gamma_1 \gamma_2)$, effectively performing a particle--hole conjugation in the fermionic space.
For example, consider the Clifford transformation of $a_c^\dagger a_b$ generated by $a_c^\dagger - a_c$, namely,
\begin{equation}\label{eq:example_half_f}
	e^{-(2k+1)\frac{\pi}{2} (a_c^\dagger - a_c)} a_c^\dagger a_b e^{(2k+1)\frac{\pi}{2} (a_c^\dagger - a_c)}.
\end{equation}
Using \cref{eq:f2m,eq:f2m_half}, we express \cref{eq:example_half_f} in the Majorana representation, arriving at
\begin{equation}\label{eq:example_half_m}
	\begin{split}
	e^{(2k+1)\frac{\pi}{2}\I\gamma_2^{(c)}}\frac{1}{4}\left(\gamma_1^{(c)}\gamma_1^{(b)} + \I \gamma_1^{(c)} \gamma_2^{(b)}\right.\\
	\left. -\I\gamma_2^{(c)} \gamma_1^{(b)} + \gamma_2^{(c)} \gamma_2^{(b)} \right)e^{-(2k+1)\frac{\pi}{2}\I\gamma_2^{(c)}}.
	\end{split}
\end{equation}
Since $[\gamma_2^{(c)}, \gamma_1^{(c)}\gamma_1^{(b)}] = [\gamma_2^{(c)}, \gamma_1^{(c)} \gamma_2^{(b)}] = 0$, the transformation of the first two Majorana strings in \cref{eq:example_half_m} is trivial.
The remaining two Majorana strings anticommute with the generator $\gamma_2^{(c)}$, and the Clifford transformation inverts their sign since $\theta=(2k+1)\frac{\pi}{2}$ (see \cref{fig:majorana_rotation}).
As a result, the transformed expression is
\begin{equation}
	\frac{1}{4}\left(\gamma_1^{(c)}\gamma_1^{(b)} + \I \gamma_1^{(c)} \gamma_2^{(b)} +\I\gamma_2^{(c)} \gamma_1^{(b)} - \gamma_2^{(c)} \gamma_2^{(b)}\right),
\end{equation}
whose fermionic representation is $a_c a_b$.

Now, let us consider the case involving pair generators.
One can readily verify that the Majorana strings arising from these pair generators commute among themselves.
This general result follows directly from the observation that each Majorana string is uniquely mapped to a single Pauli string, and Pauli strings arising from single (anti-)Hermitian fermionic operators pairwise commute \cite{Romero.2019.10.1088/2058-9565/aad3e4}.
Thus, the Clifford transformation of a single fermionic string can be expressed as a sequence of two unitary transformations of $2L_O$ Majorana strings.
Due to the additional numerical prefactor of $\tfrac{1}{2}$ appearing in the Majorana generators, these Clifford transformations will introduce new Majorana strings, unless they are trivial.
These transformations lead to the replacement of $\gamma_1^{(c)}$ by $\gamma_1^{(d)}$ and $\gamma_2^{(c)}$ by $\gamma_2^{(d)}$ (index swap) for single-excitation generators, and similarly, it replaces $\gamma_1^{(c)}$ by $\gamma_2^{(d)}$ and $\gamma_2^{(c)}$ by $\gamma_1^{(d)}$ (index swap and particle--hole conjugation) for pair annihilation/creation generators.
As an illustration, we consider the Clifford transformation of $a_c^\dagger a_b$ generated by $a_c^\dagger a_d^\dagger - a_d a_c$, namely,
\begin{equation}
	e^{-(2k+1)\frac{\pi}{2} (a_c^\dagger a_d^\dagger - a_d a_c)} a_c^\dagger a_b e^{(2k+1)\frac{\pi}{2} (a_c^\dagger a_d^\dagger - a_d a_c)}.
\end{equation}
The Majorana representation of the above transformation reads
\begin{equation}
	\begin{split}
	e^{-(2k+1)\frac{\pi}{4} \gamma_1^{(c)} \gamma_1^{(d)}} e^{(2k+1)\frac{\pi}{4} \gamma_2^{(c)} \gamma_2^{(d)}} \frac{1}{4}\left(\gamma_1^{(c)}\gamma_1^{(b)} + \I \gamma_1^{(c)} \gamma_2^{(b)}\right.\\
	\left.-\I\gamma_2^{(c)} \gamma_1^{(b)} + \gamma_2^{(c)} \gamma_2^{(b)} \right) e^{-(2k+1)\frac{\pi}{4} \gamma_2^{(c)} \gamma_2^{(d)}} e^{(2k+1)\frac{\pi}{4} \gamma_1^{(c)} \gamma_1^{(d)}}.
	\end{split}
\end{equation}
The first sequence of Clifford transformations is generated by $\gamma_2^{(c)} \gamma_2^{(d)}$, and leaves the first two Majorana strings unchanged while replacing $\gamma_2^{(c)}$ by $\gamma_2^{(d)}$ in the last two, yielding
\begin{equation}
	\begin{split}
	e^{-(2k+1)\frac{\pi}{4} \gamma_1^{(c)} \gamma_1^{(d)}} \frac{1}{4} \left(\gamma_1^{(c)}\gamma_1^{(b)} + \I \gamma_1^{(c)} \gamma_2^{(b)}\right.\\
	\left. -(-1)^{k+1}\I\gamma_2^{(d)} \gamma_1^{(b)} + (-1)^{k+1}\gamma_2^{(d)} \gamma_2^{(b)} \right) e^{(2k+1)\frac{\pi}{4} \gamma_1^{(c)} \gamma_1^{(d)}}.
	\end{split}
\end{equation}
In the second sequence of Clifford transformations, generated by $\gamma_1^{(c)} \gamma_1^{(d)}$, $\gamma_1^{(c)}$ is replaced by $\gamma_1^{(d)}$ in the first two Majorana strings while the last two remain unchanged.
As a result, the outcome of the transformation reads
\begin{equation}
	(-1)^{k}\frac{1}{4}\left(\gamma_1^{(d)}\gamma_1^{(b)} + \I \gamma_1^{(d)} \gamma_2^{(b)} + \I\gamma_2^{(d)} \gamma_1^{(b)} - \gamma_2^{(d)} \gamma_2^{(b)} \right).
\end{equation}
Expressing the above result in the language of fermions yields $(-1)^{k} a_d a_b$.

So far, we have shown that the unitaries $\exp[k\frac{\pi}{2} (a_p^\dagger - a_p)]$, $\exp[\I k\frac{\pi}{2} (a_p^\dagger + a_p)]$, $\exp[k\frac{\pi}{2} (a_q^\dagger a_p - a_p^\dagger a_q)]$, $\exp[\I k\frac{\pi}{2} (a_q^\dagger a_p + a_p^\dagger a_q)]$, $\exp[k\frac{\pi}{2} (a_p^\dagger a_q^\dagger - a_q a_p)]$, $\exp[\I k\frac{\pi}{2} (a_p^\dagger a_q^\dagger + a_q a_p)]$, and $\exp(\I \theta n_p)$, with $k\in\mathbb{Z}$, are elements of the fermionic Clifford group, denoted as $\mathcal{C}_{\mathcal{F}_M}$.
In fact, all fermionic Clifford unitaries can be expressed as products of the aforementioned gates.
To show this, we start by proving that fermionic Clifford transformations preserve the many-body rank of the string that is transformed.
For example, let as assume that the fermionic strings $a_p^\dagger a_q$ and $a_r^\dagger a_s^\dagger a_u a_t$ are connected via a Clifford transformation.
The Frobenius norms of the aforementioned operators in the $2^M$-dimensional Fock space generated by $M$ single-particle states are
\begin{equation}
	\begin{split}
	\norm{a_p^\dagger a_q}_F^2 &= \Tr(a_p^\dagger a_q a_q^\dagger a_p)\\
	&= \Tr(n_p h_q)\\
	&= 2^{M-2}
	\end{split}
\end{equation}
for $a_p^\dagger a_q$ and
\begin{equation}
	\begin{split}
	\norm{a_r^\dagger a_s^\dagger a_u a_t}_F^2 &= \Tr(a_r^\dagger a_s^\dagger a_u a_t a_t^\dagger a_u^\dagger a_s a_r)\\
	&= \Tr(n_r n_s h_t h_u)\\
	&= 2^{M-4}
	\end{split}
\end{equation}
in the case of $a_r^\dagger a_s^\dagger a_u a_t$.
However, Clifford transformations are unitary, and, thus, norm-preserving.
As a result, Clifford transformations cannot modify the many-body rank of the string that is transformed.
Fermionic strings of the same many-body rank can differ in their excitation indices (index swap) and/or the number of creation and annihilation operators (particle--hole conjugation).
These are exactly the types of transformations induced by Clifford unitaries generated by the half-body and pair operators.
We thus conclude that the entire Clifford group is comprised of the aforementioned Clifford unitaries and their products.
After inspecting \cref{table_antihermitian,table_hermitian} and considering the action of $\exp(\I\theta n_p)$, a generating set of the fermionic Clifford group is
\begin{equation}
	\mathcal{C}_{\mathcal{F}_M} = \langle e^{\I \theta}I, e^{\I \theta n_p}, e^{k\frac{\pi}{2}(a_p^\dagger - a_p)}, e^{k\frac{\pi}{2}(a_p^\dagger a_q - a_q^\dagger a_p)} \rangle.
\end{equation}

At this point, it is worth noting that, by preserving the many-body rank of the fermionic string to be transformed, fermionic Clifford transformations respect fermionic parity, defined as
\begin{equation}\label{f_parity}
	\Pi = (-1)^{\sum_k n_k},
\end{equation}
but may break particle number symmetry.
Consequently, applying Clifford transformations to a fermionic Hamiltonian yields a similarity-transformed Hamiltonian with the same number of terms, each preserving its original many-body rank.
However, when the Clifford unitaries involve half-body and/or pair-annihilation/creation operators, the transformed Hamiltonian no longer conserves the total particle number, as it includes terms that add or remove two or four electrons.
To illustrate this, we consider the Clifford transformation of the one-body Hamiltonian $\mathcal{H} = \alpha n_p + \beta n_q + \gamma (a_p^\dagger a_q + a_q^\dagger a_p)$, with $\alpha,\beta,\gamma \in \mathbb{R}$, generated by $a_p^\dagger - a_p$, namely,
	\begin{equation}
		e^{-k\frac{\pi}{2} (a_p^\dagger - a_p)} \left[\alpha n_p + \beta n_q + \gamma (a_p^\dagger a_q + a_q^\dagger a_p)\right] e^{k\frac{\pi}{2} (a_p^\dagger - a_p)}.
	\end{equation}
	We first notice that the $n_q$ term has even length and does not contain the index $p$, and, thus, its Clifford transformation is trivial.
	One can show that the remaining terms in $\mathcal{H}$ satisfy $A[O,A]A = [O,A]$, so we use $\alpha = 4$ in \cref{fST_antihermitian}.
	For $\theta = k\frac{\pi}{2}$, $k \in \mathbb{Z}$, the Clifford transformation of $n_p$ yields
	\begin{equation}
		\begin{split}
		e^{-k\frac{\pi}{2} (a_p^\dagger - a_p)} n_p e^{k\frac{\pi}{2} (a_p^\dagger - a_p)} ={}& \left[ 1 - \frac{1+(-1)^{k+1}}{2} \right] n_p\\
		&+ \frac{1+(-1)^{k+1}}{2} h_p,
		\end{split}
	\end{equation}
	while the transformations of the off-diagonal terms $a_p^\dagger a_q$ and $a_q^\dagger a_p$ read
	\begin{equation}
		\begin{split}
		e^{-k\frac{\pi}{2} (a_p^\dagger - a_p)} a_p^\dagger a_q e^{k\frac{\pi}{2} (a_p^\dagger - a_p)} ={}& \left[ 1 - \frac{1+(-1)^{k+1}}{2} \right] a_p^\dagger a_q\\
		&+ \frac{1+(-1)^{k+1}}{2} a_p a_q
		\end{split}
	\end{equation}
	and
	\begin{equation}
		\begin{split}
		e^{-k\frac{\pi}{2} (a_p^\dagger - a_p)} a_q^\dagger a_p e^{k\frac{\pi}{2} (a_p^\dagger - a_p)} ={}& \left[ 1 - \frac{1+(-1)^{k+1}}{2} \right] a_q^\dagger a_p\\
		&+ \frac{1+(-1)^{k+1}}{2} a_q^\dagger a_p^\dagger,
		\end{split}
	\end{equation}
	respectively.
	Therefore, when $k$ is even the Hamiltonian remains unchanged, whereas for odd values of $k$ it transforms to
	\begin{equation}
		\begin{split}
		e^{-(2k+1)\frac{\pi}{2} (a_p^\dagger - a_p)} \mathcal{H} e^{(2k+1)\frac{\pi}{2} (a_p^\dagger - a_p)} ={}& \alpha h_p + \beta n_q\\
		&+ \gamma (a_p a_q + a_q^\dagger a_p^\dagger).
		\end{split}
	\end{equation}
	As was anticipated, although the particle--hole conjugation induced by the Clifford transformation did not change the number of terms or their many-body rank, the transformed Hamiltonian does not conserve the number of electrons, containing terms that remove ($a_p a_q$) and add ($a_q^\dagger a_p^\dagger$) two electrons.

Before we conclude this section, we highlight the major differences between the fermionic Clifford unitaries and their Pauli and Majorana counterparts.
To begin with, the generators of Pauli and Majorana Cliffords can have arbitrary length, whereas in the fermionic case they are up to one-body.
On the same note, although Pauli and Majorana Clifford transformations can modify the length of the transformed string, their fermionic counterparts are rank-preserving.
While the rotation angles are strictly fixed for Pauli and Majorana Clifford unitaries, in the case of fermions, $\exp(\I \theta n_p)$ is Clifford for arbitrary values of the angle $\theta$.
Specifically, as shown in Section S3 of the \sm, the transformed fermionic string $O$ is multiplied by $\exp(\I\theta)$ if $O$ contains a single $a_p$ operator, multiplied by $\exp(-\I\theta)$ if it contains a single $a_p^\dagger$ operator, and remains unchanged otherwise.
In addition, using the JW mapping, one readily obtains that the $Z_p = \exp(\I\pi n_p)$, $S\equiv\sqrt{Z} = \exp(\I\frac{\pi}{2}n_p)$, and $T\equiv\sqrt[4]{Z}=\exp(\I\frac{\pi}{4}n_p)$ gates are Clifford, mapping one fermionic string to another.
This is quite intriguing since the $T$ gate is not Clifford in the case of Pauli and Majorana strings.
Nevertheless, one can show that the fermionic representation of the $H$ [\cref{eq:clifford_h}] and CNOT [\cref{eq:clifford_cnot}] gates are not Clifford, in contrast to their Pauli and Majorana counterparts.
Finally, we showed that even fermionic Clifford transformations generated by half-body operators do not violate fermionic parity.
However, this is not true for odd-length Majorana strings, which, in the non-trivial case, generate Clifford transformations that map even-length Majorana strings to odd-length ones and vice versa.

\section{Fermionic Clifford Transformations and Qubit Tapering}\label{sec:tapering}

The ability to transform between Pauli, Majorana, and fermionic operators allows one to select the representation in which a given problem simplifies or symmetries become more evident.
A recent example of the latter is qubit tapering \cite{Bravyi.2017.1701.08213,Setia.2020.10.1021/acs.jctc.0c00113}, a technique that exploits $\mathbb{Z}_2$ symmetries of the Hamiltonian to eliminate qubits from quantum simulations of many-body systems.
In what follows, we provide a bird's-eye view of qubit tapering, emphasizing the connection to fermionic Clifford transformations discussed in the previous section.

Before one is able to use a quantum device to simulate a many-fermion system, the problem must be translated from second quantization into the qubit representation through a suitable fermionic encoding scheme.
In doing so, the Hamiltonian takes the form of a linear combination of Pauli strings.
If the Hamiltonian acted trivially, \foreign{i.e.}, with the identity operator, on a subset of qubits, these qubits could be removed from the computation without introducing any approximation.
Another scenario allowing the exact elimination of qubits  occurs when the Hamiltonian acts on a subset of qubits with at most a single type of Pauli gate, \foreign{e.g.}, $\sigma_x$, since these gates would commute among themselves and with the Hamiltonian.
By expressing the Hamiltonian in the common eigenbasis of these Pauli gates,
they are replaced by their eigenvalues and the corresponding qubits are dropped from the simulation.
The above observation forms the core philosophy behind qubit tapering, which, on a high level, can be viewed as a two-step process.
In the first step, one finds the maximal abelian subgroup $\mathcal{S}$, $-I\notin\mathcal{S}$, of the Pauli group $\mathcal{P}_M$ whose elements commute with the Hamiltonian (the requirement $-I\notin\mathcal{S}$ ensures that all elements of $\mathcal{S}$ are Hermitian).
Since Pauli strings, as  Hermitian involutions,  have eigenvalues of $\pm1$, $\mathcal{S}$ is known as the $\mathbb{Z}_2$ symmetry group of the Hamiltonian.
In the second step, one exploits these symmetries to construct Clifford unitaries that transform the Hamiltonian such that it acts on a subset of qubits with, at most, a single type of Pauli gate.

\begin{figure}
\begin{tikzpicture}
	\draw (-2,0) -- (-1,0);
	\draw (1,0) -- (2,0);
	\draw[dashed] (-1,0) -- (-0.5,-0.75);
	\draw[dashed] (-1,0) -- (-0.5,1);
	\draw[dashed] (1,0) -- (0.5,-0.75);
	\draw[dashed] (1,0) -- (0.5,1);
	\draw (-0.5,-0.75) -- (0.5,-0.75);
	\draw (-0.5,1) -- (0.5,1);
	\draw[-Stealth] (-0.25,-0.75) -- (-0.25,-0.25);
	\draw[-Stealth] (0.25,-0.25) -- (0.25,-0.75);
	\node[below] at (0,-0.75) {$\ket{\sigma_g}$};
	\node[below] at (0,1) {$\ket{\sigma_u}$};
	\node[below] at (-1.5,0) {$\ket{1s}$};
	\node[below] at (1.5,0) {$\ket{1s}$};
	\node[anchor=base] at (-1.5,1.5) {H};
	\node[anchor=base] at (0,1.5) {\ce{H2}};
	\node[anchor=base] at (1.5,1.5) {H};
\end{tikzpicture}
\caption{Molecular orbital diagram of the \ce{H2} molecule in a minimum basis.}
\label{fig:h2_mo}
\end{figure}
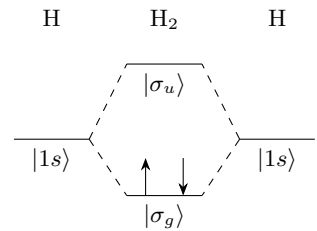

In Section S6 of the \sm, we have provided a detailed example of qubit tapering for the \ce{H2} molecule in a minimum basis set (MBS), first considered in Ref.\ \cite{Bravyi.2017.1701.08213}.
As illustrated in \cref{fig:h2_mo}, a restricted Hartree--Fock (RHF) calculation for \ce{H2}/MBS results in a doubly occupied molecular orbital of $\sigma_g$ symmetry and an unoccupied one of $\sigma_u$ symmetry.
The four spinorbitals defining the one-particle basis are denoted as $\ket{\psi_0}\equiv\ket{\sigma_g {\uparrow}}$, $\ket{\psi_1}\equiv\ket{\sigma_g {\downarrow}}$, $\ket{\psi_2}\equiv\ket{\sigma_u {\uparrow}}$, and $\ket{\psi_3}\equiv\ket{\sigma_u {\downarrow}}$.
In the occupation number representation, the various Slater determinants are expressed as $\ket{n_0,n_1,n_2,n_3}$, with $n_p$ designating the occupancy of spinorbital $\ket{\psi_p}$.
For example, $\ket{1100}$ denotes the RHF Slater determinant.
As $\sigma_g \leftrightarrow\sigma_u$ single excitations are forbidden, the unitary extension \cite{Kutzelnigg.1977.10.1007/978-1-4757-0887-5_5,Kutzelnigg.1982.10.1063/1.444231,
	Kutzelnigg.1983.10.1063/1.446313,Kutzelnigg.1984.10.1063/1.446736,Bartlett.1989.10.1016/S0009-2614(89)87372-5,
	Szalay.1995.10.1063/1.469641,Taube.2006.10.1002/qua.21198,Cooper.2010.10.1063/1.3520564,
	Evangelista.2011.10.1063/1.3598471,Harsha.2018.10.1063/1.5011033,Filip.2020.10.1063/5.0026141,
	Freericks.2022.10.3390/sym14030494,Anand.2022.10.1039/d1cs00932j} of coupled cluster theory \cite{Coester.1958.10.1016/0029-5582(58)90280-3,
	Coester.1960.10.1016/0029-5582(60)90140-1,Cizek.1966.10.1063/1.1727484,Cizek.1969.10.1002/9780470143599.ch2,
	Cizek.1971.10.1002/qua.560050402,Paldus.1972.10.1103/PhysRevA.5.50} with double excitations (UCCD) is exact in this case, allowing one to right the \ce{H2}/MBS Schr\"{o}dinger equation as
\begin{equation}\label{eq:SE_H2}
	\mathcal{H}e^{\theta A^{23}_{01}} \ket{1100} = E_0 e^{\theta A^{23}_{01}} \ket{1100},
\end{equation}
where $A^{23}_{01} \equiv a_2^\dagger a_3^\dagger a_1 a_0 - a_0^\dagger a_1^\dagger a_3 a_2$.
In second quantization, the electronic Hamiltonian of \ce{H2}/MBS is given by
\begin{equation}\label{eq:H2_sq_ham}
	\begin{split}
		\mathcal{H} ={}& h_0^0 \left( n_0 + n_1\right) + h_2^2 \left( n_2 + n_3\right) + 
		v_{01}^{01} n_0 n_1  + v_{23}^{23} n_2 n_3  \\
		&+ \left( v_{02}^{02} - v_{02}^{20} \right) \left(n_0 n_2 + n_1 n_3\right)  + v_{02}^{02} \left(n_0 n_3 + n_1 n_2\right)\\
		&+ v_{01}^{23} \left( a_{01}^{23} - a_{12}^{03} - a_{03}^{12} + a_{23}^{01} \right),
	\end{split}	
\end{equation}
with $h_p^q$ and $v_{pq}^{rs}$ denoting the one- and two-electron integrals, respectively.
Note that in writing \cref{eq:H2_sq_ham}, we took advantage of the following symmetries in the one- and two-electron integrals: $h_0^0 = h_1^1$, $h_2^2 = h_3^3$, $v_{02}^{02} = v_{20}^{20} = v_{03}^{03} = v_{30}^{30} = v_{12}^{12} = v_{21}^{21} = v_{13}^{13} = v_{31}^{31}$, and $v_{02}^{20} = v_{20}^{02} = v_{13}^{31} = v_{31}^{13} = v_{01}^{23} = v_{03}^{21} = v_{10}^{32} = v_{12}^{30} = v_{21}^{03} = v_{23}^{01} = v_{30}^{12} = v_{32}^{10}$.

Under the JW transformation, the electronic Hamiltonian shown in \cref{eq:H2_sq_ham} takes the following form in the qubit space:
\begin{equation}\label{eq:H2_q_ham}
	\begin{split}
		\mathcal{H} ={}& c_1 + c_2 \sigma_z^{(0)} + c_3 \sigma_z^{(1)} +  c_4 \sigma_z^{(2)} + c_5 \sigma_z^{(3)} \\
		&+ c_6 \sigma_z^{(0)} \sigma_z^{(1)} + c_7 \sigma_z^{(0)} \sigma_z^{(2)} + c_8 \sigma_z^{(0)} \sigma_z^{(3)}\\
		&+ c_9 \sigma_z^{(1)} \sigma_z^{(2)} + c_{10} \sigma_z^{(1)} \sigma_z^{(3)} + c_{11} \sigma_z^{(2)} \sigma_z^{(3)} \\
		&+ c_{12} \sigma_y^{(0)} \sigma_y^{(1)} \sigma_x^{(2)} \sigma_x^{(3)}
		+ c_{13} \sigma_y^{(0)} \sigma_x^{(1)} \sigma_x^{(2)} \sigma_y^{(3)}\\
		&+ c_{14} \sigma_x^{(0)} \sigma_y^{(1)} \sigma_y^{(2)} \sigma_x^{(3)}
		+ c_{15} \sigma_x^{(0)} \sigma_x^{(1)} \sigma_x^{(2)} \sigma_y^{(3)},
	\end{split}
\end{equation}
with the definitions of the coefficients in terms of one- and two-electron integrals given Table S1 in the \sm.
A quick inspection of \cref{eq:H2_q_ham} reveals that, in its current form, the Hamiltonian does not act on any of the four qubits with at most a single kind of Pauli gate.
Using the qubit tapering algorithm, the generators of the $\mathbb{Z}_2$ symmetry group of the \ce{H2}/MBS Hamiltonian
are \cite{Bravyi.2017.1701.08213}
\begin{equation}
	\mathcal{S}=\left\langle \sigma_z^{(0)}\sigma_z^{(1)}, \sigma_z^{(0)}\sigma_z^{(2)}, \sigma_z^{(0)}\sigma_z^{(3)} \right\rangle.
\end{equation}

To gain deeper physical insights into the $\mathbb{Z}_2$ symmetries revealed by  qubit tapering ,  we now employ the inverse JW transformation to translate  the generators of the $\mathbb{Z}_2$ symmetry group into the language of second quantization.
In doing so, we obtain
\begin{equation}
	\mathcal{S} = \left\langle (-1)^{n_0+n_1}, (-1)^{n_0+n_2}, (-1)^{n_0+n_3} \right\rangle,
\end{equation}
where $n_0$, $n_1$, $n_2$, and $n_3$ denote particle number operators.
Consequently, in addition to the fermionic parity [\cref{f_parity}], which is always conserved for all fermionic Hamiltonians and does not allow the coupling of states whose total particle numbers differ in even/odd parity, the electronic Hamiltonian of \ce{H2}/MBS also preserves the parities of the $\ket{n_0 n_1 \ldots}$, $\ket{n_0 \ldots n_2 \ldots}$, and $\ket{n_0 \ldots n_3}$ subspaces of the Fock space.
As shown in \cref{table_fock}, depending on the parities of the individual subspaces, the 16-dimensional Fock space of \ce{H2}/MBS can be expressed as the direct sum of eight 2-dimensional Hilbert spaces.
\begin{table*}
	\caption{
		Decomposition of the  Fock space minimum-basis \ce{H2} into 2-dimensional Hilbert spaces, resulting from the qubit tapering algorithm.
	}
	\begin{ruledtabular}
		\begin{tabular}{cc}
			$\left( (-1)^{n_0+n_1}, (-1)^{n_0+n_2}, (-1)^{n_0+n_3}\right)$ & Hilbert Space\\
			\colrule
			$({+}{+}{+})$ & $\Span(\{\ket{0000}, \ket{1111}\})$\\
			$({-}{+}{+})$ & $\Span(\{\ket{0100}, \ket{1011}\})$\\
			$({+}{-}{+})$ & $\Span(\{\ket{0010}, \ket{1101}\})$\\
			$({+}{+}{-})$ & $\Span(\{\ket{1110}, \ket{0001}\})$\\
			$({-}{-}{+})$ & $\Span(\{\ket{0110}, \ket{1001}\})$\\
			$({-}{+}{-})$ & $\Span(\{\ket{1010}, \ket{0101}\})$\\
			$({+}{-}{-})$ & $\Span(\{\ket{1100}, \ket{0011}\})$\\
			$({-}{-}{-})$ & $\Span(\{\ket{1000}, \ket{0111}\})$\\
		\end{tabular}
	\end{ruledtabular}
	\label{table_fock}
\end{table*}

Next, we examine the Pauli Clifford unitaries arising from the application of the qubit tapering technique to \ce{H2}/MBS.
As shown in Ref.\ \cite{Bravyi.2017.1701.08213}, to taper off the last three qubits, the pertinent Clifford gates are
\begin{equation}
	U_1 = \frac{1}{\sqrt{2}} (\sigma_x^{(1)} + \sigma_z^{(0)} \sigma_z^{(1)}),
\end{equation}
\begin{equation}
	U_2 = \frac{1}{\sqrt{2}} (\sigma_x^{(2)} + \sigma_z^{(0)} \sigma_z^{(2)}),
\end{equation}
and
\begin{equation}
	U_3 = \frac{1}{\sqrt{2}} (\sigma_x^{(3)} + \sigma_z^{(0)} \sigma_z^{(3)}),
\end{equation}
while their representation in second quantization takes the form
\begin{equation}\label{eq:tapering_u1}
		U_1 = \frac{1}{\sqrt{2}} e^{\I \pi n_0}\left(e^{\I \pi n_1} -\I e^{\I\frac{\pi}{2} (a_1^\dagger + a_1)} \right),
\end{equation}
\begin{equation}\label{eq:tapering_u2}
	U_2 = \frac{1}{\sqrt{2}} e^{\I \pi n_0}\left(e^{\I \pi n_2} -\I e^{\I \pi n_1}e^{\I\frac{\pi}{2} (a_2^\dagger + a_2)} \right),
\end{equation}
and
\begin{equation}\label{eq:tapering_u3}
	U_3 = \frac{1}{\sqrt{2}} e^{\I \pi n_0}\left(e^{\I \pi n_3} -\I e^{\I \pi n_1}e^{\I \pi n_2}e^{\I\frac{\pi}{2} (a_3^\dagger + a_3)} \right),
\end{equation}
respectively (see \sm).
\cref{eq:tapering_u1,eq:tapering_u2,eq:tapering_u3} show that, under the inverse JW transformation, these Pauli Clifford unitaries are mapped to linear combinations of two products of fermionic Clifford unitaries, generated by number and half-body operators.
It is worth noting, however, that a linear combination of Clifford unitaries is not, in general, a Clifford unitary itself.
Thus, although the transformation of the Hamiltonian is Clifford in the Pauli space, mapping a given Pauli string to another, in second quantization the number of fermionic strings is not necessarily preserved.
Indeed, as shown in Section S6 of the \sm, as a result of the Pauli Clifford transformation, the number of fermionic strings in the \ce{H2}/MBS electronic Hamiltonian increases from 14 to 82.
Furthermore, the transformed Hamiltonian does not conserve the number of particles nor the fermionic parity.
This is due to the presence of half-body generators in the above unitaries and the fact that the transformation is not Clifford in the fermionic space.

Having found the Pauli Clifford unitary, $U = U_1 U_2 U_3$, one then transforms the \ce{H2}/MBS Schr\"{o}dinger equation.
To that end, we multiply \cref{eq:SE_H2} from the left by $U$ and insert the identity, expressed as $I = U^\dagger U$, on both sides of the wave operator, $\exp(\theta A_{01}^{23})$, obtaining
\begin{equation}
	\bar{\mathcal{H}} e^{\theta \bar{A}_{01}^{23}} U \ket{1100} = E_0 e^{\theta \bar{A}_{01}^{23}} U \ket{1100},
\end{equation}
with the overbar denoting similarity-transformed operators.
As  shown in Ref.\ \cite{Bravyi.2017.1701.08213}, in the qubit space, the transformed Hamiltonian $\bar{\mathcal{H}}$ of \ce{H2}/MBS acts on the last three qubits by, at most, a Pauli $\sigma_x$ gate.
By replacing these $\sigma_x$ gates with their corresponding eigenvalues, the original Hamiltonian acting in the 16-dimensional Fock space is replaced by eight Hamiltonians, each operating on one of the eight 2-dimensional Hilbert spaces listed in \cref{table_fock} (see, also, \sm).
It can be shown that under the JW mapping the action of the UCCD unitary on the RHF state of $\text{H}_2$/MBS simplifies to \cite{Hempel.2018.10.1103/PhysRevX.8.031022}:
\begin{equation}
	e^{\theta A_{01}^{23}} \ket{1100} \xrightarrow{\text{JW}} e^{\I \theta \sigma_y^{(0)} \sigma_x^{(1)} \sigma_x^{(2)} \sigma_x^{(3)}} \ket{1100}.
\end{equation}
As the UCCD unitary is already acting on qubits $q_1$--$q_3$ with at most $\sigma_x$ gates, its transformation by $U$ is trivial.
The Clifford unitary $U$ transforms the RHF Slater determinant, $\ket{1100}$, to the $\ket{1{+}{-}{-}}$ state, an eigenstate of the $\sigma_x^{(1)}$, $\sigma_x^{(2)}$, and $\sigma_x^{(3)}$ Pauli gates with eigenvalues $+1$, $-1$, and $-1$, respectively.
Considering that the RHF Slater determinant has the same symmetry properties as the $^1\Sigma_g^+$ ground electronic state of \ce{H2}, we focus on the $\bar{\mathcal{H}}^{({+}{-}{-})}$ Hamiltonian.

\begin{figure}
	\resizebox{\columnwidth}{!}{
	\begin{tikzpicture}
		\draw (-0.5,-0.5) -- (0.5,-0.5);
		\draw (-0.5,0.5) -- (0.5,0.5);
		\draw[-Stealth] (-0.25,-0.5) -- (-0.25,0);
		\draw[-Stealth] (0.25,0) -- (0.25,-0.5);
		\draw (2,-0.5) -- (3,-0.5);
		\draw (2,0.5) -- (3,0.5);
		\draw[-Stealth] (2.25,0.5) -- (2.25,1);
		\draw[-Stealth] (2.75,1) -- (2.75,0.5);
		\node[anchor=base] at (-1,-0.1) {$c_0$};
		\node[anchor=base] at (1,-0.1) {$+$};
		\node[anchor=base] at (1.5,-0.1) {$c_1$};
		\node[anchor=base] at (4,-0.1) {$\xrightarrow{\text{taper}}$};
		\node[anchor=base] at (5,-0.1) {$c_0$};
		\draw (5.5,-0.5) -- (6,-0.5);
		\draw[-Stealth] (5.75,-0.5) -- (5.75, 0);
		\node[anchor=base] at (6.5,-0.1) {$+$};
		\node[anchor=base] at (7,-0.1) {$c_1$};
		\draw (7.5,-0.5) -- (8,-0.5);
		\node[anchor=base] at (0,-1) {$\ket{1100}$};
		\node[anchor=base] at (2.5,-1) {$\ket{0011}$};
		\node[anchor=base] at (5.75,-1) {$\ket{1}$};
		\node[anchor=base] at (7.75,-1) {$\ket{0}$};
	\end{tikzpicture}
	}
	\caption{Schematic representation of the ground electronic state of the \ce{H2} molecule as described by a minimum basis set. Before tapering, the state is a linear combination of two Slater determinants, $\ket{\Psi} = c_0 \ket{1100} + c_1 \ket{0011}$, while after the tapering of the last three spinorbitals it becomes $\ket{\Psi} = c_0 \ket{1} + c_1 \ket{0}$.}
	\label{fig:h2_fci_taper}
\end{figure}
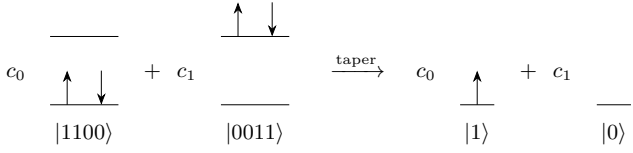
After removing the last three qubits and expressing the tapered Schr\"{o}dinger equation for \ce{H2}/MBS in second quantization, we obtain
\begin{equation}\label{tapered_SE_sq}
	\begin{split}
	\left[ h_0^0 + h_2^2 + \dfrac{1}{2}\left(v_{01}^{01} + v_{23}^{23}\right)\right.\\
	+ \left(h_2^2 - h_0^0 + \dfrac{1}{2} \left(v_{23}^{23} - v_{01}^{01}\right)\right)(I-2n)\\
	\left. \vphantom{\frac{1}{2}} + v_{01}^{23} (a^\dagger + a) \right] e^{- \theta (a^\dagger - a)} \ket{1} = E_0 e^{- \theta (a^\dagger - a)} \ket{1}.
	\end{split}
\end{equation}
A schematic illustration of the effect of the tapering procedure on the ground electronic state $\ket{\Psi}$ of \ce{H2}/MBS is provided in \cref{fig:h2_fci_taper}.
\cref{tapered_SE_sq} can also be written in terms of matrix elements of the original Hamiltonian as
\begin{equation}\label{eq:tapered_SE_sq_mel}
	\begin{split}
	\left[ \dfrac{1}{2}\left(\mathcal{H}_{00} + \mathcal{H}_{11}\right) + \dfrac{1}{2}\left(\mathcal{H}_{11} - \mathcal{H}_{00}\right)(I-2n)\right.\\
	\left. \vphantom{\frac{1}{2}}+ \mathcal{H}_{10}(a^\dagger + a) \right] e^{- \theta (a^\dagger - a)} \ket{1} = E_0 e^{- \theta (a^\dagger - a)} \ket{1},
	\end{split}
\end{equation}
with
\begin{equation}
	\mathcal{H}_{00}\equiv\mel{1100}{\mathcal{H}}{1100} = 2h_0^0 + v_{01}^{01},
\end{equation}
\begin{equation}
	\mathcal{H}_{11}\equiv\mel{0011}{\mathcal{H}}{0011} = 2h_2^2 + v_{23}^{23},
\end{equation}
and
\begin{equation}
	\mathcal{H}_{10}\equiv\mel{0011}{\mathcal{H}}{1100} = v_{01}^{23}.
\end{equation}
The presence of half-body operators in both the tapered Hamiltonian and UCCD unitary implies that neither of them conserve particle number and fermionic parity.
Nevertheless, the above problem is isospectral to that of \ce{H2}/MBS in the subspace spanned by the reference Slater determinant $\ket{1100}$ and the doubly excited Slater determinant $\ket{0011}$.
Thus, solution of \cref{eq:tapered_SE_sq_mel} will yield the exact energies in the chosen minimum basis for the $^1\Sigma_g^+$ ground electronic state of \ce{H2} and the $^1\Sigma_g^+$ excited state arising from the ionic channel (\ce{H+}+\ce{H-}).
It is interesting to note that \cref{eq:tapered_SE_sq_mel} can be naturally expressed in terms of Majorana operators, giving rise to
\begin{equation}
	\begin{split}
	\left[ \dfrac{1}{2}\left(\mathcal{H}_{00} + \mathcal{H}_{11}\right) + \dfrac{1}{2}\left(\mathcal{H}_{11} - \mathcal{H}_{00}\right)i\gamma_1\gamma_2\right.\\
	\left. \vphantom{\frac{1}{2}}+ \mathcal{H}_{10} \gamma_1 \right] e^{\I\theta \gamma_2} \ket{1} = E_0 e^{\I\theta \gamma_2} \ket{1}.
	\end{split}
\end{equation}

At this point, it is worth mentioning that it would be virtually impossible to perform an orbital tapering technique without going through the Pauli or Majorana spaces.
The philosophy of the qubit tapering technique described in Refs.\ \cite{Bravyi.2017.1701.08213,Setia.2020.10.1021/acs.jctc.0c00113} is based on identifying $\mathbb{Z}_2$ symmetries of the Hamiltonian.
Pauli strings constitute a natural representation for uncovering such symmetries, as they are Hermitian involutions and, hence, have a spectrum of $\pm 1$.
An alternative approach could rely on the Majorana representation.
At a first glance, Majorana strings might not seem suitable for finding $\mathbb{Z}_2$ symmetries because, depending on their length, they are either Hermitian involutions ($\pm 1$ eigenvalues) or anti-Hermitian skew-involutions ($\pm \I$ eigenvalues).
However, since the Hamiltonian is Hermitian, it should be expressed as a linear combination of Hermitian Majorana strings.
Indeed, the one- and two-body terms of the electronic Hamiltonian are expressed in the Majorana representation as
\begin{equation}\label{eq:1b2m}
	a^p a_q + a^q a_p = \frac{1}{2} \left( \I \gamma_1^{(p)} \gamma_2^{(q)} - \I \gamma_2^{(p)} \gamma_1^{(q)}\right) 
\end{equation}
and
\begin{equation}\label{eq:2b2m}
	\begin{split}
		a^p a^q a_s a_r + a^r a^s a_q a_p =\\
		\frac{1}{8} \left( \gamma_1^{(p)} \gamma_1^{(q)} \gamma_1^{(s)} \gamma_1^{(r)} + \gamma_1^{(p)} \gamma_2^{(q)} \gamma_2^{(s)} \gamma_1^{(r)}\right.\\
		+ \gamma_2^{(p)} \gamma_1^{(q)} \gamma_1^{(s)} \gamma_2^{(r)} + \gamma_2^{(p)} \gamma_2^{(q)} \gamma_2^{(s)} \gamma_2^{(r)}\\
		-\gamma_1^{(p)} \gamma_1^{(q)} \gamma_2^{(s)} \gamma_2^{(r)} + \gamma_1^{(p)} \gamma_2^{(q)} \gamma_1^{(s)} \gamma_2^{(r)}\\
		\left. + \gamma_2^{(p)} \gamma_1^{(q)} \gamma_2^{(s)} \gamma_1^{(r)} - \gamma_2^{(p)} \gamma_2^{(q)} \gamma_1^{(s)} \gamma_1^{(r)}\right)
	\end{split}
\end{equation}
Although the Majorana strings $\gamma_1^{(p)} \gamma_2^{(q)}$ and $\gamma_2^{(p)} \gamma_1^{(q)}$ appearing in \cref{eq:1b2m} are anti-Hermitian skew-involutions [see \cref{mstring_antihermitian}], they become involutory upon multiplication by the imaginary unit.
The Majorana strings in \cref{eq:2b2m} have a length of 4 and, according to \cref{mstring_hermitian}, are Hermitian involutions.
This analysis further emphasizes the fact that choosing a suitable representation may lead to simplifications of the problem at hand.
  
\section{From Fermionic Clifford Unitaries to Mean-Field Theories}\label{sec:mean-field}

The fermionic Clifford unitaries and transformations we have presented can be efficiently simulated classically, as is the case with their Pauli and Majorana counterparts \cite{Gottesman.1998.10.1103/PhysRevA.57.127,Gottesman.1998.quant-ph/9807006}.
This can be understood in two different ways.
From the point of view of the wavefunction, a Clifford unitary applied on a Slater determinant will, at most, rotate it to another Slater determinant in the Fock space.
From the perspective of the transformed Hamiltonian, the number and many-body rank of its terms remain invariant under Clifford transformations.

However, Clifford unitaries are very constrained as the parameter $\theta$ can only take certain values, namely, $k\frac{\pi}{2}$, $k\in\mathbb{Z}$.
Although the angle $\theta$ can take arbitrary values when the generator is a number operator, the Clifford unitary simply adds a phase to the pertinent Slater determinants.
Therefore, although useful, Clifford operators can only perform discrete transformations.
To go beyond the capabilities of Clifford transformations, one can relax the constraints in the $\theta$ angles and treat them as variational parameters.
In doing so, the transformations will generally map a given fermionic string to a linear combination of strings.
For the transformations to remain classically tractable, the number of terms in the transformed Hamiltonian should remain $\text{poly}(M)$, where $M$ is the size of the one-electron basis, and its maximum many-body rank should not increase .

We start by examining transformations generated by single excitation operators, \foreign{i.e.}, $a^\dagger_p a_q \pm a^\dagger_q a_p$.
As shown in \cref{table_transform}, although such transformations are not, in general, Clifford, they give rise to linear combinations of operators with the same many-body rank as the transformed fermionic string.
Furthermore, single excitations can be adapted to conserve all symmetries characterizing the electronic Hamiltonian of molecular systems.
Indeed, they conserve particle number and fermionic parity by construction, while the enforcement of point group, $S_z$, and $S^2$ symmetries is straightforward.
Considering that the electronic Hamiltonian is comprised of generalized single and double excitations, transformations generated by single excitations do not introduce any new fermionic strings, rendering them classically tractable.
Indeed, exponentiated single excitations map one Slater determinant to another via Thouless' theorem \cite{Thouless.1960.10.1016/0029-5582(60)90048-1}, and form the core of Hartree--Fock (HF) theory.
Different variants of HF can be obtained by relaxing the symmetry constraints imposed in single excitations.
For example, in unrestricted HF, the single excitations break $S^2$ symmetry, while in generalized HF spin-flip single excitations are also incorporated, violating both $S^2$ and $S_z$ symmetries \cite{Echenique.2007.10.1080/00268970701757875}.
Even though such transformations introduce new fermionic strings into the transformed Hamiltonian, the total number of terms still scales as $M^4$, where $M$ is the number of spin-orbitals.
As a result, these transformations remain classically tractable.
\begin{table*}
	\caption{
		Analytic expressions for the fermionic transformations generated by the half-body and pair anti-Hermitian generators.\protect\footnotemark[1]
	}
	{\resizebox{\textwidth}{!}{
			\begin{ruledtabular}
				\begin{tabular}{ccc}
					$A$ & $O$ & $e^{-\theta A}Oe^{\theta A}$\\
					\colrule
					\multirow{4}{*}[-20pt]{$a_{\textcolor{blue}{\boldsymbol{c}}}^\dagger - a_{\textcolor{blue}{\boldsymbol{c}}}$} & $a_{\textcolor{blue}{\boldsymbol{c}}}$ &
					$\cos^2(\theta) a_{\textcolor{blue}{\boldsymbol{c}}} -\sin^2(\theta) a_{\textcolor{blue}{\boldsymbol{c}}}^\dagger + \dfrac{\sin(2\theta)}{2} (h_{\textcolor{blue}{\boldsymbol{c}}} - n_{\textcolor{blue}{\boldsymbol{c}}})$\\[10pt]
					& $a^{\textcolor{blue}{\boldsymbol{c}}}$ &
					$\cos^2(\theta) a_{\textcolor{blue}{\boldsymbol{c}}}^\dagger -\sin^2(\theta) a_{\textcolor{blue}{\boldsymbol{c}}} + \dfrac{\sin(2\theta)}{2} (h_{\textcolor{blue}{\boldsymbol{c}}} - n_{\textcolor{blue}{\boldsymbol{c}}})$\\[10pt]
					& $n_{\textcolor{blue}{\boldsymbol{c}}}$ &
					$\cos^2(\theta) n_{\textcolor{blue}{\boldsymbol{c}}} +\sin^2(\theta) h_{\textcolor{blue}{\boldsymbol{c}}} + \dfrac{\sin(2\theta)}{2} (a_{\textcolor{blue}{\boldsymbol{c}}} + a_{\textcolor{blue}{\boldsymbol{c}}}^\dagger)$\\[10pt]
					& $h_{\textcolor{blue}{\boldsymbol{c}}}$ &
					$\cos^2(\theta) h_{\textcolor{blue}{\boldsymbol{c}}} +\sin^2(\theta) n_{\textcolor{blue}{\boldsymbol{c}}} - \dfrac{\sin(2\theta)}{2} (a_{\textcolor{blue}{\boldsymbol{c}}} + a_{\textcolor{blue}{\boldsymbol{c}}}^\dagger)$\\[20pt]
					\multirow{4}{*}[-20pt]{$a_{\textcolor{blue}{\boldsymbol{c}}}^\dagger a_{\textcolor{red}{\boldsymbol{d}}}^\dagger - a_{\textcolor{red}{\boldsymbol{d}}} a_{\textcolor{blue}{\boldsymbol{c}}}$\footnotemark[2]} & $a_{\textcolor{blue}{\boldsymbol{c}}}$ & $\cos(\theta) a_{\textcolor{blue}{\boldsymbol{c}}} + \sin(\theta) a_{\textcolor{red}{\boldsymbol{d}}}^\dagger$\\[10pt]
					& $a^{\textcolor{blue}{\boldsymbol{c}}}$ & $\cos(\theta) a_{\textcolor{blue}{\boldsymbol{c}}}^\dagger + \sin(\theta) a_{\textcolor{red}{\boldsymbol{d}}}$\\[10pt]
					& $n_{\textcolor{blue}{\boldsymbol{c}}}$ &
					$\cos^2(\theta) n_{\textcolor{blue}{\boldsymbol{c}}} + \sin^2(\theta) h_{\textcolor{red}{\boldsymbol{d}}} + \dfrac{\sin(2\theta)}{2} (a_{\textcolor{blue}{\boldsymbol{c}}}^\dagger a_{\textcolor{red}{\boldsymbol{d}}}^\dagger + a_{\textcolor{red}{\boldsymbol{d}}} a_{\textcolor{blue}{\boldsymbol{c}}})$\\[10pt]
					& $h_{\textcolor{blue}{\boldsymbol{c}}}$ &
					$\cos^2(\theta) h_{\textcolor{blue}{\boldsymbol{c}}} + \sin^2(\theta) n_{\textcolor{red}{\boldsymbol{d}}} - \dfrac{\sin(2\theta)}{2} (a_{\textcolor{blue}{\boldsymbol{c}}}^\dagger a_{\textcolor{red}{\boldsymbol{d}}}^\dagger + a_{\textcolor{red}{\boldsymbol{d}}} a_{\textcolor{blue}{\boldsymbol{c}}})$\\[20pt]
					\multirow{4}{*}[-20pt]{$a_{\textcolor{blue}{\boldsymbol{c}}}^\dagger a_{\textcolor{red}{\boldsymbol{d}}} - a_{\textcolor{red}{\boldsymbol{d}}}^\dagger a_{\textcolor{blue}{\boldsymbol{c}}}$\footnotemark[2]} & $a_{\textcolor{blue}{\boldsymbol{c}}}$ & $\cos(\theta)a_{\textcolor{blue}{\boldsymbol{c}}} + \sin(\theta) a_{\textcolor{red}{\boldsymbol{d}}}$\\[10pt]
					& $a^{\textcolor{blue}{\boldsymbol{c}}}$ & $\cos(\theta)a_{\textcolor{blue}{\boldsymbol{c}}}^\dagger + \sin(\theta) a_{\textcolor{red}{\boldsymbol{d}}}^\dagger$\\[10pt]
					& $n_{\textcolor{blue}{\boldsymbol{c}}}$ &
					$\cos^2(\theta) n_{\textcolor{blue}{\boldsymbol{c}}} + \sin^2(\theta) n_{\textcolor{red}{\boldsymbol{d}}} + \dfrac{\sin(2\theta)}{2} (a_{\textcolor{blue}{\boldsymbol{c}}}^\dagger a_{\textcolor{red}{\boldsymbol{d}}} + a_{\textcolor{red}{\boldsymbol{d}}}^\dagger a_{\textcolor{blue}{\boldsymbol{c}}})$\\[10pt]
					& $h_{\textcolor{blue}{\boldsymbol{c}}}$ &
					$\cos^2(\theta) h_{\textcolor{blue}{\boldsymbol{c}}} + \sin^2(\theta) h_{\textcolor{red}{\boldsymbol{d}}} - \dfrac{\sin(2\theta)}{2} (a_{\textcolor{blue}{\boldsymbol{c}}}^\dagger a_{\textcolor{red}{\boldsymbol{d}}} + a_{\textcolor{red}{\boldsymbol{d}}}^\dagger a_{\textcolor{blue}{\boldsymbol{c}}})$\\
				\end{tabular}
			\end{ruledtabular}
	}}
	\footnotetext[1]{The analytic expressions for cases in which $a_c^\dagger a_d^\dagger - a_d a_c$ or $a_c^\dagger a_d - a_d^\dagger a_c$ share two common indices with the string $O$ can be obtained by performing the transformation in two steps, each step involving one common index.}
	\footnotetext[2]{If the common index is $d$ instead of $c$, the sign of any expression whose sign factor depends on $k$ must be inverted.}
	\label{table_transform}
\end{table*}

Next, we proceed to unitaries generated by pair creation/annihilation operators, \foreign{i.e.}, $a_p^\dagger a_q^\dagger \pm a_q a_p$.
Similar to the singles case, these unitaries give rise to rank-preserving transformations.
Although pair creation/annihilation operators automatically conserve fermionic parity, they violate particle number.
As a result, the excitation character of the transformed operator is not preserved, mapping, for example, a creation operator to a linear combination of an annihilation and a creation operators.
This introduces particle-number-violating terms in the transformed Hamiltonian.
Nevertheless, the number of terms still scales as $M^4$, and the transformation remains classically tractable.
These generators give rise to Bogoliubov transformations \cite{Bogoljubov.1958.10.1007/BF02745585,Valatin.1958.10.1007/BF02745589}, and are employed in the Hartree--Fock--Bogoliubov (HFB) approach (see, also, \cite{Thouless.1960.10.1016/0029-5582(60)90048-1}), the mean-field theory of choice when effective attractive interactions between electrons become prevalent, as is the case in superconducting Cooper pairs \cite{Cooper.1956.10.1103/PhysRev.104.1189,Bardeen.1957.10.1103/PhysRev.106.162,Bardeen.1957.10.1103/PhysRev.108.1175}.
By allowing the $S^2$ symmetry to be broken, one arrives at unrestricted HFB, while the simultaneous violation of $S_z$ and $S^2$ symmetries leads to generalized HFB (see, for example, \cite{Scuseria.2011.10.1063/1.3643338}).

Finally, we examine transformations generated by half-body operators, \foreign{i.e.}, $a_p^\dagger \pm a_p$.
In contrast to singles and pair creation/annihilation generators, half-body operators not only violate particle number but also fermionic parity.
Consequently, such transformations do not necessarily preserve the many-body rank of the operator that they are applied to (see, for example, \cref{table_transform}).
However, as demonstrated in Section S4 of the \sm, applying the unitary transformation generated by the operators $a_p^\dagger \pm a_p$ to a fermionic string $O$ of even length produces new terms whose many-body rank is decreased by at most one-half.
Furthermore, if $O$ has odd length, the transformation generally yields terms whose many-body rank is increased by at most one-half, unless the common index appears in a number operator, in which case the many-body rank is decreased by at most one-half.
Since fermionic Hamiltonians conserve fermionic parity, they contain operator strings of even length or, equivalently, integer many-body rank.
As a result, such transformations cannot increase the many-body rank of fermionic Hamiltonians.
For example, when unitary transformations generated by half-body operators act on the electronic Hamiltonian, they introduce half- and one--and--a--half-body terms, as well as particle-nonconserving one- and two-body terms.
Notably, the transformed Hamiltonian does not contain higher-than-two-body interactions, retaining a number of terms that scales as $M^4$.
Therefore, such transformations can be efficiently applied classically and, in fact, are employed in the Hartree--Fock--Bogoliubov--Fukutome (HFBF) mean-field theory \cite{Fukutome.1977.10.1143/PTP.57.1554,Moussa.2012.1208.1086,Henderson.2024.10.1063/5.0188155}.
This more exotic variant of HF theory is the natural starting point for fermionic Hamiltonians that do not preserve fermionic parity.
Although this symmetry is respected by all physical fermionic Hamiltonians, it is violated by spin Hamiltonians that have been transformed to the fermionic space \cite{Henderson.2024.10.1063/5.0242219}.

To gain a deeper understanding of the algebraic structures generated by the sets of single excitation ($\{A^p_q\}_{p<q}$), pair creation/annihilation ($\{A^{pq}\}_{p<q}$), and half-body ($\{A^p\}$) operators, we examine their commutators.
	Starting with the single excitation operators, we obtain
	\begin{equation}
		\left[A^p_q, A^r_s\right] = \delta_{qr} A^p_s - \delta_{qs} A^p_r - \delta_{pr} A^q_s +\delta_{ps} A^q_r,
	\end{equation}
	which is exactly the commutator relation satisfied by the generators of the special orthogonal ($\mathcal{SO}$) Lie group.
	Thus, given an $M$-dimensional single-particle basis, the $\frac{M(M-1)}{2}$ single excitation operators $A^p_q$ span the $\mathfrak{so}(M)$ Lie algebra.
	This should not be surprising, as the functional form $e^{\theta A^p_q}$ implies that the unitaries are also special orthogonal transformations.
	We, thus, anticipate that the remaining algebraic structures will be related to the special orthogonal Lie algebra as well.
		
	Moving on to the case of pair creation/annihilation operators, we find that they are not closed under the commutator,
	\begin{equation}
		\left[A^{pq}, A^{rs}\right] = \left[A^p_q, A^r_s\right],
	\end{equation}
	and, as a result, $\{A^{pq}\}$ does not form a Lie algebra by itself.
	Nevertheless, considering that
	\begin{equation}
		\left[A^p_q, A^{rs}\right] = \delta_{qr} A^{ps} - \delta_{qs} A^{pr} - \delta_{pr} A^{qs} + \delta_{ps} A^{qr},
	\end{equation}
	closure can be obtained in the extended set $\{A^p_q, A^{pq}\}$.
	Indeed, as shown in Section S7 of the \sm, the $\frac{M(M-1)}{2}$ single excitation operators $A^p_q$ plus the $\frac{M(M-1)}{2}$ pair creation/annihilation operators $A^{pq}$ form a basis for the $M(M-1)$-dimensional Lie algebra direct sum $\mathfrak{so}(M) \oplus \mathfrak{so}(M)$.

Similar to the case of pair creation/annihilation operators, the half-body operators do not form a Lie algebra on their own, as their commutator introduces both single excitation and pair creation/annihilation operators,
	\begin{equation}
		\left[A^p, A^q\right] = 2(A^{pq} - A^p_q).
	\end{equation}
	Closure is attained by the extended set $\{A^p_q, A^{pq}, A^p\}$, with the remaining commutation relations reading
	\begin{equation}
		\left[A^p_q, A^r\right] = \delta_{qr} A^p - \delta_{pr} A^q
	\end{equation}
	and
	\begin{equation}
		\left[A^{pq}, A^r\right] = -\left[A^p_q, A^r\right].
	\end{equation}
	As shown in the \sm, the set $\{A^p_q, A^{pq}, A^p\}$ spans the $M^2$-dimensional Lie algebra direct sum $\mathfrak{so}(M) \oplus \mathfrak{so}(M+1)$.

At this point, it is worth noting that the group theoretical structure of many-fermion systems arising from the canonical anticommutation relations was originally studied by Fukutome more than 40 years ago \cite{Fukutome.1981.10.1143/PTP.65.809}.
Using standard, \foreign{i.e.}, not anti-Hermitian, operators, Fukutome showed that the sets $\{a_p^\dagger a_q -\frac{1}{2}\delta_{pq}\}$, $\{a_p^\dagger a_q -\frac{1}{2}\delta_{pq}, a_p a_q, a_p^\dagger a_q^\dagger\}$, and $\{a_p, a_p^\dagger, a_p^\dagger a_q -\frac{1}{2}\delta_{pq}, a_p a_q, a_p^\dagger a_q^\dagger\}$ span the $\mathfrak{u}(n)$, $\mathfrak{so}(2n)$, and $\mathfrak{so}(2n+1)$ Lie algebras.
Note that the single excitation operators are shifted by $-\frac{1}{2}\delta_{pq}$ to eliminate constant terms from the pertinent commutators.
The various Lie algebras of fermionic operators are related as follows: $\mathfrak{so}(n) \subset\mathfrak{u}(n)$, $\mathfrak{so}(n) \oplus \mathfrak{so}(n) \subset \mathfrak{so}(2n)$, and $\mathfrak{so}(n) \oplus \mathfrak{so}(n+1) \subset \mathfrak{so}(2n+1)$.

Before concluding, we would like to mention that unitaries and transformations generated by linear combinations of half-body operators can be expressed in closed form.
Indeed, as shown in Section S5 of the \sm, when the anti-Hermitian generator is
\begin{equation}\label{eq:gen_half_sum}
	A = \sum_p \theta_p (a_p^\dagger - a_p),
\end{equation}
the corresponding unitary can be written as
\begin{equation}\label{eq:half_closed}
	e^{A} = \cos(\sqrt{\sum_p \theta_p^2}) + \frac{\sin(\sqrt{\sum_p \theta_p^2})}{\sqrt{\sum_p \theta_p^2}} \left( \sum_p \theta_p (a_p^\dagger - a_p) \right),
\end{equation}
with the corresponding transformation taking the form
\begin{equation}\label{eq:st_sum}
	\begin{split}
	e^{-A}Oe^{A} ={}& O\\
	&+\frac{1}{2} \frac{\sin(2\sqrt{\sum_p \theta_p^2})}{\sqrt{\sum_p \theta_p^2}} \sum_q \theta_q [O, a_q^\dagger - a_q]\\
	&+ \frac{1}{2} \frac{\sin^2(\sqrt{\sum_p \theta_p^2})}{\sum_p \theta_p^2}\sum_q\sum_r \theta_q \theta_r [[O, a_q^\dagger - a_q], a_r^\dagger - a_r].
	\end{split}
\end{equation}
Although \cref{eq:half_closed} can be found in the literature \cite{Fukutome.1981.10.1143/PTP.65.809}, to the best of our knowledge, \cref{eq:st_sum} is a new analytical result.
Using \cref{eq:st_sum}, we can examine the transformation of the elementary annihilation and creation operators.
For example, we obtain that the unitary transformation of the annihilation operator $a_q$ generated by \cref{eq:gen_half_sum} is given by
\begin{equation}\label{eq:canonical}
	\begin{split}
	e^{-\sum_{p} \theta_p (a_p^\dagger - a_p)} a_q e^{\sum_{p} \theta_p (a_p^\dagger - a_p)} =\\
	\left[1 - \frac{\sin^2\left(\sqrt{\sum_p\theta_p^2}\right)}{\sum_p \theta_p^2}\left(\sum_p\theta_p^2 + \sum_p^\prime \theta_p^2 \right) \right] a_q\\
	-\frac{\sin^2\left(\sqrt{\sum_p\theta_p^2}\right)}{\sum_p \theta_p^2} \theta_q^2 a_q^\dagger\\
	+\frac{1}{2} \frac{\sin\left(2\sqrt{\sum_p\theta_p^2}\right)}{\sqrt{\sum_p\theta_p^2}} \theta_q \left(h_q -n_q\right)\\
	-\frac{\sin^2\left(\sqrt{\sum_p\theta_p^2}\right)}{\sum_p \theta_p^2} \sum_p^\prime \theta_q \theta_p \left(a_p^\dagger - a_p\right)\\
	+\frac{\sin\left(2\sqrt{\sum_p\theta_p^2}\right)}{\sqrt{\sum_p\theta_p^2}} \sum_p^\prime \theta_p a_q \left(a_p^\dagger - a_p\right)
	\end{split}.
\end{equation}
A few remarks are in order regarding the transformation shown in \cref{eq:canonical}.
To begin with, being unitary, the transformation preserves the fermionic commutation relations and, thus, is canonical.
Furthermore, we observe that a single annihilation operator is mapped to a linear combination of annihilation and creation operators, reminiscent of Bogoliubov transformations.
In addition, it introduces one-body terms, namely, $h_q - n_q$ and $a_q(a_p^\dagger - a_p)$, violating fermionic parity.
It is interesting to note that this transformation does not generate zero-body terms.
A fermionic-parity-violating transformation of a single annihilation/creation operator that introduces a zero-body term is possible.
However, for this transformation to be canonical, the zero-body term must be a Grassmann number \cite{Henderson.2024.10.1063/5.0188155}.
Furthermore, it is worth noting that even in the simpler case $A = \theta_p (a_p^\dagger - a_p) + \theta_q (a_q^\dagger - a_q)$, no angles $\theta_p$ and $\theta_q$ exist to render the transformation of \cref{eq:st_sum} Clifford, as might have been anticipated.

\section{Conclusions}
In this work, we have presented Clifford transformations for fermions.
We demonstrated that the generators of fermionic Clifford unitaries are anti-Hermitian and Hermitian linear combinations of half-body and pair operators with the parameter $\theta$ taking values $k\frac{\pi}{2}$, $k\in\mathbb{Z}$.
Furthermore, we fully characterized the action of such transformations.
In particular, we showed that Clifford transformations generated by half-body operators result in particle--hole conjugation, replacing creation operators by their annihilation counterparts and vice versa.
Clifford transformations based on single excitation operators result in an index swap and can be used to perform spin-flip transformations, replacing spin-up operators by their spin-down counterparts and vice versa.
The outcome of Clifford transformations generated by pair creation/annihilation operators is to perform simultaneously index swap and particle--hole conjugation.

Contrary to the Pauli and Majorana cases, we showed that continuous fermionic Clifford transformations exist, as long as the generator is a single number operator.
In the nontrivial case, such transformations multiply the fermionic string that is transformed by a phase.
We demonstrated that the $T$ gate is Clifford in the case of fermions, in contrast to its action on Pauli and Majorana strings.
Additionally, we showed that fermionic Clifford transformations respect fermionic parity, even in the case where the generator is a half-body operator.

As a practical application of fermionic Clifford transformations, we examined the qubit tapering technique as applied to the minimum-basis-set \ce{H2} molecule.
We showed that the resulting Pauli Clifford gates translate, under the Jordan--Wigner map, to linear combinations of products of fermionic Clifford unitaries generated by number and half-body operators.
Although the final transformation is Clifford in the Pauli space, it is not in second quantization, resulting in a Hamiltonian with substantially more terms.
After the removal of redundant qubits/spinorbitals, the resulting second-quantized Hamiltonian violated not only particle number but also the fermionic parity.

Finally, we showed that if the angles in the fermionic Clifford unitaries are replaced by continuous variables, the various flavors of mean-field theories are obtained.
Indeed, the single excitation operators give rise to restricted, unrestricted, and generalized Hartree--Fock.
The inclusion of pair creation/annihilation operators results in the Hartree--Fock--Bogoliubov scheme, while incorporation of half-body operators produces the most general variant, namely, Hartree--Fock--Bogoliubov--Fukutome.

\section*{Acknowledgments}

This work was supported by the U.S.\ Department of Energy under Award No.\ DE-SC0019374.

\clearpage

\renewcommand{\theequation}{S\arabic{equation}}
\setcounter{equation}{0}

\renewcommand{\thetable}{S\arabic{table}}
\setcounter{table}{0}

\renewcommand{\thefigure}{S\arabic{figure}}
\setcounter{figure}{0}

\renewcommand{\thesection}{S\arabic{section}}
\setcounter{section}{0}

\renewcommand{\thepage}{S\arabic{page}}
\setcounter{page}{1}

\definecolor{colorA}{rgb}{0.00, 0.45, 0.70}  
\definecolor{colorB}{rgb}{0.00, 0.62, 0.45}  
\definecolor{colorC}{rgb}{0.35, 0.70, 0.90}  
\definecolor{colorD}{rgb}{0.00, 0.75, 0.85}  
\definecolor{colorE}{rgb}{0.44, 0.88, 0.56}  
\definecolor{colorF}{rgb}{0.68, 1.00, 0.18}  
\definecolor{colorG}{rgb}{0.95, 0.90, 0.25}  
\definecolor{colorH}{rgb}{1.00, 0.80, 0.00}  
\definecolor{colorI}{rgb}{1.00, 0.60, 0.00}  
\definecolor{colorJ}{rgb}{0.80, 0.40, 0.00}  
\definecolor{colorK}{rgb}{0.85, 0.10, 0.10}  
\definecolor{colorL}{rgb}{0.94, 0.50, 0.72}  
\definecolor{colorM}{rgb}{0.60, 0.30, 0.70}  
\definecolor{colorN}{rgb}{0.55, 0.35, 0.80}  
\definecolor{colorO}{rgb}{0.50, 0.50, 0.50}  

\onecolumngrid
\fontsize{12}{18}\selectfont
\begin{center}
	\textbf{\large Supplemental Material:\\
		Clifford Transformations for Fermionic Quantum Systems: From Paulis to Majoranas to Fermions
	}\\[.2cm]
	Ilias Magoulas$^{1,*}$ and Francesco A.\ Evangelista$^{1}$\\[.1cm]
	{\itshape ${}^1$Department of Chemistry and Cherry Emerson Center for Scientific Computation,\\ 
		Emory University, Atlanta, Georgia 30322, USA\\}
	${}^*$Corresponding author; e-mail: ilias.magoulas@emory.edu.
\end{center}

\newpage
\noindent

The Supplemental Material is organized as follows.
For the sake of completeness, in \cref{ssec:pauli,ssec:majorana}, we review the Pauli and Majorana algebras, respectively, and their connections to Clifford transformations.
In \cref{ssec:fermion}, we review the fermionic algebra and provide the key derivations for the Clifford transformations characterizing the fermionic monoid.
In \cref{ssec:transform_single_half}, we analyze generic unitary transformations generated by fermionic half-body generators. For strings of even length, these transformations introduce new terms whose many-body rank is reduced by at most 0.5.
For strings of odd length, the many-body rank of the new terms increase by at most 0.5, unless the common index appears in a number operator, in which case the rank decreases by at most 0.5.
In \cref{ssec:transform_sum_half}, we derive closed-form expressions for unitaries generated by linear combinations of half-body fermionic operators and for their induced transformations.
In \cref{ssec:taper}, we illustrate the role of fermionic Clifford unitaries and transformations in spinorbital tapering, using the minimum-basis-set representation of the $\text{H}_2$ molecule as an example.
Finally, in \cref{ssec:Lie}, we provide proofs for the Lie algebra classifications mentioned in the main text.

\section{Pauli Algebra}\label{ssec:pauli}

The Pauli matrices $\sigma_x$, $\sigma_y$, and $\sigma_z$ are Hermitian involutions, \foreign{i.e.},
\begin{equation}\label{seq:sigma_involution}
	\sigma_i^\dagger = \sigma_i^{-1} = \sigma_i,
\end{equation}
and they pairwise anticommute,
\begin{equation}\label{seq:pauli_anticommute}
	\{\sigma_i, \sigma_j\} = 2\delta_{ij}, \quad i,j \in \{x,y,z\}.
\end{equation}
The set $\{\sigma_0, \sigma_x, \sigma_y, \sigma_z\}$, where $\sigma_0 \equiv I$ is the identity matrix, is not closed under matrix multiplication, as illustrated in panel (a) of \cref{sfig:pauli_cayley}.
As seen in the aforementioned Cayley table, the product of two Pauli matrices can be generally expressed as
\begin{equation}\label{seq:pauli_product}
	\sigma_i \sigma_j = \I^\alpha \sigma_k, \quad \alpha \in \mathbb{Z}_4, i,j,k \in \{0,x,y,z\}.
\end{equation}
In light of \cref{seq:pauli_product}, and as shown in \cref{sfig:pauli_cayley}(b), closure can be achieved by extending the set to $\mathcal{P}_1 = \left\{\I^\alpha \sigma_i \mid \alpha\in\mathbb{Z}_4, i\in\{0,x,y,z\}\right\}$, known as the single-qubit Pauli group.

\begin{figure}
	\centering
	\subfloat[]{
		\begin{tabular}{c|cccc}
			\(\cdot\) & $\sigma_0$ & \cellcolor{colorA} $\sigma_x$ & \cellcolor{colorB} $\sigma_y$ & \cellcolor{colorC} $\sigma_z$ \\ \hline
			$\sigma_0$ & $\sigma_0$ &\cellcolor{colorA} $\sigma_x$ & \cellcolor{colorB} $\sigma_y$ & \cellcolor{colorC} $\sigma_z$ \\
			\cellcolor{colorA} $\sigma_x$ & \cellcolor{colorA} $\sigma_x$ & $\sigma_0$ & \cellcolor{colorK} $\I\sigma_z$ & \cellcolor{colorN} $-\I\sigma_y$  \\
			\cellcolor{colorB} $\sigma_y$ & \cellcolor{colorB} $\sigma_y$ & \cellcolor{colorO} $-\I\sigma_z$ & $\sigma_0$ & \cellcolor{colorI} $\I\sigma_x$  \\
			\cellcolor{colorC} $\sigma_z$ & \cellcolor{colorC} $\sigma_z$ & \cellcolor{colorJ} $\I\sigma_y$ & \cellcolor{colorM} $-\I\sigma_x$ & $\sigma_0$
		\end{tabular}
	}\\
	\subfloat[]{\resizebox{\textwidth}{!}{
			\begin{tabular}{c|cccccccccccccccc}
				\(\cdot\) & $\sigma_0$ & \cellcolor{colorA} $\sigma_x$ & \cellcolor{colorB} $\sigma_y$ & \cellcolor{colorC} $\sigma_z$ & \cellcolor{colorD} $-\sigma_0$ & \cellcolor{colorE} $-\sigma_x$ & \cellcolor{colorF} $-\sigma_y$ & \cellcolor{colorG} $-\sigma_z$ & \cellcolor{colorH} $\I\sigma_0$ & \cellcolor{colorI} $\I\sigma_x$ & \cellcolor{colorJ} $\I\sigma_y$ & \cellcolor{colorK} $\I\sigma_z$ & \cellcolor{colorL} $-\I\sigma_0$ & \cellcolor{colorM} $-\I\sigma_x$ & \cellcolor{colorN} $-\I\sigma_y$ & \cellcolor{colorO} $-\I\sigma_z$ \\ \hline
				$\sigma_0$ & $\sigma_0$ & \cellcolor{colorA} $\sigma_x$ & \cellcolor{colorB} $\sigma_y$ & \cellcolor{colorC} $\sigma_z$ & \cellcolor{colorD} $-\sigma_0$ & \cellcolor{colorE} $-\sigma_x$ & \cellcolor{colorF} $-\sigma_y$ & \cellcolor{colorG} $-\sigma_z$ & \cellcolor{colorH} $\I\sigma_0$ & \cellcolor{colorI} $\I\sigma_x$ & \cellcolor{colorJ} $\I\sigma_y$ & \cellcolor{colorK} $\I\sigma_z$ & \cellcolor{colorL} $-\I\sigma_0$ & \cellcolor{colorM} $-\I\sigma_x$ & \cellcolor{colorN} $-\I\sigma_y$ & \cellcolor{colorO} $-\I\sigma_z$ \\
				\cellcolor{colorA} $\sigma_x$ & \cellcolor{colorA} $\sigma_x$ & $\sigma_0$ & \cellcolor{colorK} $\I\sigma_z$ & \cellcolor{colorN} $-\I\sigma_y$ & \cellcolor{colorE} $-\sigma_x$ & \cellcolor{colorD} $-\sigma_0$ & \cellcolor{colorO} $-\I\sigma_z$ & \cellcolor{colorJ} $\I\sigma_y$ & \cellcolor{colorI} $\I\sigma_x$ & \cellcolor{colorH} $\I\sigma_0$ & \cellcolor{colorG} $-\sigma_z$ & \cellcolor{colorB} $\sigma_y$ & \cellcolor{colorM} $-\I\sigma_x$ & \cellcolor{colorL} $-\I\sigma_0$ & \cellcolor{colorC} $\sigma_z$ & \cellcolor{colorF} $-\sigma_y$ \\
				\cellcolor{colorB} $\sigma_y$ & \cellcolor{colorB} $\sigma_y$ & \cellcolor{colorO} $-\I\sigma_z$ & $\sigma_0$ & \cellcolor{colorI} $\I\sigma_x$ & \cellcolor{colorF} $-\sigma_y$ & \cellcolor{colorK} $\I\sigma_z$ & \cellcolor{colorD} $-\sigma_0$ & \cellcolor{colorM} $-\I\sigma_x$ & \cellcolor{colorJ} $\I\sigma_y$ & \cellcolor{colorC} $\sigma_z$ & \cellcolor{colorH} $\I\sigma_0$ & \cellcolor{colorE} $-\sigma_x$ & \cellcolor{colorN} $-\I\sigma_y$ & \cellcolor{colorG} $-\sigma_z$ & \cellcolor{colorL} $-\I\sigma_0$ & \cellcolor{colorA} $\sigma_x$ \\
				\cellcolor{colorC} $\sigma_z$ & \cellcolor{colorC} $\sigma_z$ & \cellcolor{colorJ} $\I\sigma_y$ & \cellcolor{colorM} $-\I\sigma_x$ & $\sigma_0$ & \cellcolor{colorG} $-\sigma_z$ & \cellcolor{colorN} $-\I\sigma_y$ & \cellcolor{colorI} $\I\sigma_x$ & \cellcolor{colorD} $-\sigma_0$ & \cellcolor{colorK} $\I\sigma_z$ & \cellcolor{colorF} $-\sigma_y$ & \cellcolor{colorA} $\sigma_x$ & \cellcolor{colorH} $\I\sigma_0$ & \cellcolor{colorO} $-\I\sigma_z$ & \cellcolor{colorB} $\sigma_y$ & \cellcolor{colorE} $-\sigma_x$ & \cellcolor{colorL} $-\I\sigma_0$ \\
				\cellcolor{colorD} $-\sigma_0$ & \cellcolor{colorD} $-\sigma_0$ & \cellcolor{colorE} $-\sigma_x$ & \cellcolor{colorF} $-\sigma_y$ & \cellcolor{colorG} $-\sigma_z$ & $\sigma_0$ & \cellcolor{colorA} $\sigma_x$ & \cellcolor{colorB} $\sigma_y$ & \cellcolor{colorC} $\sigma_z$ & \cellcolor{colorL} $-\I\sigma_0$ & \cellcolor{colorM} $-\I\sigma_x$ & \cellcolor{colorN} $-\I\sigma_y$ & \cellcolor{colorO} $-\I\sigma_z$ & \cellcolor{colorH} $\I\sigma_0$ & \cellcolor{colorI} $\I\sigma_x$ & \cellcolor{colorJ} $\I\sigma_y$ & \cellcolor{colorK} $\I\sigma_z$ \\
				\cellcolor{colorE} $-\sigma_x$ & \cellcolor{colorE} $-\sigma_x$ & \cellcolor{colorD} $-\sigma_0$ & \cellcolor{colorO} $-\I\sigma_z$ & \cellcolor{colorJ} $\I\sigma_y$ & \cellcolor{colorA} $\sigma_x$ & $\sigma_0$ & \cellcolor{colorK} $\I\sigma_z$ & \cellcolor{colorN} $-\I\sigma_y$ & \cellcolor{colorM} $-\I\sigma_x$ & \cellcolor{colorL} $-\I\sigma_0$ & \cellcolor{colorC} $\sigma_z$ & \cellcolor{colorF} $-\sigma_y$ & \cellcolor{colorI} $\I\sigma_x$ & \cellcolor{colorH} $\I\sigma_0$ & \cellcolor{colorG} $-\sigma_z$ & \cellcolor{colorB} $\sigma_y$ \\
				\cellcolor{colorF} $-\sigma_y$ & \cellcolor{colorF} $-\sigma_y$ & \cellcolor{colorK} $\I\sigma_z$ & \cellcolor{colorD} $-\sigma_0$ & \cellcolor{colorM} $-\I\sigma_x$ & \cellcolor{colorB} $\sigma_y$ & \cellcolor{colorO} $-\I\sigma_z$ & $\sigma_0$ & \cellcolor{colorI} $\I\sigma_x$ & \cellcolor{colorN} $-\I\sigma_y$ & \cellcolor{colorG} $-\sigma_z$ & \cellcolor{colorL} $-\I\sigma_0$ & \cellcolor{colorA} $\sigma_x$ & \cellcolor{colorJ} $\I\sigma_y$ & \cellcolor{colorC} $\sigma_z$ & \cellcolor{colorH} $\I\sigma_0$ & \cellcolor{colorE} $-\sigma_x$ \\
				\cellcolor{colorG} $-\sigma_z$ & \cellcolor{colorG} $-\sigma_z$ & \cellcolor{colorN} $-\I\sigma_y$ & \cellcolor{colorI} $\I\sigma_x$ & \cellcolor{colorD} $-\sigma_0$ & \cellcolor{colorC} $\sigma_z$ & \cellcolor{colorJ} $\I\sigma_y$ & \cellcolor{colorM} $-\I\sigma_x$ & $\sigma_0$ & \cellcolor{colorO} $-\I\sigma_z$ & \cellcolor{colorB} $\sigma_y$ & \cellcolor{colorE} $-\sigma_x$ & \cellcolor{colorL} $-\I\sigma_0$ & \cellcolor{colorK} $\I\sigma_z$ & \cellcolor{colorF} $-\sigma_y$ & \cellcolor{colorA} $\sigma_x$ & \cellcolor{colorH} $\I\sigma_0$ \\
				\cellcolor{colorH} $\I\sigma_0$ & \cellcolor{colorH} $\I\sigma_0$ & \cellcolor{colorI} $\I\sigma_x$ & \cellcolor{colorJ} $\I\sigma_y$ & \cellcolor{colorK} $\I\sigma_z$ & \cellcolor{colorL} $-\I\sigma_0$ & \cellcolor{colorM} $-\I\sigma_x$ & \cellcolor{colorN} $-\I\sigma_y$ & \cellcolor{colorO} $-\I\sigma_z$ & \cellcolor{colorD} $-\sigma_0$ & \cellcolor{colorE} $-\sigma_x$ & \cellcolor{colorF} $-\sigma_y$ & \cellcolor{colorG} $-\sigma_z$ & $\sigma_0$ & \cellcolor{colorA} $\sigma_x$ & \cellcolor{colorB} $\sigma_y$ & \cellcolor{colorC} $\sigma_z$ \\
				\cellcolor{colorI} $\I\sigma_x$ & \cellcolor{colorI} $\I\sigma_x$ & \cellcolor{colorH} $\I\sigma_0$ & \cellcolor{colorG} $-\sigma_z$ & \cellcolor{colorB} $\sigma_y$ & \cellcolor{colorM} $-\I\sigma_x$ & \cellcolor{colorL} $-\I\sigma_0$ & \cellcolor{colorC} $\sigma_z$ & \cellcolor{colorF} $-\sigma_y$ & \cellcolor{colorE} $-\sigma_x$ & \cellcolor{colorD} $-\sigma_0$ & \cellcolor{colorO} $-\I\sigma_z$ & \cellcolor{colorJ} $\I\sigma_y$ & \cellcolor{colorA} $\sigma_x$ & $\sigma_0$ & \cellcolor{colorK} $\I\sigma_z$ & \cellcolor{colorN} $-\I\sigma_y$ \\
				\cellcolor{colorJ} $\I\sigma_y$ & \cellcolor{colorJ} $\I\sigma_y$ & \cellcolor{colorC} $\sigma_z$ & \cellcolor{colorH} $\I\sigma_0$ & \cellcolor{colorE} $-\sigma_x$ & \cellcolor{colorN} $-\I\sigma_y$ & \cellcolor{colorG} $-\sigma_z$ & \cellcolor{colorL} $-\I\sigma_0$ & \cellcolor{colorA} $\sigma_x$ & \cellcolor{colorF} $-\sigma_y$ & \cellcolor{colorK} $\I\sigma_z$ & \cellcolor{colorD} $-\sigma_0$ & \cellcolor{colorM} $-\I\sigma_x$ & \cellcolor{colorB} $\sigma_y$ & \cellcolor{colorO} $-\I\sigma_z$ & $\sigma_0$ & \cellcolor{colorI} $\I\sigma_x$ \\
				\cellcolor{colorK} $\I\sigma_z$ & \cellcolor{colorK} $\I\sigma_z$ & \cellcolor{colorF} $-\sigma_y$ & \cellcolor{colorA} $\sigma_x$ & \cellcolor{colorH} $\I\sigma_0$ & \cellcolor{colorO} $-\I\sigma_z$ & \cellcolor{colorB} $\sigma_y$ & \cellcolor{colorE} $-\sigma_x$ & \cellcolor{colorL} $-\I\sigma_0$ & \cellcolor{colorG} $-\sigma_z$ & \cellcolor{colorN} $-\I\sigma_y$ & \cellcolor{colorI} $\I\sigma_x$ & \cellcolor{colorD} $-\sigma_0$ & \cellcolor{colorC} $\sigma_z$ & \cellcolor{colorJ} $\I\sigma_y$ & \cellcolor{colorM} $-\I\sigma_x$ & $\sigma_0$ \\
				\cellcolor{colorL} $-\I\sigma_0$ & \cellcolor{colorL} $-\I\sigma_0$ & \cellcolor{colorM} $-\I\sigma_x$ & \cellcolor{colorN} $-\I\sigma_y$ & \cellcolor{colorO} $-\I\sigma_z$ & \cellcolor{colorH} $\I\sigma_0$ & \cellcolor{colorI} $\I\sigma_x$ & \cellcolor{colorJ} $\I\sigma_y$ & \cellcolor{colorK} $\I\sigma_z$ & $\sigma_0$ & \cellcolor{colorA} $\sigma_x$ & \cellcolor{colorB} $\sigma_y$ & \cellcolor{colorC} $\sigma_z$ & \cellcolor{colorD} $-\sigma_0$ & \cellcolor{colorE} $-\sigma_x$ & \cellcolor{colorF} $-\sigma_y$ & \cellcolor{colorG} $-\sigma_z$ \\
				\cellcolor{colorM} $-\I\sigma_x$ & \cellcolor{colorM} $-\I\sigma_x$ & \cellcolor{colorL} $-\I\sigma_0$ & \cellcolor{colorC} $\sigma_z$ & \cellcolor{colorF} $-\sigma_y$ & \cellcolor{colorI} $\I\sigma_x$ & \cellcolor{colorH} $\I\sigma_0$ & \cellcolor{colorG} $-\sigma_z$ & \cellcolor{colorB} $\sigma_y$ & \cellcolor{colorA} $\sigma_x$ & $\sigma_0$ & \cellcolor{colorK} $\I\sigma_z$ & \cellcolor{colorN} $-\I\sigma_y$ & \cellcolor{colorE} $-\sigma_x$ & \cellcolor{colorD} $-\sigma_0$ & \cellcolor{colorO} $-\I\sigma_z$ & \cellcolor{colorJ} $\I\sigma_y$ \\
				\cellcolor{colorN} $-\I\sigma_y$ & \cellcolor{colorN} $-\I\sigma_y$ & \cellcolor{colorG} $-\sigma_z$ & \cellcolor{colorL} $-\I\sigma_0$ & \cellcolor{colorA} $\sigma_x$ & \cellcolor{colorJ} $\I\sigma_y$ & \cellcolor{colorC} $\sigma_z$ & \cellcolor{colorH} $\I\sigma_0$ & \cellcolor{colorE} $-\sigma_x$ & \cellcolor{colorB} $\sigma_y$ & \cellcolor{colorO} $-\I\sigma_z$ & $\sigma_0$ & \cellcolor{colorI} $\I\sigma_x$ & \cellcolor{colorF} $-\sigma_y$ & \cellcolor{colorK} $\I\sigma_z$ & \cellcolor{colorD} $-\sigma_0$ & \cellcolor{colorM} $-\I\sigma_x$ \\
				\cellcolor{colorO} $-\I\sigma_z$ & \cellcolor{colorO} $-\I\sigma_z$ & \cellcolor{colorB} $\sigma_y$ & \cellcolor{colorE} $-\sigma_x$ & \cellcolor{colorL} $-\I\sigma_0$ & \cellcolor{colorK} $\I\sigma_z$ & \cellcolor{colorF} $-\sigma_y$ & \cellcolor{colorA} $\sigma_x$ & \cellcolor{colorH} $\I\sigma_0$ & \cellcolor{colorC} $\sigma_z$ & \cellcolor{colorJ} $\I\sigma_y$ & \cellcolor{colorM} $-\I\sigma_x$ & $\sigma_0$ & \cellcolor{colorG} $-\sigma_z$ & \cellcolor{colorN} $-\I\sigma_y$ & \cellcolor{colorI} $\I\sigma_x$ & \cellcolor{colorD} $-\sigma_0$
			\end{tabular}
		}
	}
	\caption{\label{sfig:pauli_cayley}
		Cayley tables defining the Pauli algebra. (a) Cayley table for the Pauli matrices. (b) Cayley table for the single-qubit Pauli group. Cells are color-coded to highlight closure under multiplication: (a) is not closed, while (b) is closed.
	}
\end{figure}

Let $P$ be an $M$-qubit Pauli string, \foreign{i.e.}, an $M$-fold tensor product of Pauli matrices.
\cref{seq:sigma_involution} together with the commutativity of Pauli matrices when acting on different qubits directly imply that \textbf{Pauli strings are Hermitian involutions},
\begin{equation}
	P^\dagger = P^{-1} = P.
\end{equation}
Using the above property, one can derive closed-form expressions for the complex exponential of a Pauli string, obtaining
\begin{equation}\label{seq:pauli_exp_closed}
	\begin{split}
		e^{\I\theta P} &= \sum_{k = 0}^{\infty} \frac{\I^k \theta^k}{k!} P^k \\
		&= \sum_{k = 0}^\infty \I^{2k} \frac{\theta^{2k}}{(2k)!} P^{2k} + \sum_{k = 0}^\infty \I^{2k + 1} 
		\frac{\theta^{2k + 1}}{(2k + 1)!} P^{2k + 1} \\
		&= \sum_{k = 0}^\infty (-1)^{k} \frac{\theta^{2k}}{(2k)!} I + \I \sum_{k = 0}^\infty (-1)^{k} 
		\frac{\theta^{2k + 1}}{(2k + 1)!} P \\
		&= \cos(\theta) I + \I \sin(\theta) P,
	\end{split}
\end{equation}
where $\theta \in \mathbb{R}$.

Given two Pauli strings $P_1$ and $P_2$,
by combining \cref{seq:pauli_anticommute} with the fact that Pauli matrices acting on different qubits commute, we immediately obtain
\begin{equation}\label{seq:pauli_string_product}
	P_1 P_2 = (-1)^k P_2 P_1,
\end{equation}
where $k$ denotes the number of pairwise non-trivial multiplications.
Consequently, \textbf{two Pauli strings either commute or anticommute} depending on whether the number of pairwise non-trivial multiplications is even or odd.
Combining this result with \cref{seq:pauli_exp_closed}, one immediately obtains a closed-form expression for the similarity transformation of a Pauli string $O$ by the complex exponential of another Pauli string $P$:
\begin{equation}\label{seq:pST}
	e^{-\I \theta P} O e^{\I \theta P} = \begin{cases}
		O, &[O,P] = 0 \\
		\cos(2\theta) O + \I \sin(2\theta)OP, &\{O,P\} = 0
	\end{cases}.
\end{equation}
In the non-trivial case, the transformation performs a continuous rotation of the Pauli string $O$ into $\I OP$.
As a result,
\textbf{the transformation is Clifford for $\boldsymbol{\theta = k\frac{\pi}{4}, k \in \mathbb{Z}}$}, while for $\theta = (2k+1)\frac{\pi}{4}$ it maps the original Pauli string $O$ into $\pm\I OP$.

Since we are primarily interested in quantum computing applications, where gates must be unitary, we limit the discussion to complex exponentials of Pauli strings.
In principle, closed-form expressions for the exponential and corresponding similarity transformation can also be obtained for real exponentials of Pauli strings, $e^{\theta P}$.
In this case, one simply replaces the trigonometric functions in \cref{seq:pauli_exp_closed,seq:pST} by their hyperbolic counterparts.

\section{Majorana Algebra}\label{ssec:majorana}

We now perform a similar analysis for Majorana fermions. For a single-particle fermionic state, we define two Majorana operators as
\begin{equation}\label{seq:gamma_1}
	\gamma_1 = a^\dagger + a
\end{equation}
and
\begin{equation}\label{seq:gamma_2}
	\gamma_2 = \I(a^\dagger - a),
\end{equation}
where $a$ and $a^\dagger$ are the usual fermionic annihilation and creation operators, respectively. As was the 
case with Pauli matrices, Majorana operators are Hermitian involutions,
\begin{equation}
	\begin{split}
		\gamma_1^2 &= (a^\dagger + a)^2 \\
		&= (\cancelto{0}{{a^\dagger}^2} + a^\dagger a + aa^\dagger  + \cancelto{0}{a^2}) \\
		&= 1
	\end{split}
\end{equation}
and
\begin{equation}
	\begin{split}
		\gamma_2^2 &= \I^2 (a^\dagger - a)^2 \\
		&= - (\cancelto{0}{{a^\dagger}^2} - a^\dagger a -aa^\dagger + \cancelto{0}{a^2}) \\
		&= 1,
	\end{split}
\end{equation}
where we used the $\{a, a^\dagger\} = 1$ fermionic anticommutation relation and the fact that $a^2 = {a^\dagger}^2 
= 0$. The product of the two Majorana operators yields
\begin{equation}
	\begin{split}
		\gamma_1 \gamma_2 &= \I (a^\dagger + a) (a^\dagger - a) \\
		&= \I (\cancelto{0}{{a^\dagger}^2} -a^\dagger a + a a^\dagger - \cancelto{0}{a^2})\\
		&= \I (I - 2 a^\dagger a)
	\end{split}
\end{equation}
and
\begin{equation}
	\begin{split}
		\gamma_2 \gamma_1 &= \I (a^\dagger - a) (a^\dagger + a) \\
		&= \I (\cancelto{0}{{a^\dagger}^2} + a^\dagger a - a a^\dagger - \cancelto{0}{a^2}) \\
		&= \I(2a^\dagger a - I) \\
		&= - \gamma_1 \gamma_2.
	\end{split}
\end{equation}
To facilitate a direct comparison with the Pauli group, we introduce an auxiliary operator $\gamma_3$ defined as
\begin{equation}
	\gamma_3 = I-2a^\dagger a,
\end{equation}
which, similar to $\gamma_1$ and $\gamma_2$, is a Hermitian involution,
\begin{equation}
	\gamma_3^2 = I.
\end{equation}
Furthermore, it can be shown that the three Majorana operators pairwise anticommute,
\begin{equation}
	\{\gamma_i, \gamma_j\} = 2 \delta_{ij}, \quad i,j \in \{1,2,3\}.
\end{equation}
As illustrated in \cref{sfig:majorana_cayley}, while the set $\{\gamma_0, \gamma_1, \gamma_2, \gamma_3\}$, with $\gamma_0 \equiv I$, is not closed under operator multiplication, closure is achieved by extending it to $\mathcal{M}_2 = \left\{\I^\alpha \gamma_i \mid \alpha\in\mathbb{Z}_4, i\in\{0,1,2,3\}\right\}$, called the Majorana group for two Majorana modes.
A comparison of \cref{sfig:pauli_cayley}(b) and \cref{sfig:majorana_cayley}(b) reveals an isomorphism between the single-qubit Pauli group and the Majorana group for two Majorana modes, denoted as $\mathcal{M}_2 \cong \mathcal{P}_1$.
Under this isomorphism, the Majorana operators are mapped to Pauli matrices as follows: $\gamma_1 \mapsto \sigma_x$, $\gamma_2 \mapsto \sigma_y$, and $\gamma_3 \mapsto \sigma_z$.

\begin{figure}
	\centering
	\subfloat[]{
		\begin{tabular}{c|cccc}
			\(\cdot\) & $\gamma_0$ & \cellcolor{colorA} $\gamma_1$ & \cellcolor{colorB} $\gamma_2$ & \cellcolor{colorC} $\gamma_3$ \\ \hline
			$\gamma_0$ & $\gamma_0$ &\cellcolor{colorA} $\gamma_1$ & \cellcolor{colorB} $\gamma_2$ & \cellcolor{colorC} $\gamma_3$ \\
			\cellcolor{colorA} $\gamma_1$ & \cellcolor{colorA} $\gamma_1$ & $\gamma_0$ & \cellcolor{colorK} $\I\gamma_3$ & \cellcolor{colorN} $-\I\gamma_2$  \\
			\cellcolor{colorB} $\gamma_2$ & \cellcolor{colorB} $\gamma_2$ & \cellcolor{colorO} $-\I\gamma_3$ & $\gamma_0$ & \cellcolor{colorI} $\I\gamma_1$  \\
			\cellcolor{colorC} $\gamma_3$ & \cellcolor{colorC} $\gamma_3$ & \cellcolor{colorJ} $\I\gamma_2$ & \cellcolor{colorM} $-\I\gamma_1$ & $\gamma_0$
		\end{tabular}
	}\\
	\subfloat[]{\resizebox{\textwidth}{!}{
			\begin{tabular}{c|cccccccccccccccc}
				\(\cdot\) & $\gamma_0$ & \cellcolor{colorA} $\gamma_1$ & \cellcolor{colorB} $\gamma_2$ & \cellcolor{colorC} $\gamma_3$ & \cellcolor{colorD} $-\gamma_0$ & \cellcolor{colorE} $-\gamma_1$ & \cellcolor{colorF} $-\gamma_2$ & \cellcolor{colorG} $-\gamma_3$ & \cellcolor{colorH} $\I\gamma_0$ & \cellcolor{colorI} $\I\gamma_1$ & \cellcolor{colorJ} $\I\gamma_2$ & \cellcolor{colorK} $\I\gamma_3$ & \cellcolor{colorL} $-\I\gamma_0$ & \cellcolor{colorM} $-\I\gamma_1$ & \cellcolor{colorN} $-\I\gamma_2$ & \cellcolor{colorO} $-\I\gamma_3$ \\ \hline
				$\gamma_0$ & $\gamma_0$ & \cellcolor{colorA} $\gamma_1$ & \cellcolor{colorB} $\gamma_2$ & \cellcolor{colorC} $\gamma_3$ & \cellcolor{colorD} $-\gamma_0$ & \cellcolor{colorE} $-\gamma_1$ & \cellcolor{colorF} $-\gamma_2$ & \cellcolor{colorG} $-\gamma_3$ & \cellcolor{colorH} $\I\gamma_0$ & \cellcolor{colorI} $\I\gamma_1$ & \cellcolor{colorJ} $\I\gamma_2$ & \cellcolor{colorK} $\I\gamma_3$ & \cellcolor{colorL} $-\I\gamma_0$ & \cellcolor{colorM} $-\I\gamma_1$ & \cellcolor{colorN} $-\I\gamma_2$ & \cellcolor{colorO} $-\I\gamma_3$ \\
				\cellcolor{colorA} $\gamma_1$ & \cellcolor{colorA} $\gamma_1$ & $\gamma_0$ & \cellcolor{colorK} $\I\gamma_3$ & \cellcolor{colorN} $-\I\gamma_2$ & \cellcolor{colorE} $-\gamma_1$ & \cellcolor{colorD} $-\gamma_0$ & \cellcolor{colorO} $-\I\gamma_3$ & \cellcolor{colorJ} $\I\gamma_2$ & \cellcolor{colorI} $\I\gamma_1$ & \cellcolor{colorH} $\I\gamma_0$ & \cellcolor{colorG} $-\gamma_3$ & \cellcolor{colorB} $\gamma_2$ & \cellcolor{colorM} $-\I\gamma_1$ & \cellcolor{colorL} $-\I\gamma_0$ & \cellcolor{colorC} $\gamma_3$ & \cellcolor{colorF} $-\gamma_2$ \\
				\cellcolor{colorB} $\gamma_2$ & \cellcolor{colorB} $\gamma_2$ & \cellcolor{colorO} $-\I\gamma_3$ & $\gamma_0$ & \cellcolor{colorI} $\I\gamma_1$ & \cellcolor{colorF} $-\gamma_2$ & \cellcolor{colorK} $\I\gamma_3$ & \cellcolor{colorD} $-\gamma_0$ & \cellcolor{colorM} $-\I\gamma_1$ & \cellcolor{colorJ} $\I\gamma_2$ & \cellcolor{colorC} $\gamma_3$ & \cellcolor{colorH} $\I\gamma_0$ & \cellcolor{colorE} $-\gamma_1$ & \cellcolor{colorN} $-\I\gamma_2$ & \cellcolor{colorG} $-\gamma_3$ & \cellcolor{colorL} $-\I\gamma_0$ & \cellcolor{colorA} $\gamma_1$ \\
				\cellcolor{colorC} $\gamma_3$ & \cellcolor{colorC} $\gamma_3$ & \cellcolor{colorJ} $\I\gamma_2$ & \cellcolor{colorM} $-\I\gamma_1$ & $\gamma_0$ & \cellcolor{colorG} $-\gamma_3$ & \cellcolor{colorN} $-\I\gamma_2$ & \cellcolor{colorI} $\I\gamma_1$ & \cellcolor{colorD} $-\gamma_0$ & \cellcolor{colorK} $\I\gamma_3$ & \cellcolor{colorF} $-\gamma_2$ & \cellcolor{colorA} $\gamma_1$ & \cellcolor{colorH} $\I\gamma_0$ & \cellcolor{colorO} $-\I\gamma_3$ & \cellcolor{colorB} $\gamma_2$ & \cellcolor{colorE} $-\gamma_1$ & \cellcolor{colorL} $-\I\gamma_0$ \\
				\cellcolor{colorD} $-\gamma_0$ & \cellcolor{colorD} $-\gamma_0$ & \cellcolor{colorE} $-\gamma_1$ & \cellcolor{colorF} $-\gamma_2$ & \cellcolor{colorG} $-\gamma_3$ & $\gamma_0$ & \cellcolor{colorA} $\gamma_1$ & \cellcolor{colorB} $\gamma_2$ & \cellcolor{colorC} $\gamma_3$ & \cellcolor{colorL} $-\I\gamma_0$ & \cellcolor{colorM} $-\I\gamma_1$ & \cellcolor{colorN} $-\I\gamma_2$ & \cellcolor{colorO} $-\I\gamma_3$ & \cellcolor{colorH} $\I\gamma_0$ & \cellcolor{colorI} $\I\gamma_1$ & \cellcolor{colorJ} $\I\gamma_2$ & \cellcolor{colorK} $\I\gamma_3$ \\
				\cellcolor{colorE} $-\gamma_1$ & \cellcolor{colorE} $-\gamma_1$ & \cellcolor{colorD} $-\gamma_0$ & \cellcolor{colorO} $-\I\gamma_3$ & \cellcolor{colorJ} $\I\gamma_2$ & \cellcolor{colorA} $\gamma_1$ & $\gamma_0$ & \cellcolor{colorK} $\I\gamma_3$ & \cellcolor{colorN} $-\I\gamma_2$ & \cellcolor{colorM} $-\I\gamma_1$ & \cellcolor{colorL} $-\I\gamma_0$ & \cellcolor{colorC} $\gamma_3$ & \cellcolor{colorF} $-\gamma_2$ & \cellcolor{colorI} $\I\gamma_1$ & \cellcolor{colorH} $\I\gamma_0$ & \cellcolor{colorG} $-\gamma_3$ & \cellcolor{colorB} $\gamma_2$ \\
				\cellcolor{colorF} $-\gamma_2$ & \cellcolor{colorF} $-\gamma_2$ & \cellcolor{colorK} $\I\gamma_3$ & \cellcolor{colorD} $-\gamma_0$ & \cellcolor{colorM} $-\I\gamma_1$ & \cellcolor{colorB} $\gamma_2$ & \cellcolor{colorO} $-\I\gamma_3$ & $\gamma_0$ & \cellcolor{colorI} $\I\gamma_1$ & \cellcolor{colorN} $-\I\gamma_2$ & \cellcolor{colorG} $-\gamma_3$ & \cellcolor{colorL} $-\I\gamma_0$ & \cellcolor{colorA} $\gamma_1$ & \cellcolor{colorJ} $\I\gamma_2$ & \cellcolor{colorC} $\gamma_3$ & \cellcolor{colorH} $\I\gamma_0$ & \cellcolor{colorE} $-\gamma_1$ \\
				\cellcolor{colorG} $-\gamma_3$ & \cellcolor{colorG} $-\gamma_3$ & \cellcolor{colorN} $-\I\gamma_2$ & \cellcolor{colorI} $\I\gamma_1$ & \cellcolor{colorD} $-\gamma_0$ & \cellcolor{colorC} $\gamma_3$ & \cellcolor{colorJ} $\I\gamma_2$ & \cellcolor{colorM} $-\I\gamma_1$ & $\gamma_0$ & \cellcolor{colorO} $-\I\gamma_3$ & \cellcolor{colorB} $\gamma_2$ & \cellcolor{colorE} $-\gamma_1$ & \cellcolor{colorL} $-\I\gamma_0$ & \cellcolor{colorK} $\I\gamma_3$ & \cellcolor{colorF} $-\gamma_2$ & \cellcolor{colorA} $\gamma_1$ & \cellcolor{colorH} $\I\gamma_0$ \\
				\cellcolor{colorH} $\I\gamma_0$ & \cellcolor{colorH} $\I\gamma_0$ & \cellcolor{colorI} $\I\gamma_1$ & \cellcolor{colorJ} $\I\gamma_2$ & \cellcolor{colorK} $\I\gamma_3$ & \cellcolor{colorL} $-\I\gamma_0$ & \cellcolor{colorM} $-\I\gamma_1$ & \cellcolor{colorN} $-\I\gamma_2$ & \cellcolor{colorO} $-\I\gamma_3$ & \cellcolor{colorD} $-\gamma_0$ & \cellcolor{colorE} $-\gamma_1$ & \cellcolor{colorF} $-\gamma_2$ & \cellcolor{colorG} $-\gamma_3$ & $\gamma_0$ & \cellcolor{colorA} $\gamma_1$ & \cellcolor{colorB} $\gamma_2$ & \cellcolor{colorC} $\gamma_3$ \\
				\cellcolor{colorI} $\I\gamma_1$ & \cellcolor{colorI} $\I\gamma_1$ & \cellcolor{colorH} $\I\gamma_0$ & \cellcolor{colorG} $-\gamma_3$ & \cellcolor{colorB} $\gamma_2$ & \cellcolor{colorM} $-\I\gamma_1$ & \cellcolor{colorL} $-\I\gamma_0$ & \cellcolor{colorC} $\gamma_3$ & \cellcolor{colorF} $-\gamma_2$ & \cellcolor{colorE} $-\gamma_1$ & \cellcolor{colorD} $-\gamma_0$ & \cellcolor{colorO} $-\I\gamma_3$ & \cellcolor{colorJ} $\I\gamma_2$ & \cellcolor{colorA} $\gamma_1$ & $\gamma_0$ & \cellcolor{colorK} $\I\gamma_3$ & \cellcolor{colorN} $-\I\gamma_2$ \\
				\cellcolor{colorJ} $\I\gamma_2$ & \cellcolor{colorJ} $\I\gamma_2$ & \cellcolor{colorC} $\gamma_3$ & \cellcolor{colorH} $\I\gamma_0$ & \cellcolor{colorE} $-\gamma_1$ & \cellcolor{colorN} $-\I\gamma_2$ & \cellcolor{colorG} $-\gamma_3$ & \cellcolor{colorL} $-\I\gamma_0$ & \cellcolor{colorA} $\gamma_1$ & \cellcolor{colorF} $-\gamma_2$ & \cellcolor{colorK} $\I\gamma_3$ & \cellcolor{colorD} $-\gamma_0$ & \cellcolor{colorM} $-\I\gamma_1$ & \cellcolor{colorB} $\gamma_2$ & \cellcolor{colorO} $-\I\gamma_3$ & $\gamma_0$ & \cellcolor{colorI} $\I\gamma_1$ \\
				\cellcolor{colorK} $\I\gamma_3$ & \cellcolor{colorK} $\I\gamma_3$ & \cellcolor{colorF} $-\gamma_2$ & \cellcolor{colorA} $\gamma_1$ & \cellcolor{colorH} $\I\gamma_0$ & \cellcolor{colorO} $-\I\gamma_3$ & \cellcolor{colorB} $\gamma_2$ & \cellcolor{colorE} $-\gamma_1$ & \cellcolor{colorL} $-\I\gamma_0$ & \cellcolor{colorG} $-\gamma_3$ & \cellcolor{colorN} $-\I\gamma_2$ & \cellcolor{colorI} $\I\gamma_1$ & \cellcolor{colorD} $-\gamma_0$ & \cellcolor{colorC} $\gamma_3$ & \cellcolor{colorJ} $\I\gamma_2$ & \cellcolor{colorM} $-\I\gamma_1$ & $\gamma_0$ \\
				\cellcolor{colorL} $-\I\gamma_0$ & \cellcolor{colorL} $-\I\gamma_0$ & \cellcolor{colorM} $-\I\gamma_1$ & \cellcolor{colorN} $-\I\gamma_2$ & \cellcolor{colorO} $-\I\gamma_3$ & \cellcolor{colorH} $\I\gamma_0$ & \cellcolor{colorI} $\I\gamma_1$ & \cellcolor{colorJ} $\I\gamma_2$ & \cellcolor{colorK} $\I\gamma_3$ & $\gamma_0$ & \cellcolor{colorA} $\gamma_1$ & \cellcolor{colorB} $\gamma_2$ & \cellcolor{colorC} $\gamma_3$ & \cellcolor{colorD} $-\gamma_0$ & \cellcolor{colorE} $-\gamma_1$ & \cellcolor{colorF} $-\gamma_2$ & \cellcolor{colorG} $-\gamma_3$ \\
				\cellcolor{colorM} $-\I\gamma_1$ & \cellcolor{colorM} $-\I\gamma_1$ & \cellcolor{colorL} $-\I\gamma_0$ & \cellcolor{colorC} $\gamma_3$ & \cellcolor{colorF} $-\gamma_2$ & \cellcolor{colorI} $\I\gamma_1$ & \cellcolor{colorH} $\I\gamma_0$ & \cellcolor{colorG} $-\gamma_3$ & \cellcolor{colorB} $\gamma_2$ & \cellcolor{colorA} $\gamma_1$ & $\gamma_0$ & \cellcolor{colorK} $\I\gamma_3$ & \cellcolor{colorN} $-\I\gamma_2$ & \cellcolor{colorE} $-\gamma_1$ & \cellcolor{colorD} $-\gamma_0$ & \cellcolor{colorO} $-\I\gamma_3$ & \cellcolor{colorJ} $\I\gamma_2$ \\
				\cellcolor{colorN} $-\I\gamma_2$ & \cellcolor{colorN} $-\I\gamma_2$ & \cellcolor{colorG} $-\gamma_3$ & \cellcolor{colorL} $-\I\gamma_0$ & \cellcolor{colorA} $\gamma_1$ & \cellcolor{colorJ} $\I\gamma_2$ & \cellcolor{colorC} $\gamma_3$ & \cellcolor{colorH} $\I\gamma_0$ & \cellcolor{colorE} $-\gamma_1$ & \cellcolor{colorB} $\gamma_2$ & \cellcolor{colorO} $-\I\gamma_3$ & $\gamma_0$ & \cellcolor{colorI} $\I\gamma_1$ & \cellcolor{colorF} $-\gamma_2$ & \cellcolor{colorK} $\I\gamma_3$ & \cellcolor{colorD} $-\gamma_0$ & \cellcolor{colorM} $-\I\gamma_1$ \\
				\cellcolor{colorO} $-\I\gamma_3$ & \cellcolor{colorO} $-\I\gamma_3$ & \cellcolor{colorB} $\gamma_2$ & \cellcolor{colorE} $-\gamma_1$ & \cellcolor{colorL} $-\I\gamma_0$ & \cellcolor{colorK} $\I\gamma_3$ & \cellcolor{colorF} $-\gamma_2$ & \cellcolor{colorA} $\gamma_1$ & \cellcolor{colorH} $\I\gamma_0$ & \cellcolor{colorC} $\gamma_3$ & \cellcolor{colorJ} $\I\gamma_2$ & \cellcolor{colorM} $-\I\gamma_1$ & $\gamma_0$ & \cellcolor{colorG} $-\gamma_3$ & \cellcolor{colorN} $-\I\gamma_2$ & \cellcolor{colorI} $\I\gamma_1$ & \cellcolor{colorD} $-\gamma_0$
			\end{tabular}
		}
	}
	\caption{\label{sfig:majorana_cayley}
		Cayley tables defining the Majorana algebra. (a) Cayley table for the Majorana operators. (b) Cayley table for the Majorana group for two Majorana Modes. Cells are color-coded to highlight closure under multiplication: (a) is not closed, while (b) is closed.
	}
\end{figure} 

For a system of $M$ fermionic modes, there exist $2M$ Majorana modes.
Following the standard approach, we define a Majorana string $\Gamma$ as a product of $\gamma_1$ and $\gamma_2$ operators acting on the various single-particle states. For the Majorana operators acting on the $j$th single-particle state, we adopt the notation
\begin{equation}
	\gamma_{p_{2j}} = \begin{cases}
		I, &\text{mode inactive}\\
		\gamma_1^{(j)} = a_j^\dagger + a_j, &\text{mode active}
	\end{cases}
\end{equation}
and
\begin{equation}
	\gamma_{p_{2j+1}} = \begin{cases}
		I, &\text{mode inactive}\\
		\gamma_2^{(j)} = \I (a_j^\dagger - a_j), &\text{mode active}
	\end{cases},
\end{equation}
where $a_j^\dagger$ and $a_j$ are the conventional fermionic creation and annihilation operators acting on the $j$th single-particle state.
The length $L$ of a Majorana string is defined as the number of its non-identity elements. Therefore, the maximum possible length is $2M$, which corresponds, up to a phase, to an $M$-fold product of $\gamma_3$ operators, each acting on a different single-particle state.

Due to their fermionic nature, Majorana operators acting on different single-particle states anticommute,
\begin{equation}\label{seq:acomm_m}
	\{\gamma_p^{(k)},\gamma_q^{(l)}\} = 2\delta_{pq} \delta_{kl}, \quad p,q \in \{1,2\}, k,l\in \{0,\ldots, M-1\}.
\end{equation}
Using \cref{seq:acomm_m}, it is straightforward to show that
\begin{equation}
	\Gamma^2 = (-1)^{\sum_{l = 1}^{L-1}l}I
\end{equation}
and
\begin{equation}
	\Gamma^\dagger = (-1)^{\sum_{l = 1}^{L-1}l}\Gamma,
\end{equation}
where $L$ denotes the length of the Majorana string $\Gamma$.
Thus, in stark contrast to the Pauli case, \textbf{Majorana strings come in two flavors,
	either Hermitian involutions,}
\begin{equation}
	\Gamma^2 = I \quad \text{and} \quad \Gamma^\dagger = \Gamma \quad \text{if} \quad L = 1,4,5,8,9,\ldots
\end{equation}
\textbf{or anti-Hermitian skew-involutions,}
\begin{equation}
	\Gamma^2 = -I \quad \text{and} \quad \Gamma^\dagger = -\Gamma \quad \text{if} \quad L = 2,3,6,7,10,11,\ldots.
\end{equation}
Since our primary interest is in quantum computing applications, we focus on forming unitary operators by exponentiating Majorana strings. There are two distinct cases, depending on the nature of $\Gamma$:
\begin{enumerate}
	\item For a Hermitian Majorana string $\Gamma$, we have:
	\begin{equation}\label{seq:m_exp_hermitian_closed}
		\begin{split}
			e^{\I \theta \Gamma} &= \sum_{k = 0}^\infty \frac{(-1)^{k(1+2m)} \theta^{2k}}{(2k)!}I + \I \sum_{k = 	
				0}^{\infty} \frac{(-1)^{k(1 + 2m)} \theta^{2k+1}}{(2k+1)!} \Gamma \\
			&= \sum_{k = 0}^\infty \frac{(-1)^k \theta^{2k}}{(2k)!}I + \I \sum_{k = 	
				0}^{\infty} \frac{(-1)^k \theta^{2k+1}}{(2k+1)!} \Gamma \\
			&= \cos(\theta) I + \I\sin(\theta) \Gamma.
		\end{split}
	\end{equation}
	
	\item For an anti-Hermitian Majorana string $\Gamma$, the corresponding unitary exponential is given by
	\begin{equation}\label{seq:m_exp_antihermitian_closed}
		\begin{split}
			e^{\theta \Gamma} &= \sum_{k = 0}^{\infty}\frac{\theta^k \Gamma^k}{k!} \\
			&= \sum_{k = 0}^{\infty}\frac{\theta^{2k} \Gamma^{2k}}{(2k)!} + \sum_{k = 0}^{\infty}\frac{\theta^{2k+1} 
				\Gamma^{2k+1}}{(2k+1)!} \\
			&= \sum_{k = 0}^{\infty}\frac{(-1)^k \theta^{2k}}{(2k)!} I + \sum_{k = 0}^{\infty}\frac{(-1)^k 
				\theta^{2k+1}}{(2k+1)!} \Gamma \\
			&= \cos(\theta) I + \sin(\theta) \Gamma.
		\end{split}
	\end{equation}
\end{enumerate}

Let $\Gamma_1$ and $\Gamma_2$ be two Majorana strings with lengths $L_1$ and $L_2$, respectively.
From \cref{seq:acomm_m}, we directly obtain that
\begin{equation}\label{seq:majorana_product}
	\Gamma_1 \Gamma_2 = (-1)^{L_1 L_2 - c} \Gamma_2 \Gamma_1,
\end{equation}
where $c$ denotes the number of operators that appear in both strings.
Thus, we conclude that \textbf{two Majorana strings either commute or anticommute}, depending 
on their lengths and the number of common elements.
Using this result, along with the closed-form expressions in \cref{seq:m_exp_antihermitian_closed,seq:m_exp_hermitian_closed}, we readily arrive at the following closed-form expressions for the unitary transformation of a Majorana string $O$ by the exponential of the Majorana string $\Gamma$:
\begin{equation}\label{seq:mST_hermitian}
	e^{-\I\theta\Gamma}Oe^{\I\theta\Gamma} =
	\begin{cases}
		O, &[O,\Gamma] = 0\\
		\cos(2\theta)O + \I\sin(2\theta)O\Gamma, &\{O,\Gamma\}=0
	\end{cases},
	\quad L = 1,4,5,8,9,\ldots
\end{equation}
and
\begin{equation}\label{seq:mST_antihermitian}
	e^{-\theta\Gamma}Oe^{\theta\Gamma} =
	\begin{cases}
		O, &[O,\Gamma] = 0\\
		\cos(2\theta)O + \sin(2\theta)O\Gamma, &\{O,\Gamma\}=0
	\end{cases},
	\quad L = 2,3,6,7,10,11,\ldots
\end{equation}
In the non-trivial case, the above transformations perform a continuous rotation of the Pauli string $O$ into $(\I) O\Gamma$ and 
\textbf{become Clifford for $\boldsymbol{\theta = k\frac{\pi}{4}, k \in \mathbb{Z}}$}.

As in the Pauli case, the corresponding expressions for a Hermitian exponential can be obtained by replacing trigonometric functions by their hyperbolic counterparts.

\section{Fermionic Algebra}\label{ssec:fermion}

Now we move on to the case of fermions. As before, given a single-particle state, the corresponding annihilation
and creation operators are denoted as $a$ and $a^\dagger$, respectively. In contrast to the previously discussed Pauli 
matrices and Majorana operators, which are Hermitian involutions, the creation and annihilation operators are nilpotent,
\begin{equation}\label{f_nilpotent}
	a^2 = {a^\dagger}^2 = 0.
\end{equation}
As shown in \cref{sfig:fermion_cayley}, although the set $\{I, a, a^\dagger\}$ is not closed under multiplication, closure is attained by the augmented set
$\mathcal{F}_1 = \{I, a, a^\dagger, n, h, 0\}$, where
\begin{equation}
	n= a^\dagger a
\end{equation}
and
\begin{equation}
	h = a a^\dagger
\end{equation}
denote the particle and hole number operators, respectively.
In contrast to $\mathcal{P}_1$ and $\mathcal{M}_2$, $\mathcal{F}_1$ is not a group since, with the exception of the identity, its elements are not invertible.
Consequently, we refer to $\mathcal{F}_1$ as the fermionic monoid for a single fermionic mode.
\begin{figure}
	\centering
	\subfloat[]{
		\begin{tabular}{c|ccc}
			\(\cdot\) & $I$ & \cellcolor{colorA} $a$ & \cellcolor{colorB} $a^\dagger$ \\ \hline
			$I$ & $I$ &\cellcolor{colorA} $a$ & \cellcolor{colorB} $a^\dagger$ \\
			\cellcolor{colorA} $a$ & \cellcolor{colorA} $a$ & \cellcolor{colorO} 0 & \cellcolor{colorD} $h$  \\
			\cellcolor{colorB} $a^\dagger$ & \cellcolor{colorB} $a^\dagger$ & \cellcolor{colorC} $n$ & \cellcolor{colorO} $0$
		\end{tabular}
	}\\
	\subfloat[]{
		\begin{tabular}{c|cccccc}
			\(\cdot\) & $I$ & \cellcolor{colorA} $a$ & \cellcolor{colorB} $a^\dagger$ & \cellcolor{colorC} $n$ & \cellcolor{colorD} $h$ & \cellcolor{colorO} 0 \\ \hline
			$I$ & $I$ &\cellcolor{colorA} $a$ & \cellcolor{colorB} $a^\dagger$ & \cellcolor{colorC} $n$ & \cellcolor{colorD} $h$ & \cellcolor{colorO} 0 \\
			\cellcolor{colorA} $a$ & \cellcolor{colorA} $a$ & \cellcolor{colorO} 0 & \cellcolor{colorD} $h$ & \cellcolor{colorA} $a$ & \cellcolor{colorO} 0 & \cellcolor{colorO} 0 \\
			\cellcolor{colorB} $a^\dagger$ & \cellcolor{colorB} $a^\dagger$ & \cellcolor{colorC} $n$ & \cellcolor{colorO} $0$ & \cellcolor{colorO} 0 & \cellcolor{colorB} $a^\dagger$ & \cellcolor{colorO} 0\\
			\cellcolor{colorC} $n$ & \cellcolor{colorC} $n$ & \cellcolor{colorO} 0 & \cellcolor{colorB} $a^\dagger$ & \cellcolor{colorC} $n$ & \cellcolor{colorO} 0 & \cellcolor{colorO} 0 \\
			\cellcolor{colorD} $h$ & \cellcolor{colorD} $h$ & \cellcolor{colorA} $a$ & \cellcolor{colorO} 0 & \cellcolor{colorO} 0 & \cellcolor{colorD} $h$ & \cellcolor{colorO} 0\\
			\cellcolor{colorO} 0 & \cellcolor{colorO} 0 & \cellcolor{colorO} 0 & \cellcolor{colorO} 0 & \cellcolor{colorO} 0 & \cellcolor{colorO} 0 & \cellcolor{colorO} 0
		\end{tabular}
	}
	\caption{\label{sfig:fermion_cayley}
		Cayley tables defining the Fermionic algebra. (a) Cayley table for the fermionic creation and annihilation operators. (b) Cayley table for the single-state fermionic monoid. Cells are color-coded to highlight closure under multiplication: (a) is not closed, while (b) is closed.
	}
\end{figure}

The elements of the fermionic monoid for $M$ fermionic modes, $\mathcal{F}_M$, are $M$-fold products of $f \in \mathcal{F}_1$ operators each acting on a different single particle state, called fermionic strings.
The corresponding algebra is governed by the fermionic anticommutation relations, which read
\begin{equation}
	\{a_p, a_q\} = \{a_p^\dagger, a_q^\dagger\} = 0, \quad \{a_p, a_q^\dagger\} = \delta_{pq}.
\end{equation}
The following commutation relations will also be useful in the subsequent discussions:
\begin{equation}\label{seq:com_number}
	[a_p, n_q] = a_p \delta_{pq}, \quad [n_p, a_q^\dagger] = a_p^\dagger \delta_{pq}, \quad [h_p, a_q] = a_p \delta_{pq}, \quad [a_p^\dagger, h_q] = a_p^\dagger \delta_{pq}.
\end{equation}
In this work, a generic fermionic string $F$ is expressed as
\begin{equation}\label{seq:f_string}
	\begin{split}
		F &= a_{p_1}^\dagger \cdots a_{p_k}^\dagger a_{q_l} \cdots a_{q_1} h_{r_1} \cdots h_{r_m} n_{s_1} \cdots n_{s_n}\\
		&= a_{q_1 \ldots q_l}^{p_1 \ldots p_k} h_{r_1 \ldots r_m} n_{s_1 \ldots s_n},
	\end{split}
\end{equation} 
where all indices are assumed to be distinct.
The length $L$ of a fermionic string is defined as the number of creation and annihilation operators that it contains, including those arising from number operators.
Equivalently, the length is equal to twice the many-body rank of the fermionic operator.
If $F$ is comprised exclusively of number operators, then it is Hermitian and idempotent.
Otherwise, it is nilpotent.
Using these properties, we readily obtain the following closed-form expressions for exponentials of a single fermionic string:
\begin{equation}
	e^{\theta F} = \begin{cases}
		I + (e^\theta - 1) F, &F^2 = F\\
		I + \theta F, &F^2 = 0
	\end{cases}.
\end{equation}

In contrast to the Majorana case, fermionic strings are not, in general, Hermitian nor anti-Hermitian.
As a result, it is not possible to construct a unitary operator by exponentiating a single fermionic string.
Nevertheless, a fermionic string can always be expressed as a sum of two components, one Hermitian and the other anti-Hermitian, \foreign{i.e.},
\begin{equation}
	\begin{split}
		F &= \frac{1}{2}(F-F^\dagger) + \frac{1}{2}(F+F^\dagger)\\
		&= \frac{1}{2}A + \frac{1}{2}H,
	\end{split}
\end{equation}
where $A = F - F^\dagger$ and $H = F + F^\dagger$ denote the anti-Hermitian and Hermitian components, respectively.
Similar to the Majorana case, elementary unitary exponentials can be constructed as real exponentials of $A$, $e^{\theta A}$, or complex exponentials of $H$, $e^{\I\theta H}$.
Assuming that $F$ is not a product of number operators, it is straightforward to show that $A^3 = -A$ and $H^3 = H$.
Using these relations, one can Taylor-expand the corresponding exponentials and arrive at the following closed-form expressions:
\begin{equation}\label{seq:f_exp_closed_antihermitian}
	e^{\theta A} = I + \sin(\theta)A + [1-\cos(\theta)]A^2
\end{equation}
and
\begin{equation}\label{seq:f_exp_closed_hermitian}
	e^{\I \theta H} = I + \I \sin(\theta) H + [\cos(\theta) - 1]H^2.
\end{equation}
Note that
\begin{equation}\label{seq:projection_operator}
	H^2 = -A^2 = FF^\dagger + F^\dagger F = (h_{p_1 \ldots p_k} n_{q_1 \ldots q_l} + h_{q_1 \ldots q_l} n_{p_1 \ldots p_k}) h_{r_1 \ldots r_m} n_{s_1 \ldots s_n}
\end{equation}
is a projection operator onto the domains of operators $F$ and $F^\dagger$.
If $F$ is a product of number operators, denoted as $N$, then $A = 0$ and $H = N$ since $N$ is already Hermitian.
Taking advantage of the fact that $N$ is idempotent, we obtain
\begin{equation}\label{seq:f_exp_closed_n}
	e^{\I \theta N} = I + (e^{\I \theta} - 1)N.
\end{equation}

As was the case with the Pauli and Majorana Clifford transformations, we are seeking similarity transformations
generated by (anti-)Hermitian combinations of a single fermionic string $F$ that preserve $\mathcal{F}_M$.
To find under which conditions such a transformation is possible, we start from the following closed-form expressions that we recently derived (Ref.\ [64] of the main text):
\begin{equation}\label{seq:fST_antihermitian}
	e^{-\theta A} O e^{\theta A} = O + \frac{\sin\left( \sqrt{\alpha} \theta \right)}{\sqrt{\alpha}} [O,A] + \frac{1-\cos\left( \sqrt{\alpha} \theta \right)}{\alpha} [[O,A],A]
\end{equation}
and
\begin{equation}\label{seq:fST_hermitian}
	e^{-\I \theta H} O e^{\I \theta H} = O + \I \frac{\sin\left(\sqrt{\beta} \theta\right)}{\sqrt{\beta}} [O,H] + \frac{\cos\left( \sqrt{\beta} \theta \right) - 1}{\beta}[[O,H],H],
\end{equation}
with $\alpha$ and $\beta$ taking values of either 1 ($A[O,A]A = 0$, $H[O,H]H = 0$) or 4 ($A[O,A]A = [O,A]$, $H[O,H]H = [O,H]$) depending on the structures of the $O$ and $F$ fermionic strings.
A quick comparison of \cref{seq:fST_antihermitian,seq:fST_hermitian} with the corresponding expressions for Pauli, \cref{seq:pST}, and Majorana, \cref{seq:mST_hermitian,seq:mST_antihermitian}, strings immediately reveals the complexity of the fermionic case.
Using the definitions of $A$ and $H$, it is straightforward to show that, in general, the above transformations convert $O$ to a linear combination of 13 fermionic strings.
Therefore, without placing any restrictions on the form of $F$, it is not possible to find a particular value of $\theta$ for which the above transformations become Clifford.

Since both kinds of transformations have similar structure, we focus on the one based on an anti-Hermitian generator,
\cref{seq:fST_antihermitian}, with the string $F$ given by \cref{seq:f_string} and $O$ expressed as
\begin{equation}\label{seq:o_string}
	O= a_{u_1 \ldots u_f}^{t_1 \ldots t_e} h_{v_1 \ldots v_g} n_{w_1 \ldots w_h}.
\end{equation}
We begin by focusing on constraints arising when $F$ and $O$ have no common indices.
In this case, depending on their lengths, the two strings will either commute or anticommute.
Specifically, we have
\begin{equation}
	O F = (-1)^{L_F L_O} F O,
\end{equation}
with $L_F$ and $L_O$ denoting the lengths of the $F$ and $O$ strings.
If at least one of the strings has even length, then they commute and the transformation is trivial for any value of $\theta$. Consequently, if the generator $A$ has even length, no restrictions arise when it shares no indices with $O$.
If both strings have odd length, then they anticommute, and using \cref{seq:f_exp_closed_antihermitian}, we obtain
\begin{equation}\label{seq:ST-half}
	\begin{split}
		e^{-\theta A} O e^{\theta A} &= O + \sin(2\theta)OA + [1-\cos(2\theta)]OA^2\\
		&= O + \sin(2\theta)O(F-F^\dagger) - 2\sin^2(\theta) O (FF^\dagger + F^\dagger F)\\
		&= \left[1 - 2\sin^2(\theta)(FF^\dagger + F^\dagger F) \right] O + \sin(2\theta)O(F-F^\dagger).
	\end{split}
\end{equation}
We see that even the anticommuting case is more complicated than for Majorana operators, as, in general, the transformation results in 5 strings rather than 2.
Unless we enforce some constraints on the form of $F$, no value of $\theta$ exists that can result in a single string.
Indeed, the number of terms can be reduced to 3 if the 
\begin{equation}
	FF^\dagger + F^\dagger F = (h_{p_1 \ldots p_k} n_{q_1 \ldots q_l} + h_{q_1 \ldots q_l} n_{p_1 \ldots p_k}) h_{r_1 \ldots r_m} n_{s_1 \ldots s_n} = 1
\end{equation}
condition holds.
The above equality can only be satisfied if $F = a_p^\dagger$ (since $A$ is anti-Hermitian, the $F = a_p$ case is not independent).
In this case, if $\theta = k\frac{\pi}{2}$, $k\in\mathbb{Z}$, the transformation at most inverts the sign of the original string.
The above analysis already places severe constraints on transformations generated by strings of odd length:
\textbf{the only possible, if any, fermionic Clifford unitaries generated by odd-length strings must be of the form $\boldsymbol{A^p = a_p^\dagger - a_p}$ with $\boldsymbol{\theta = k\frac{\pi}{2}}$, $\boldsymbol{k \in \mathbb{Z}}$}.

Next, we proceed to the case where $O$ and $F$ have one common index, denoted as $c$.
There are 9 possibilities to consider, depending on whether the common index is in the creation/annihilation, particle number, or hole number component of $O$ and correspondingly for $F$.
Note that since $A$ contains both $F$ and $F^\dagger$, we do not need to differentiate between the cases where the common index is creation or annihilation.

We begin with the simpler case in which $F$ and $O$ share an index in their number operator parts.
Since the only potential Clifford unitaries generated by odd-length strings have the form $A^p = a_p^\dagger - a_p$, $F$ must have an even length.
First, we assume that $F$ and $O$ have a common particle number operator, \foreign{i.e.},
\begin{equation}\label{seq:f_nc}
	\begin{split}
		F &= a_{q_1 \ldots q_l}^{p_1 \ldots p_k} h_{r_1 \ldots r_m} n_{s_1 \ldots s_{i-1} c s_{i+1} \ldots s_n}\\
		&= a_{q_1 \ldots q_l}^{p_1 \ldots p_k} h_{r_1 \ldots r_m} n_{s_1 \ldots s_{i-1} s_{i+1} \ldots s_n} n_c\\
		&= \tilde{F} n_c
	\end{split}
\end{equation}
and
\begin{equation}\label{seq:o_nc}
	\begin{split}
		O &= a_{u_1 \ldots u_f}^{t_1 \ldots t_e} h_{v_1 \ldots v_g} n_{w_1 \ldots w_{j-1} c w_{j+1} \ldots w_h}\\
		&= a_{u_1 \ldots u_f}^{t_1 \ldots t_e} h_{v_1 \ldots v_g} n_{w_1 \ldots w_{j-1} w_{j+1} \ldots w_h} n_c\\
		&= \tilde{O} n_c.
	\end{split}
\end{equation}
Since $\tilde{A} = \tilde{F} - \tilde{F}^\dagger$, $\tilde{O}$, and $n_c$ share no indices and number operators are idempotent, the commutators appearing in \cref{seq:fST_antihermitian} become
\begin{equation}
	[O, A] = [O, \tilde{A}]
\end{equation}
and
\begin{equation}
	[[O, A],A] = [[O, \tilde{A}], \tilde{A}].
\end{equation}
This reduces to the previously examined case of operators with no common indices.
Hence, no constraints arise for generators of even length. 
The same is true when $F$ and $O$ have a common hole number operator. 
Next, we assume that a particle number operator in $F$ shares an index with a hole number operator in $O$, \foreign{i.e.}, $F$ is as in \cref{seq:f_nc} and $O$ is given by
\begin{equation}
	\begin{split}
		O &= a_{u_1 \ldots u_f}^{t_1 \ldots t_e} h_{v_1 \ldots v_{j-1} c v_{j+1} \ldots v_g} n_{w_1 \ldots w_h}\\
		&= a_{u_1 \ldots u_f}^{t_1 \ldots t_e} h_{v_1 \ldots v_{j-1} v_{j+1} \ldots v_g} n_{w_1 \ldots w_h} h_c\\
		&= \tilde{O} h_c.
	\end{split}
\end{equation}
It is straightforward to show that the transformation is trivial since
\begin{equation}
	[O,A] = \tilde{O}\tilde{A} \cancelto{0}{h_c n_c} -\tilde{A} \tilde{O} \cancelto{0}{n_c h_c} = 0.
\end{equation}
As before, no constraints arise for generators $A$ with even length.
The same applies to the case where a hole number operator in $F$ shares an index with a particle number operator in $O$.

Subsequently, we consider the scenario where the common index involves a number operator in one string and a creation/annihilation operator on the other.
We begin with the case where a particle number operator in $F$ has the same index as a creation operator in $O$.
As before, $F$ must have even length since the only viable generators of odd length have the form $A^p = a_p^\dagger - a_p$.
The fermionic string $F$ is written as \cref{seq:f_nc} and $O$ is given by
\begin{equation}\label{seq:o_string_ad}
	O = a_{u_1 \ldots u_f}^{t_1 \ldots t_{j-1} c t_{j+1} \ldots t_e} h_{v_1 \ldots v_g} n_{w_1 \ldots w_h}.
\end{equation}
The single commutator yields
\begin{equation}
	\begin{split}
		[O,A] &= [O, \tilde{A}n_c]\\
		&= \cancelto{0}{[O,\tilde{A}]}n_c + \tilde{A}[O,n_c]\\
		&= \cancelto{0}{\tilde{A}On_c} - \tilde{A} n_c O\\
		&= -O \tilde{A},
	\end{split}
\end{equation}
where we used \cref{seq:com_number} and the fact that $[O,\tilde{A}] = 0$ since $\tilde{A}$ has an even length and $O$ and $\tilde{A}$ have no indices in common.
In the same manner, the double commutator becomes
\begin{equation}
	\begin{split}
		[[O,A],A] &= - [O\tilde{A}, \tilde{A}n_c]\\
		&= \tilde{A} O \tilde{A}\\
		&= O \tilde{A}^2.
	\end{split}
\end{equation}
In this case, $A[O,A]A = 0$, implying that $\alpha = 1$ in \cref{seq:fST_antihermitian}.
As a result, the final transformation reads
\begin{equation}\label{seq:fST_antihermitian_a_n}
	e^{-\theta A} O e^{\theta A} = O - \sin(\theta) O \tilde{A} + [1-\cos(\theta)]O\tilde{A}^2.
\end{equation}
A quick inspection of \cref{seq:fST_antihermitian_a_n} reveals that, aside from the trivial $\theta = 0$ case, it is not possible to transform $O$ to a single fermionic string.
Similar results can be obtained when the common index involves an annihilation operator in $O$, and when the shared index involves a hole number operator in $F$.
This places the first constraint on even-length generators, \foreign{i.e.}, \textbf{fermionic Clifford unitaries cannot contain number operators when the generator is anti-Hermitian}.

Next, we move to the scenario where the common index involves a number operator in $O$ and a creation/annihilation operator in $F$.
Without loss of generality, we assume that a particle number operator in $O$ shares an index with a creation operator in $F$.
Thus, we write $O$ as in \cref{seq:o_nc} and $F$ is given by
\begin{equation}
	F = a_{q_1 \ldots q_l}^{p_1 \ldots p_{i-1} c p_{i+1} \ldots p_k},
\end{equation}
where we used the fact that strings containing number operators cannot generate fermionic Clifford transformations.
The expressions for the single and double commutators become
\begin{equation}
	\begin{split}
		[O,A] &= \tilde{O}F + F^\dagger \tilde{O}\\
		&=\tilde{O}\left[ F + (-1)^{L_F L_O} F^\dagger \right]
	\end{split}
\end{equation}
and
\begin{equation}
	\begin{split}
		[[O,A],A] &= \cancelto{0}{[\tilde{O}F,F]} + [F^\dagger,\tilde{O}F] + (-1)^{L_F L_O} [\tilde{O}F^\dagger,F] + (-1)^{L_F L_O}\cancelto{0}{[F^\dagger, \tilde{O}F^\dagger]}\\
		&= 2\tilde{O} \left[ (-1)^{L_F L_O}F^\dagger F - FF^\dagger \right],
	\end{split}
\end{equation}
respectively.
In this case, it can be shown that $A[O,A]A = [O,A]$, meaning that $\alpha = 4$ in \cref{seq:fST_antihermitian}:
\begin{equation}\label{seq:fST_antihermitian_n_a}
	\begin{split}
		e^{-\theta A} O e^{\theta A} &= O + \frac{1}{2} \sin(2\theta)\tilde{O}\left[F + (-1)^{L_F L_O} F^\dagger \right]
		+ \sin^2(\theta)\tilde{O}\left[(-1)^{L_F L_O} F^\dagger F - F F^\dagger \right]\\
		&= O + \frac{1}{2} \sin(2\theta)\tilde{O}\left[F + (-1)^{L_F L_O} F^\dagger \right]
		+\sin^2(\theta) \tilde{O} \left[ (-1)^{L_F L_O} h_c \tilde{F}^\dagger \tilde{F} - n_c \tilde{F} \tilde{F}^\dagger \right]\\
		&= O \left(1 - \sin^2(\theta) \tilde{F} \tilde{F}^\dagger \right) + (-1)^{L_F L_O} \sin^2(\theta) \tilde{O} h_c \tilde{F}^\dagger \tilde{F} + \frac{1}{2} \sin(2\theta)\tilde{O}\left[F + (-1)^{L_F L_O} F^\dagger \right],
	\end{split}
\end{equation}
where $\tilde{F}$ denotes the string $F$ with the common index removed.
At this point, we need to distinguish between strings $F$ with even and odd lengths.
According to the restrictions imposed to odd-length generators, $F = a_p^\dagger$, $\tilde{F} = 1$, and $\theta = k\frac{\pi}{2}$, $k\in\mathbb{Z}$. For even values of $k$ the transformation of \cref{seq:fST_antihermitian_n_a} is trivial, whereas for odd values of $k$ it reduces to
\begin{equation}
	e^{-(2k+1)\frac{\pi}{2} A^p} O e^{(2k+1)\frac{\pi}{2} A^p} = (-1)^{L_O} \tilde{O}h_c.
\end{equation}
As a result, this type of transformation at most converts the original string $O$ to the new string $\tilde{O} h_c$, an action that is consistent with a Clifford transformation.
Therefore, no additional constraints arise in this case.

If the string $F$ has an even length, \cref{seq:fST_antihermitian_n_a} becomes
\begin{equation}
	\begin{split}
		e^{-\theta A} O e^{\theta A} &= O \left(1 - \sin^2(\theta) \tilde{F} \tilde{F}^\dagger \right) + \sin^2(\theta) \tilde{O} h_c \tilde{F}^\dagger \tilde{F} + \frac{1}{2} \sin(2\theta)\tilde{O}\left[F + F^\dagger \right]\\
		&= O \left[1 - \sin^2(\theta) \left(\tilde{F} \tilde{F}^\dagger + \tilde{F}^\dagger \tilde{F} \right) \right] + \sin^2(\theta)\tilde{O}\tilde{F}^\dagger \tilde{F} + \frac{1}{2} \sin(2\theta)\tilde{O}\left[F + F^\dagger \right],
	\end{split}
\end{equation}
where in the last step we used the fact that $h_c = 1 - n_c$.
For $F$ to be a viable generator of Clifford unitaries, the above transformation needs to at most convert the original string to a single new one.
The $O$ string can be eliminated if
\begin{equation}
	\tilde{F}\tilde{F}^\dagger + \tilde{F}^\dagger \tilde{F} = 1.
\end{equation}
The above relation can be satisfied if $\tilde{F} = a_d^\dagger$ or $\tilde{F} = a_d$, giving rise to the $A^{cd} = a_c^\dagger a_d^\dagger - a_d a_c$ and $A_d^c = a_c^\dagger a_d - a_d^\dagger a_c$ generators, respectively, where we assumed, without loss of generality, that $c < d$.
A single fermionic string can be produced for $\theta = (2k+1)\frac{\pi}{2}$, $k \in \mathbb{Z}$.
The final transformations read
\begin{equation}
	e^{-(2k+1)\frac{\pi}{2}A^{cd}} O e^{(2k+1)\frac{\pi}{2} A^{cd}} = \tilde{O}h_d
\end{equation}
and
\begin{equation}
	e^{-(2k+1)\frac{\pi}{2}A_d^c} O e^{(2k+1)\frac{\pi}{2} A_d^c} = \tilde{O}n_d.
\end{equation}
Furthermore, for $2k\pi$, $k\in\mathbb{Z}$, the transformation is trivial.
Consequently, the following constraints need to be imposed when the generators have even length:
\textbf{ the only possible, if any, fermionic Clifford unitaries generated by even-length strings must be pair operators of the form $\boldsymbol{A^{pq} = a_p^\dagger a_q^\dagger - a_q a_p}$ and $\boldsymbol{A_p^q = a_q^\dagger a_p - a_p^\dagger a_q}$ with $\boldsymbol{\theta = k\frac{\pi}{2}}$, $\boldsymbol{k\in\mathbb{Z}}$}.

Now we proceed to the final case for one common index, namely, when $F$ and $O$ share an index among their creation and annihilation operators.
If $F$ and $O$ have a common creation operator, we write $O$ as in \cref{seq:o_string_ad} and express $F$ as
\begin{equation}
	A = a_c^\dagger f - f^\dagger a_c,
\end{equation}
with $f = 1$ for $A^c$, $f = a_d^\dagger$ for $A^{cd}$, and $f = a_d$ in the case of $A_d^c$. Without loss of generality, we have assumed that $c<d$.
We need to examine whether the corresponding transformations with $\theta = k\frac{\pi}{2}$, $k\in\mathbb{Z}$, remain viable, \foreign{i.e.}, whether they generate at most a single fermionic string.

The single and double commutators yield
\begin{equation}
	\begin{split}
		[O, A] &= f^\dagger a_c O - O f^\dagger a_c\\
		&= (-1)^{L_f+j+1} \tilde{O} f^\dagger \left((-1)^{L_f L_O} h_c + (-1)^{L_O} n_c\right)
	\end{split}
\end{equation}
and
\begin{equation}
	\begin{split}
		[[O,A], A] &= (-1)^{L_f+j+1} [\tilde{O} f^\dagger \left((-1)^{L_f L_O} h_c + (-1)^{L_O} n_c\right), a_c^\dagger f - f^\dagger a_c]\\
		&= (-1)^{L_f + j+ 1} \tilde{O} \left( (-1)^{L_O + L_f} a_c^\dagger (ff^\dagger + f^\dagger f) + (-1)^{L_f L_O + 1} 2{f^\dagger}^2 a_c \right),
	\end{split}
\end{equation}
where $L_f = 0$ in the case of $f = 1$ and $L_f = 1$ for $f = a_d^\dagger$, $f = a_d$.
It can be shown that $A[O,A]A = (1-L_f)[O,A]$ and, thus, the parameter $\alpha$ in \cref{seq:fST_antihermitian} equals 4 in the case of $A^c$ and 1 for $A^{cd}$ and $A_d^c$. The corresponding transformations read
\begin{equation}
	\begin{split}
		e^{-\theta A^c} O e^{\theta A^c} &= O + \frac{1}{2} \sin(2\theta) (-1)^{j+1} \tilde{O} \left(h_c + (-1)^{L_O} n_c\right) + \sin^2(\theta) (-1)^{j+ 1} \tilde{O} \left( (-1)^{L_O} a_c^\dagger - a_c \right)\\
		&=(1-\sin^2(\theta))O + (-1)^j\sin^2(\theta) \tilde{O}a_c + \frac{1}{2} \sin(2\theta) (-1)^{j+1} \tilde{O} (h_c + (-1)^{L_O} n_c),
	\end{split}
\end{equation}
\begin{equation}
	\begin{split}
		e^{-\theta A^{cd}} O e^{\theta A^{cd}} &= O +\sin(\theta)(-1)^{L_O + j} \tilde{O} a_d \left(h_c + n_c\right) + \left(1 - \cos(\theta)\right) (-1)^{L_O + j+ 1} \tilde{O} a_c^\dagger (n_d + h_d)\\
		&= \cos(\theta) O + (-1)^{L_O + j} \sin(\theta) \tilde{O} a_d,
	\end{split}
\end{equation}
and
\begin{equation}
	\begin{split}
		e^{-\theta A_d^c} O e^{\theta A_d^c} &= O + \sin(\theta) (-1)^{L_O+j} \tilde{O} a_d^\dagger \left( h_c + n_c\right) + \left( 1 - \cos(\theta)\right) (-1)^{L_O+j+1} \tilde{O} a_c^\dagger (h_d + n_d)\\
		&= \cos(\theta) O + (-1)^{L_O+j} \tilde{O}a_d^\dagger.
	\end{split}
\end{equation}
For $\theta = k\frac{\pi}{2}$, $k\in\mathbb{Z}$, the above transformations at most convert $O$ to $\tilde{O}a_c$, $\tilde{O}a_d$, and $\tilde{O}a_d^\dagger$ for $A^c$, $A^{cd}$, and $A_d^c$, respectively.
Similar results can be obtained when the common indices involve two annihilation operators, and a creation operator on one string and an annihilation on the other.
We have exhausted all possible cases regarding common indices for $A^c$.
Thus, we have shown that \textbf{unitary operators of the form $\boldsymbol{e^{k\frac{\pi}{2}(a_p^\dagger - a_p)}}$, $\boldsymbol{k\in\mathbb{Z}}$, are Clifford}.
Before we conclude whether the $A^{cd}$ and $A_d^c$ generators can act as Clifford, we need to consider the case where both of their indices appear in the fermionic string $O$.
It is straightforward to show that this case reduces to two transformations with a single common index, which result in a single fermionic string.
Consequently, \textbf{unitary operators of the form $\boldsymbol{e^{k\frac{\pi}{2}(a_p^\dagger a_q^\dagger - a_q a_p)}}$ and $\boldsymbol{e^{k\frac{\pi}{2}(a_q^\dagger a_p - a_p^\dagger a_q)}}$, $\boldsymbol{k\in\mathbb{Z}}$, are Clifford}.

Working in a similar manner, one can show that \textbf{Clifford transformations can also be obtained for Hermitian generators of the form $\boldsymbol{a^\dagger_p + a_p}$, $\boldsymbol{a^\dagger_p a^\dagger_q + a_q a_p}$, and $\boldsymbol{a^\dagger_p a_q + a^\dagger_q a_p}$ with $\boldsymbol{\theta = k\frac{\pi}{2}}$, $\boldsymbol{k\in\mathbb{Z}}$}.
However, the case where the generator is a number operator requires particular attention.
Without loss of generality, we assume that the generator of the unitary is $n_p$.
Let $O$ be a fermionic string that contains the index $p$.
Using \cref{seq:f_exp_closed_n}, we arrive at the following expression for the unitary transformation of $O$ by $n_p$:
\begin{equation}
	\begin{split}
		e^{-\I\theta n_p} O e^{\I \theta n_p} &= \left[ I + (e^{-\I \theta} - 1)n_p \right] O \left[ I + (e^{\I \theta} - 1)n_p \right] \\
		&= O + (e^{\I\theta} - 1) O n_p + (e^{-\I\theta} - 1) n_p O + 2\left[1 - \cos(\theta)\right] n_p O n _p.
	\end{split}
\end{equation}
Using \cref{seq:com_number} and the fact that number operators commute with second-quantized operators that share no indices, we obtain
\begin{equation}
	e^{-\I\theta n_p} O e^{\I \theta n_p} =
	\begin{cases}
		e^{\I\theta}O, &O=\cdots a_p \cdots\\
		e^{-\I\theta}O, &O=\cdots a^\dagger_p \cdots\\
		O, &O= \cdots n_p \cdots \text{ or } O = \cdots h_p \cdots
	\end{cases}.
\end{equation}
As a result, \textbf{$\boldsymbol{e^{\I \theta n_p}}$ is a Clifford unitary for any value of $\boldsymbol{\theta}$}.

Note that the fermionic Clifford transformations not only map one fermionic string to another, but also preserve the many-body rank and, thus, the fermionic parity.
This is true even for the $A^c$ generator, which, in general, violates fermionic parity.

\section{Unitary Transformations Generated by $\boldsymbol{a_p^\dagger \pm a_p}$}\label{ssec:transform_single_half}

In this section, we prove that applying the unitary transformation generated by the operators $a_p^\dagger \pm a_p$ to a fermionic string $O$ of even length produces new terms whose many-body rank is decreased by at most 0.5.
We also show that if $O$ has odd length, the transformation generally yields terms whose many-body rank is increased by at most 0.5, unless the common index appears in a number operator, in which case the many-body rank is decreased by at most 0.5.

Without loss of generality, we focus on the anti-Hermitian generator $a_p^\dagger - a_p$.
First, a direct application of \cref{seq:fST_antihermitian} leads to the following expressions for the elementary annihilation, creation, and number operators:
\begin{equation}\label{seq:ST-half_ap}
	e^{-\theta (a_p^\dagger - a_p)} a_p e^{\theta (a_p^\dagger-a_p)} = \cos^2(\theta)a_p - \sin^2(\theta) a_p^\dagger + \frac{1}{2}\sin(2\theta) (h_p - n_p),
\end{equation}
\begin{equation}
	e^{-\theta (a_p^\dagger - a_p)} a_p^\dagger e^{\theta (a_p^\dagger-a_p)} = \cos^2(\theta)a_p^\dagger - \sin^2(\theta) a_p + \frac{1}{2}\sin(2\theta) (h_p - n_p),
\end{equation}
\begin{equation}\label{seq:ST-half_np}
	e^{-\theta (a_p^\dagger - a_p)} n_p e^{\theta (a_p^\dagger-a_p)} = \cos^2(\theta)n_p + \sin^2(\theta) h_p + \frac{1}{2}\sin(2\theta) (a_p^\dagger + a_p),
\end{equation}
and
\begin{equation}
	e^{-\theta (a_p^\dagger - a_p)} h_p e^{\theta (a_p^\dagger-a_p)} = \cos^2(\theta)h_p + \sin^2(\theta) n_p - \frac{1}{2}\sin(2\theta) (a_p^\dagger + a_p).
\end{equation}
Subsequently, we examine the case in which the fermionic string $O$ to be transformed has an even length.
If the generator and $O$ share no common indices, then they commute and the transformation is trivial.
Next, we assume that the index $p$ of the generator appears only once, either as $a_p$ or as $a_p^\dagger$, in $O$.
Without loss of generality, we focus on the $a_p$ case, and write $O$ as
\begin{equation}
	\begin{split}
		O &= f_0 f_1 \cdots f_{p-1} a_{p} f_{p+1} \cdots f_{2k-2} f_{2k-1}\\
		&= F_{0,p-1} a_p F_{p+1,2k-1},
	\end{split}
\end{equation}
with $f_q$ being either $a_q$ or $a_q^\dagger$, $k\in\mathbb{N}$, and 
\begin{equation}
	F_{0,p-1} = f_0 f_1 \cdots f_{p-1}
\end{equation}
and
\begin{equation}
	F_{p+1,2k-1} = f_{p+1} \cdots f_{2k-2} f_{2k-1}.
\end{equation}
The unitary transformation reads
\begin{equation}
	\begin{split}
		\bar{O} &= e^{-\theta (a_p^\dagger - a_p)} O e^{\theta (a_p^\dagger-a_p)}\\
		&= e^{-\theta (a_p^\dagger - a_p)} F_{0,p-1} a_{p} F_{p+1,2k-1} e^{\theta (a_p^\dagger-a_p)}\\
		&= e^{-\theta (a_p^\dagger - a_p)} F_{0,p-1} e^{\theta (a_p^\dagger-a_p)} e^{-\theta (a_p^\dagger - a_p)} a_{p} e^{\theta (a_p^\dagger-a_p)} e^{-\theta (a_p^\dagger - a_p)} F_{p+1,2k-1} e^{\theta (a_p^\dagger-a_p)}\\
		&= \bar{F}_{0,p-1} \left[\cos^2(\theta) a_p -\sin^2(\theta) a_p^\dagger + \frac{1}{2} \sin(2\theta) (h_p - n_p) \right] \bar{F}_{p+1,2k-1},
	\end{split}
\end{equation}
where in the last step we used \cref{seq:ST-half_ap}.
Before we are able to proceed any further, we need to distinguish two cases depending on the lengths of the fermionic strings $F_{0,p-1}$ and $F_{p+1,2k-1}$.

In the first case, the $F_{0,p-1}$ and $F_{p+1,2k-1}$ strings have even and odd lengths, respectively.
Consequently, the transformation of $F_{0,p-1}$ is trivial, $\bar{F}_{0,p-1} = F_{0,p-1}$, while the transformation of $F_{p+1,2k-1}$ follows \cref{seq:ST-half}, yielding
\begin{equation}
	\begin{split}
		\bar{F}_{p+1,2k-1} &= F_{p+1,2k-1} \left[1-2\sin^2(\theta) + \sin(2\theta) (a_p^\dagger - a_p)\right]\\
		&= \left[1-2\sin^2(\theta) + \sin(2\theta) (a_p - a_p^\dagger)\right] F_{p+1,2k-1}.
	\end{split}
\end{equation}
After carrying out the operator multiplications and simplifying, the transformed fermionic string becomes
\begin{equation}
	\bar{O} = F_{0,p-1} \left[\cos^2(\theta) a_p + \sin^2(\theta) a_p^\dagger - \frac{1}{2} \sin(2\theta) \right] F_{p+1,2k-1}.
\end{equation}

In the second case, $F_{0,p-1}$ is odd and $F_{p+1,2k-1}$ is even.
Working in a similar manner, the transformed operator yields
\begin{equation}
	\bar{O} = F_{0,p-1} \left[\cos^2(\theta) a_p + \sin^2(\theta) a_p^\dagger + \frac{1}{2} \sin(2\theta) \right] F_{p+1,2k-1}.
\end{equation}
Similar conclusions can be obtained when a creation operator in $O$ shares an index with the generator $a_p^\dagger - a_p$.
As a result, \textbf{when the even-length string $\boldsymbol{O}$ has an annihilation ($\boldsymbol{a_p}$) or creation ($\boldsymbol{a_p^\dagger}$) operator in common with the generator $\boldsymbol{a_p^\dagger -a_p}$, the transformation will introduce terms which do not include the common index $\boldsymbol{p}$, having a many-body rank that is decreased by 0.5 compared to $\boldsymbol{O}$}.

If the common index appears in a number operator in $O$, either $n_p$ or $h_p$, the situation is straightforward.
The number operator commutes with all other operators appearing in $O$.
After moving the number operator to the rightmost position, the fermionic string is expressed as
\begin{equation}
	O = F n_p,
\end{equation}
where, without loss of generality, we assumed that the common index appears in a particle number operator.
The fact that both $O$ and $n_p$ have an even length implies that the length of $F$ is also even.
As a result, $\exp[-\theta (a_p^\dagger - a_p)]$ commutes with $F$ since they have no common indices.
Using \cref{seq:ST-half_np}, we immediately obtain
\begin{equation}
	\bar{O} = F \left[ \cos^2(\theta)n_p + \sin^2(\theta) h_p + \frac{1}{2}\sin(2\theta) (a_p^\dagger + a_p) \right].
\end{equation}
Similar conclusions can be obtained when a hole number operator in $O$ shares an index with the generator $a_p^\dagger - a_p$.
Therefore, \textbf{when the common index between the generator $\boldsymbol{a_p^\dagger -a_p}$ and the even-length fermionic string $\boldsymbol{O}$ appears in a number operator ($\boldsymbol{n_p}$ or $\boldsymbol{h_p}$) in the latter, the transformation will introduce terms in which $\boldsymbol{n_p}$/$\boldsymbol{h_p}$ is either replaced by $\boldsymbol{a_p}$ or $\boldsymbol{a_p^\dagger}$, respectively, having a many-body rank that is decreased by 0.5 compared to $\boldsymbol{O}$}.

From the above analysis, we see that \textbf{the non-trivial transformation of an even-length fermionic string $\boldsymbol{O}$ by the unitary $\boldsymbol{\exp[\theta (a_p^\dagger - a_p)]}$ introduces, in general, fermionic strings whose many-body rank is reduced by 0.5 compared to that of $\boldsymbol{O}$}. The same is true when the hermitian generator $a_p^\dagger + a_p$ is employed instead.

Next, we proceed to the case where the fermionic string $O$ to be transformed has odd length.
If the string $O$ does not share an index with the generator $a_p^\dagger - a_p$, the transformation follows \cref{seq:ST-half}, resulting in
\begin{equation}
	\bar{O} = O \left[1-2\sin^2(\theta) + \sin(2\theta) (a_p^\dagger - a_p)\right].
\end{equation}
Thus, \textbf{when the odd-length fermionic string $\boldsymbol{O}$ and the generator $\boldsymbol{a_p^\dagger - a_p}$ have no common indices, the transformation introduces terms that include either $\boldsymbol{a_p^\dagger}$ or $\boldsymbol{a_p}$, having a many-body rank that is increased by 0.5 compared to $\boldsymbol{O}$}.

Next, we assume that the index $p$ of the generator appears only once, either as $a_p$ or $a_p^\dagger$, in $O$.
As before, we focus on the $a_p$ case, and write the transformed string as
\begin{equation}
	\bar{O} = \bar{F}_{0,p-1} \left[ \cos^2(\theta)a_p - \sin^2(\theta) a_p^\dagger + \frac{1}{2}\sin(2\theta) (h_p - n_p) \right] \bar{F}_{p+1,2k},
\end{equation}
with $k\in\mathbb{N}_0$, and 
\begin{equation}
	F_{p+1,2k} = f_{p+1} \cdots f_{2k}.
\end{equation}
There are two cases to consider.
In the first, both $F_{0,p-1}$ and $F_{p+1,2k}$ have even length, hence their transformations are trivial.
As a result, one readily obtains
\begin{equation}
	\bar{O} = F_{0,p-1} \left[ \cos^2(\theta)a_p - \sin^2(\theta) a_p^\dagger + \frac{1}{2}\sin(2\theta) (h_p - n_p) \right] F_{p+1,2k}.
\end{equation}
In the second case, both $F_{0,p-1}$ and $F_{p+1,2k}$ have odd length. Employing \cref{seq:ST-half,seq:ST-half_ap} and carrying out the operator multiplications and simplifying, yields the following expression for the transformed string $O$:
\begin{equation}
	\bar{O} = F_{0,p-1} \left[ \cos^2(\theta)a_p - \sin^2(\theta) a_p^\dagger + \frac{1}{2}\sin(2\theta) (n_p - h_p) \right] F_{p+1,2k}.
\end{equation}
Similar conclusions can be obtained when a creation operator in $O$ shares an index with the generator $a_p^\dagger - a_p$.
As a result, \textbf{when the odd-length string $\boldsymbol{O}$ has an annihilation ($\boldsymbol{a_p}$) or creation ($\boldsymbol{a^\dagger_p}$) operator  in common with the generator $\boldsymbol{a_p^\dagger -a_p}$, the transformation will introduce terms in which $\boldsymbol{a_p}$/$\boldsymbol{a^\dagger_p}$ is replaced by either $\boldsymbol{n_p}$ or $\boldsymbol{h_p}$, having a many-body rank that is increased by 0.5 compared to $\boldsymbol{O}$}.

Finally, we consider the scenario in which the common index appears in a number operator, \foreign{e.g.}, $n_p$, in $O$.
After bringing the number operator to the rightmost position, the fermionic string is given by
\begin{equation}
	O = F n_p,
\end{equation}
with $F$ necessarily having an odd length. Using \cref{seq:ST-half,seq:ST-half_ap} and simplifying results in
\begin{equation}
	\bar{O} = F \left[ \cos^2(\theta)n_p - \sin^2(\theta) h_p + \frac{1}{2}\sin(2\theta) (a_p^\dagger - a_p) \right].
\end{equation}
Similar conclusions can be obtained when a hole number operator in $O$ shares an index with the generator $a_p^\dagger - a_p$.
Therefore, \textbf{when the common index between the generator $\boldsymbol{a_p^\dagger -a_p}$ and the odd-length fermionic string $\boldsymbol{O}$ appears in a number operator ($\boldsymbol{n_p}$ or $\boldsymbol{h_p})$ in the latter, the transformation will introduce terms in which $\boldsymbol{n_p}$/$\boldsymbol{h_p}$ is either replaced by $\boldsymbol{a_p}$ or $\boldsymbol{a_p^\dagger}$, respectively, having a many-body rank that is decreased by 0.5 compared to $\boldsymbol{O}$}.

From the above analysis, we see that \textbf{the transformation of an odd-length fermionic string $\boldsymbol{O}$ by the unitary $\boldsymbol{\exp[\theta (a_p^\dagger - a_p)]}$ introduces, in general, fermionic strings whose many-body rank is increased by 0.5 compared to that of $\boldsymbol{O}$. The only exception is when the common index appears in a number operator in $\boldsymbol{O}$, in which case the transformation introduces terms with many-body rank decreased by 0.5}. The same is true when the hermitian generator $a_p^\dagger + a_p$ is employed instead.

\section{Closed-Form Expressions for Fermionic Unitaries Generated by Linear Combinations of Half-Body Operators}\label{ssec:transform_sum_half}

In certain cases, it is possible to derive closed-form expressions for fermionic unitaries generated by linear combinations of (anti)-Hermitian operators.

First, we consider a unitary where the anti-Hermitian generator has the form
\begin{equation}
	A = \sum_p {\theta_p} A^p = \sum_p {\theta_p} (a_p^\dagger - a_p),
\end{equation}
with $\theta_p \in \mathbb{R}$.
Before we explore the corresponding exponential, we examine the structures of $A^2$ and $A^3$:
\begin{equation}
	\begin{split}
		A^2 &= \left( \sum_p {\theta_p} (a_p^\dagger - a_p) \right)^2\\
		&= \sum_p \theta_p^2 (a_p^\dagger - a_p)^2 + \sum_p\sum_{q>p} \theta_p \theta_q \left[ \cancel{(a_p^\dagger - a_p) (a_q^\dagger - a_q)} + \cancel{(a_q^\dagger - a_q) (a_p^\dagger - a_p)} \right]\\
		&= -\sum_p \theta_p^2 (n_p + h_p)\\
		&= -\sum_p \theta_p^2,\\
		&= -c
	\end{split}
\end{equation}
and
\begin{equation}
	A^3 = -c A,
\end{equation}
where
\begin{equation}
	c = \sum_p \theta_p^2.
\end{equation}
From the above, we arrive at the following recursion relations for the powers of $A$:
\begin{equation}
	A^n = \begin{cases}
		(-1)^k c^k, &n=2k, k\in\mathbb{Z}\\
		(-1)^k c^k A, &n=2k+1, k\in\mathbb{Z}
	\end{cases}.
\end{equation}
By Taylor expansion, we obtain
\begin{equation}
	\begin{split}
		e^{A} &= \sum_{n=0}^\infty \frac{A^n}{n!}\\
		&= \sum_{k=0}^\infty \frac{A^{2k}}{(2k)!} + \sum_{k=0}^\infty \frac{A^{2k+1}}{(2k+1)!}\\
		&= \sum_{k=0}^\infty \frac{(-1)^k c^k}{(2k)!} + \sum_{k=0}^\infty \frac{(-1)^k c^k}{(2k+1)!} A\\
		&= \cos(\sqrt{c}) + \frac{\sin(\sqrt{c})}{\sqrt{c}}A\\
		&= \cos(\sqrt{\sum_p \theta_p^2}) + \frac{\sin(\sqrt{\sum_p \theta_p^2})}{\sqrt{\sum_p \theta_p^2}} \left( \sum_p \theta_p (a_p^\dagger - a_p) \right).
	\end{split}
\end{equation}
Working in a similar manner, one can prove that
\begin{equation}
	e^{\I \sum_p \theta_p (a_p^\dagger + a_p)} = \cos\left(\sqrt{\sum_p \theta_p^2}\right) + \I \frac{\sin(\sqrt{\sum_p \theta_p^2})}{\sqrt{\sum_p \theta_p^2}} \left( \sum_p \theta_p (a_p^\dagger + a_p) \right).
\end{equation}
Using the definitions of Majorana operators, \cref{seq:gamma_1,seq:gamma_2}, the above expressions can also be written in the form
\begin{equation}
	e^{-\I\theta \sum_p \theta_p\gamma_2^{(p)}} = \cos(\sqrt{\sum_p \theta_p^2}) -\I \frac{\sin(\sqrt{\sum_p \theta_p^2})}{\sqrt{\sum_p \theta_p^2}} \left( \sum_p \theta_p \gamma_2^{(p)} \right)
\end{equation}
and
\begin{equation}
	e^{\I\theta \sum_p \theta_p\gamma_1^{(p)}} = \cos(\sqrt{\sum_p \theta_p^2}) +\I \frac{\sin(\sqrt{\sum_p \theta_p^2})}{\sqrt{\sum_p \theta_p^2}} \left( \sum_p \theta_p \gamma_1^{(p)} \right).
\end{equation}

Using the above closed-form expressions for the exponentials, we can readily derive the closed-form expressions for the corresponding similarity transformations. Focusing on the fermionic case with an anti-Hermitian generator, we obtain
\begin{equation}
	e^{-A} O e^{A} = \cos^2\left(\sqrt{c}\right) O + \frac{\cos(\sqrt{c}) \sin(\sqrt{c})}{\sqrt{c}} [O, A] - \frac{\sin^2(\sqrt{c})}{c} A O A.
\end{equation}
Although the above expression is closed, it is not expressed exclusively in terms of nested commutators of $O$ and $A$, as one would anticipate in light of the Hausdorff expansion.
By examining the structures of $AOA$,
\begin{equation}
	\begin{split}
		AOA &= \sum_p \sum_q \theta_p \theta_q A^p O A^q\\
		&= \sum_p \theta_p^2 A^p O A^p + \sum_p \sum_{q>p} \theta_p \theta_q (A^p O A^q + A^q OA^p),
	\end{split}
\end{equation}
and the nested commutator $[[O,A]A]$,
\begin{equation}
	\begin{split}
		[[O,A]A] &= \sum_p \sum_q \theta_p \theta_q [[O,A^p],A^q]\\
		&= \sum_p \sum_q \theta_p \theta_q [OA^p - A^pO,A^q]\\
		&= \sum_p \sum_q \theta_p \theta_q ([OA^p,A^q] + [A^q,A^pO])\\
		&= \sum_p \sum_q \theta_p \theta_q (OA^pA^q - A^qOA^p + A^qA^pO - A^pOA^q)\\
		&= \sum_p \theta_p^2 (O{A^p}^2 + {A^p}^2O - 2A^pOA^p)\\
		&\phantom{=1}+ \sum_p \sum_{q>p} \theta_p \theta_q (\cancel{OA^pA^q} + \cancel{OA^qA^p} + \textcolor{blue}{\cancel{\textcolor{black}{A^pA^qO}}} + \textcolor{blue}{\cancel{\textcolor{black}{A^qA^p O}}} -2A^pOA^q - 2A^qOA^p)\\
		&=-2\sum_p\theta_p^2O -2\left[\sum_p \theta_p^2 A^pOA^p + \sum_p \sum_{q>p} \theta_p \theta_q\left(A^pOA^q + A^qOA^p \right)\right]\\
		&=-2cO -2AOA,
	\end{split}
\end{equation}
we can write the similarity transformation in the following form:
\begin{equation}\label{seq:sum_half_ST}
	\begin{split}
		e^{-A} O e^{A} &= O + \frac{1}{2}\frac{\sin(2\sqrt{c})}{\sqrt{c}} [O,A] + \frac{1}{2}\frac{\sin^2(\sqrt{c})}{c} [[O,A],A]\\
		e^{-\sum_p \theta_p (a_p^\dagger - a_p)} O a^{\sum_p \theta_p (a_p^\dagger - a_p)} &= O + \frac{1}{2} \frac{\sin(2\sqrt{\sum_p \theta_p^2})}{\sqrt{\sum_p \theta_p^2}} \sum_q \theta_q [O, a_q^\dagger - a_q]\\
		&\phantom{{}=O}+ \frac{1}{2} \frac{\sin^2(\sqrt{\sum_p \theta_p^2})}{\sum_p \theta_p^2}\sum_q\sum_r \theta_q \theta_r [[O, a_q^\dagger - a_q], a_r^\dagger - a_r].
	\end{split}
\end{equation}
As might have been anticipated, when all $\theta$ parameters but one are set to zero, the above expression reduces to that of a single generator.
In principle, \cref{seq:sum_half_ST} serves as the starting point for determining the conditions that need to be satisfied for the transformation to be Clifford.

\section{Fermionic Clifford Transformations and Spinorbital Tapering}\label{ssec:taper}

The main steps of the qubit-reduction technique known as qubit tapering are as follows:
\begin{enumerate}
	\item Obtain second-quantized Hamiltonian $\mathscr{H}$ of system of interest.
	\item Transform $\mathscr{H}$ to qubit space using the desired fermionic encoding.
	\item Find the group $\mathcal{G}$ of all Pauli strings that commute with the transformed Hamiltonian.
	\item Find maximal abelian subgroup $\mathcal{S}$ of $\mathcal{G}$, such that $-I\notin\mathcal{S}$.
	\item Find the generators of $\mathcal{S}$. 
	\item Construct Clifford gates that transform each generator of $\mathcal{S}$ to a single-qubit Pauli gate.
	\item Transform the Hamiltonian with the Clifford gates.
	\item Choose desired symmetry sector and taper redundant qubits.
\end{enumerate}
Following Ref.\ [29] of the main text, we now illustrate this general qubit tapering procedure using the minimum-basis-set (MBS) representation of the $\text{H}_2$ molecule as a concrete example.

The second-quantized electronic Hamiltonian of $\text{H}_2$/MBS is
\begin{equation}
	\begin{split}
		\mathscr{H} =& h_0^0 \left( n_0 + n_1\right) + h_2^2 \left( n_2 + n_3\right) + 
		v_{01}^{01} n_0 n_1  + v_{23}^{23} n_2 n_3  \\
		&+ \left( v_{02}^{02} - v_{02}^{20} \right) n_0 n_2  + v_{03}^{03} n_0 n_3 
		+ v_{12}^{12} n_1 n_2  + \left( v_{13}^{13} - v_{13}^{31} \right) n_1 n_3  \\
		&+ v_{01}^{23} \left( a_{01}^{23} - a_{12}^{03} - a_{03}^{12} + a_{23}^{01} \right),
	\end{split}	
\end{equation}
where $h_q^p$ and $v_{rs}^{pq}$ denote the one- and two-electron integrals, and the spinorbital indices correspond to $\ket{0}\equiv\ket{\sigma_g {\uparrow}}$, $\ket{1}\equiv\ket{\sigma_g {\downarrow}}$, $\ket{2}\equiv\ket{\sigma_u {\uparrow}}$, and $\ket{3}\equiv\ket{\sigma_u {\downarrow}}$.
To translate the Hamiltonian into the qubit space, we employ the Jordan--Wigner (JW) transformation, which maps fermionic Slater determinants to qubit states as
\begin{equation}
	\ket{n_0 n_1 \ldots n_{M-1}} \xrightarrow{\text{JW}} \ket{q_0 q_1 \ldots q_{M-1}},
\end{equation}
where $n_p$ is the occupation number of the $p$th spinorbital and $q_p$ denotes the state of the $p$th qubit.
Under the JW transformation, the annihilation and creation operators become
\begin{equation}\label{seq:jw_ann}
	a_p \xrightarrow{\text{JW}} \frac{1}{2}(X_p + \I Y_p) \prod_{q = 0}^{p-1} Z_q
\end{equation}
and
\begin{equation}\label{seq:jw_cr}
	a_p^\dagger \xrightarrow{\text{JW}} \frac{1}{2}(X_p - \I Y_p) \prod_{q = 0}^{p-1} Z_q,
\end{equation}
respectively.
Here, $X_p$, $Y_p$, and $Z_p$ respectively denote the Pauli matrices $\sigma_x$, $\sigma_y$, and $\sigma_z$ acting on the $p$th qubit.
Using \cref{seq:jw_ann,seq:jw_cr}, the particle and hole number operators transform as
\begin{equation}
	n_p \xrightarrow{\text{JW}} \frac{1}{2}(I-Z_p)
\end{equation}
and
\begin{equation}
	h_p \xrightarrow{\text{JW}} \frac{1}{2}(I+Z_p),
\end{equation}
respectively.
The JW-transformed Hamiltonian of $\text{H}_2$/MBS reads
\begin{equation}\label{seq:H_Pauli}
	\begin{split}
		\mathscr{H} ={}& c_1 + c_2 Z_0 + c_3 Z_1 +  c_4 Z_2 + c_5 Z_3 \\
		&+ c_6 Z_0 Z_1 + c_7 Z_0 Z_2 + c_8 Z_0 Z_3 + c_9 Z_1 Z_2 + c_{10} Z_1 Z_3 + c_{11} Z_2 Z_3 \\
		&+ c_{12} Y_0 Y_1 X_2 X_3
		+ c_{13} Y_0 X_1 X_2 Y_3 + c_{14} X_0 Y_1 Y_2 X_3
		+ c_{15} X_0 X_1 Y_2 Y_3,
	\end{split}
\end{equation}
with the various coefficients defined in \cref{stable:int}.
\begin{table*}
	\begin{center}
		\begin{threeparttable}
			\caption{Explicit expressions of the coefficients multiplying the various Pauli strings in 
				the electronic Hamiltonian of the $\text{H}_2$/MBS system, \cref{seq:H_Pauli}, in 
				terms of one- and two-electron integrals.}\label{stable:int}
			\begin{tabular*}{\textwidth}{l @{\extracolsep{\fill}} c}
				\toprule
				Coefficient & Expression\tnote{a} \\
				\midrule
				$c_1$ & $h_0^0 + h_2^2 + \tfrac{1}{4} \left( v_{01}^{01} + v_{23}^{23} \right) + 
				v_{02}^{02} - \tfrac{1}{2} v_{02}^{20}$ \\
				$c_2$ & $- \tfrac{1}{2} h_0^0 - \tfrac{1}{4} \left( v_{01}^{01} - 2 v_{02}^{02} + 
				v_{02}^{20} \right)$ \\
				$c_3$ & $- \tfrac{1}{2} h_0^0 - \tfrac{1}{4} \left( v_{01}^{01} - 2 v_{02}^{02} + 
				v_{02}^{20} \right)$ \\
				$c_4$ & $- \tfrac{1}{2} h_2^2 - \tfrac{1}{4} \left( v_{23}^{23} - 2 v_{02}^{02} + 
				v_{02}^{20} \right)$ \\
				$c_5$ & $- \tfrac{1}{2} h_2^2 - \tfrac{1}{4} \left( v_{23}^{23} - 2 v_{02}^{02} + 
				v_{02}^{20} \right)$ \\
				$c_6$ & $\tfrac{1}{4} v_{01}^{01}$ \\
				$c_7$ & $\tfrac{1}{4} \left( v_{02}^{02} - v_{02}^{20} \right)$ \\
				$c_8$ & $\tfrac{1}{4} v_{02}^{02}$ \\
				$c_9$ & $\tfrac{1}{4} v_{02}^{02}$ \\
				$c_{10}$ & $\tfrac{1}{4} \left( v_{02}^{02} - v_{02}^{20} \right)$ \\
				$c_{11}$ & $\tfrac{1}{4} v_{23}^{23}$ \\
				$c_{12}$ & $-\tfrac{1}{4} v_{01}^{23}$ \\
				$c_{13}$ & $\tfrac{1}{4} v_{01}^{23}$ \\
				$c_{14}$ & $\tfrac{1}{4} v_{01}^{23}$ \\
				$c_{15}$ & $-\tfrac{1}{4} v_{01}^{23}$ \\
				\bottomrule
			\end{tabular*}
			\begin{tablenotes}
				\item[a] In writing the expressions, we took advantage of the following symmetries in the two-electron integrals for $H_2$/MBS: $v_{02}^{02} = v_{20}^{20} 
				= v_{03}^{03} = v_{30}^{30} = v_{12}^{12} = v_{21}^{21} = v_{13}^{13} = v_{31}^{31}$ and $v_{02}^{20} = v_{20}^{02} 
				= v_{13}^{31} = v_{31}^{13} = v_{01}^{23} = v_{03}^{21} = v_{10}^{32} = v_{12}^{30} = 
				v_{21}^{03} = v_{23}^{01} = v_{30}^{12} = v_{32}^{10}$.
			\end{tablenotes}
		\end{threeparttable}
	\end{center}
\end{table*}

Next, we need to find the set of Pauli strings that commute with the Hamiltonian.
Formally, this is accomplished by utilizing the fact that, upon neglecting overall phases, the Pauli strings form an elementary abelian 2-group, constituting a binary vector space equipped with a symplectic inner product.
Consequently, one expresses the Pauli strings appearing in the Hamiltonian as binary vectors, 
constructs the corresponding parity check matrix, reduces it to row-echelon form, finds its kernel, and then translates these binary vectors back to Pauli strings.
The resulting Pauli strings commute with the Hamiltonian.
However, the $\text{H}_2$/MBS problem is sufficiently small that the commuting Pauli strings can be identified by directly inspecting the qubit Hamiltonian, \cref{seq:H_Pauli}.
Indeed, we observe that the qubit Hamiltonian includes single-qubit terms consisting solely of $Z$ gates acting on each qubit.
Aside from the identity, these single-qubit terms commute exclusively with Pauli strings composed entirely of $Z$ gates.
Additionally, the qubit Hamiltonian of $\text{H}_2$/MBS contains terms involving combinations of $X$ and $Y$ gates acting simultaneously on all four qubits.
These terms commute only with $Z$-strings of even length.
Therefore, the Pauli strings that commute with the qubit Hamiltonian of $\text{H}_2$/MBS form the group $\mathcal{S} = \{I, Z_0Z_1, Z_0Z_2, Z_0Z_3, Z_1Z_2, Z_1Z_3, Z_2Z_3, Z_0Z_1Z_2Z_3\}$.
Note that, following the qubit tapering procedure, $-I\notin\mathcal{S}$.
The maximal abelian subgroup of $\mathcal{S}$ is itself, and, thus, represents the $\mathbb{Z}_2$ symmetry group of the qubit Hamiltonian.
A generating set of $\mathcal{S}$ is $\mathcal{S}=\left\langle Z_0Z_1, Z_0Z_2, Z_0Z_3 \right\rangle$.
Identifying these generators is crucial, as each allows us to reduce the qubit count by one.

Having found the generators of the symmetry group $\mathcal{S}$ of the $\text{H}_2$/MBS qubit Hamiltonian, we next construct the corresponding Clifford unitaries.
To taper off the last three qubits from the simulation, we follow the standard qubit tapering procedure and construct the Clifford unitaries as
\begin{align}
	U_1 =& \frac{1}{\sqrt{2}} (X_1 + Z_0 Z_1), \label{seq:cliff1}\\
	U_2 =& \frac{1}{\sqrt{2}} (X_2 + Z_0 Z_2), \label{seq:cliff2}\\ \shortintertext{and} 
	U_3 =& \frac{1}{\sqrt{2}} (X_3 + Z_0 Z_3).\label{seq:cliff3}
\end{align}
These particular Clifford gates are chosen because they map each symmetry generator directly onto a single-qubit $X$ gate acting on a different qubit.
As a result, the Hamiltonian can have a common eigenbasis with $X_1$, $X_2$, and $X_3$ Pauli gates, enabling the removal of three qubits from the simulation.
Note that $U_1$, $U_2$, and $U_3$ are also Hermitian involutions and pairwise commute.

In the next step of the qubit tapering algorithm, we use the $\mathscr{U} = U_1 U_2 U_3$ Clifford unitary to transform the Hamiltonian, obtaining
\begin{equation}\label{seq:st_h}
	\begin{split}
		\bar{\mathscr{H}} &= \mathscr{U} \mathscr{H} \mathscr{U}\\
		&= c_1 + c_2 Z_0 + c_3 Z_0 X_1 + c_4 Z_0 X_2 + c_5 Z_0 X_3\\
		&\phantom{{}=} + c_6 X_1 + c_7 X_2 + c_8 X_3 + c_9 X_1 X_2 + c_{10} X_1 X_3 + c_{11} X_2 X_3\\
		&\phantom{{}=} - c_{12} X_0 X_2 X_3 - c_{13} X_0 X_1 X_2 - c_{14} X_0 X_3 - c_{15} X_0 X_1.
	\end{split}
\end{equation}
One can verify that, on the last three qubits, the transformed Hamiltonian applies at most an $X$ gate.
Consequently, if we express the Hamiltonian in the common eigenbasis of the $X_1$, $X_2$, and $X_3$ operators, we can replace them by their corresponding eigenvalues, resulting in single-qubit Hamiltonians of the form
\begin{equation}
	\resizebox{\textwidth}{!}{$
		\begin{split}
			\bar{\mathscr{H}}^{(q_1 q_2 q_3)} =& c_1 + (-1)^{\delta_{{q_1}-}} c_6 + (-1)^{\delta_{{q_2}-}} c_7 + (-1)^{\delta_{{q_3}-}} c_8 + (-1)^{\delta_{{q_1}-} + \delta_{{q_2}-}} c_9 + (-1)^{\delta_{{q_1}-} + \delta_{{q_3}-}} c_{10} + (-1)^{\delta_{{q_2}-} + \delta_{{q_3}-}} c_{11} \\
			&+ \left[c_2 + (-1)^{\delta_{{q_1}-}} c_3 + (-1)^{\delta_{{q_2}-}} c_4 + (-1)^{\delta_{{q_3}-}} c_5\right] Z_0 \\
			&- \left[(-1)^{\delta_{{q_2}-} + \delta_{{q_3}-}} c_{12} + (-1)^{\delta_{{q_1}-} + \delta_{{q_2}-}} c_{13} + (-1)^{\delta_{{q_3}-}} c_{14} + (-1)^{\delta_{{q_1}-}} c_{15}\right] X_0 \\
			=& c_1 + (-1)^{\delta_{{q_1}-}} c_6 + \left((-1)^{\delta_{{q_2}-}} + (-1)^{\delta_{{q_1}-} + \delta_{{q_3}-}}\right) c_7 + \left((-1)^{\delta_{{q_3}-}} + (-1)^{\delta_{{q_1}-} + \delta_{{q_2}-}}\right) c_8 + (-1)^{\delta_{{q_2}-} + \delta_{{q_3}-}} c_{11} \\
			&+ \left[\left(1 + (-1)^{\delta_{{q_1}-}}\right) c_2 + \left((-1)^{\delta_{{q_2}-}} + (-1)^{\delta_{{q_3}-}}\right) c_4\right] Z_0 \\
			&+  -\left[(-1)^{\delta_{{q_2}-} + \delta_{{q_3}-}} - (-1)^{\delta_{{q_1}-} + \delta_{{q_2}-}} - (-1)^{\delta_{{q_3}-}}  + (-1)^{\delta_{{q_1}-}}\right] c_{12} X_0,
		\end{split}
		$}
\end{equation}
where in the last step we used the fact that $c_2 = c_3$, $c_4 = c_5$, $c_7 = c_{10}$, $c_8 = c_9$, and $c_{12} = -c_{13} = -c_{14} = c_{15}$ (see \cref{stable:int}).
Here, $q_1$--$q_3$ take values of either $+$ or $-$, corresponding to the $\ket{+}\equiv \frac{1}{\sqrt{2}}(\ket{0}+\ket{1})$ and $\ket{-}\equiv \frac{1}{\sqrt{2}}(\ket{0}-\ket{1})$ eigenvectors of the Pauli $X$ gate with eigenvalues of $+1$ and $-1$ respectively.
Hence, the Kronecker delta $\delta_{{q_p}-}$ is 0 if $q_p$ is in the $\ket{+}$ state and 1 if it is in the $\ket{-}$ state.

As a result of the qubit tapering procedure, the original Hamiltonian is brought to block diagonal form, with each block corresponding to a different set of eigenvalues of the $X_1$, $X_2$, and $X_3$ Pauli gates.
Each tapered qubit halves the dimension of the resulting $\mathbb{Z}_2$--symmetry-adapted Hamiltonians relative to the parent one.
Thus, in the case of $\text{H}_2$/MBS, the original 16-dimensional Hamiltonian in the Fock space is replaced by eight 2-dimensional $\mathbb{Z}_2$--symmetry-adapted Hamiltonians:
\begin{align}
	\bar{\mathscr{H}}^{(+++)} &= (c_1 + c_6 + 2c_7 + 2c_8 + c_{11}) + (2c_2 + 2c_4) Z \\
	\bar{\mathscr{H}}^{(-++)} &= (c_1 - c_6 + c_{11}) + 2c_4 Z \\
	\bar{\mathscr{H}}^{(+-+)} &= (c_1 + c_6 - c_{11}) + 2c_2 Z \\
	\bar{\mathscr{H}}^{(++-)} &= (c_1 + c_6 - c_{11}) + 2c_2 Z \\
	\bar{\mathscr{H}}^{(--+)} &= (c_1 - c_6 - 2c_7 + 2c_8 - c_{11}) + 4c_{12} X\\
	\bar{\mathscr{H}}^{(-+-)} &= (c_1 - c_6 + 2c_7 - 2c_8 - c_{11}) \\
	\bar{\mathscr{H}}^{(+--)} &= (c_1 + c_6 - 2c_7 - 2c_8 + c_{11}) + (2c_2 - 2c_4) Z - 4c_{12} X \label{seq:st_H_+--} \\ \shortintertext{and}
	\bar{\mathscr{H}}^{(---)} &= (c_1 - c_6 + c_{11}) - 2c_4 Z,
\end{align}
where we dropped the ``0'' subscript since there is only one qubit.
Notably, six of these Hamiltonians are already diagonal in the standard computational basis of $Z$ eigenvectors.

Before calculating the ground-state energy of $\text{H}_2$/MBS, we must first determine its $\mathbb{Z}_2$ symmetry properties, and, thus, to which of the eight symmetry-adapted Hamiltonians it belongs to.
This can be accomplished by applying the Clifford gate $\mathscr{U}$ on the Hartree--Fock Slater determinant, $\ket{1100}$.
Indeed, this transforms the $\ket{1100}$ state to $\ket{1{+}{-}{-}}$, meaning that the ground state of $\text{H}_2$/MBS is an eigenstate of $\bar{\mathscr{H}}^{(+--)}$.
By tapering off the last three qubits, the reference determinant for the quantum simulation becomes $\ket{1}$.

The last step is to express the UCC doubles (UCCD) unitary, which provides an exact parameterization of the ground electronic state of $\text{H}_2$/MBS, in the qubit space.
By using the JW mapping, we obtain
\begin{equation}
	\begin{split}
		e^{\theta (a_{01}^{23} - a_{23}^{01})} \xrightarrow{\text{JW}}{}&
		e^{\I\tfrac{\theta}{8} Y_0 X_1 X_2 X_3}
		e^{\I\tfrac{\theta}{8} X_0 Y_1 X_2 X_3}
		e^{- \I\tfrac{\theta}{8} X_0 X_1 Y_2 X_3}
		e^{- \I\tfrac{\theta}{8} X_0 X_1 X_2 Y_3}\\
		&\times e^{- \I\tfrac{\theta}{8} X_0 Y_1 Y_2 Y_3}
		e^{- \I\tfrac{\theta}{8} Y_0 X_1 Y_2 Y_3}
		e^{\I\tfrac{\theta}{8} Y_0 Y_1 X_2 Y_3}
		e^{\I\tfrac{\theta}{8} Y_0 Y_1 Y_2 X_3},
	\end{split}
\end{equation}
where we used the fact that Pauli strings originating from the same anti-Hermitian excitation operator commute (see Ref.\ [66] of the main text).
For this system, the above unitary can be simplified.
By examining the action of the individual exponentials on the Hartree--Fock state of $\text{H}_2$/MBS, it can be shown that they can be converted to one another.
Thus, the UCCD wavefunction ansatz for $\text{H}_2$/MBS becomes 
\begin{equation}
	e^{\theta (a_{01}^{23} - a_{23}^{01})} \ket{1100} \xrightarrow{\text{JW}} e^{\I\theta Y_0 X_1 X_2 X_3} \ket{1100}
\end{equation}
(see, also, Ref.\ [86] of the main text).
In principle, one would need to transform the UCCD unitary to the common eigenbasis of $X_1$, $X_2$, and $X_3$, as was done with the Hamiltonian.
However, the UCCD unitary is already acting on qubits $q_1$--$q_3$ with an $X$ gate and, thus, the Clifford transformation by $\mathscr{U}$ is trivial.
Upon acting on the transformed Hartree--Fock state, the UCCD unitary becomes
\begin{equation}
	e^{\I \theta Y_0 X_1 X_2 X_3} \ket{1{+}{-}{-}} = e^{\I \theta Y_0} \ket{1{+}{-}{-}}.
\end{equation}
As a result, the Schr\"{o}dinger equation for the ground electronic state of $\text{H}_2$/MBS in the $({+}{-}{-})$ symmetry sector of the Hamiltonian reads
\begin{equation}\label{seq:tapered_SE_qubit}
	\begin{split}
		\left[c_1 + c_6 - 2c_7 - 2c_8 + c_{11} + (2c_2 - 2c_4) Z - 4c_{12} X\right] e^{\I \theta Y} \ket{1} &= E_0 e^{\I \theta Y} \ket{1} \Rightarrow\\
		\left[ h_0^0 + h_2^2 + \frac{1}{2}\left(v_{01}^{01} + v_{23}^{23}\right) + \left(h_2^2 - h_0^0 + \frac{1}{2} \left(v_{23}^{23} - v_{01}^{01}\right)\right)Z + v_{01}^{23} X \right] e^{\I \theta Y} \ket{1} &= E_0 e^{\I \theta Y} \ket{1},
	\end{split}
\end{equation}
where $E_0$ denotes the energy of the ground electronic state of $\text{H}_2$/MBS.

To gain a deeper physical understanding, we will translate a few key results of the qubit tapering algorithm for $\text{H}_2$/MBS to the language of fermions. To that end, we use the inverse JW mapping, defined by
\begin{equation}
	X_p \xrightarrow{\text{JW}^{-1}} (a_p^\dagger + a_p) \prod_{q=0}^{p-1} (I - 2n_q),
\end{equation}
\begin{equation}
	Y_p \xrightarrow{\text{JW}^{-1}} \I(a_p^\dagger - a_p) \prod_{q=0}^{p-1} (I - 2n_q),
\end{equation}
and
\begin{equation}
	Z_p \xrightarrow{\text{JW}^{-1}} I-2n_p.
\end{equation}
Using the inverse JW mapping, we find that the generators of the $\mathbb{Z}_2$ symmetry group for the electronic Hamiltonian of the $\text{H}_2$/MBS system take the form
\begin{equation}
	\mathcal{S} = \left\langle (I - 2n_0)(I-2n_1), (I - 2n_0)(I-2n_2), (I - 2n_0)(I-2n_3) \right\rangle.
\end{equation}
Before we proceed any further, one can show that
\begin{equation}
	I - 2N = e^{\I \pi N} = (-1)^N,
\end{equation}
where $N$ is a product of elementary number operators and we used \cref{seq:f_exp_closed_n}.
This allows us to express the generators in the form
\begin{equation}
	\mathcal{S} = \left\langle (-1)^{n_0+n_1}, (-1)^{n_0+n_2}, (-1)^{n_0+n_3} \right\rangle.
\end{equation}
Next, we express the Clifford gates shown in \cref{seq:cliff1,seq:cliff2,seq:cliff3} in second quantization, based on the inverse JW map.
Starting with $U_1$, we obtain
\begin{equation}
	\begin{split}
		U_1 &= \frac{1}{\sqrt{2}} \left[\left(a_1^\dagger + a_1\right)(I - 2n_0) + (I - 2n_0)(I-2n_1)\right] (I - 2n_0)\\
		&= \frac{1}{\sqrt{2}} (I - 2n_0)\left(a_1^\dagger + a_1 + I-2n_1\right)\\
		&= \frac{1}{\sqrt{2}} (-1)^{n_0}\left( -\I e^{\I\frac{\pi}{2} (a_1^\dagger + a_1)} + (-1)^{n_1} \right).
	\end{split}
\end{equation}
In the last step we used \cref{seq:f_exp_closed_hermitian} and the fact that $(a_p^\dagger+a_p)^2 = I$ to obtain
\begin{equation}
	a_p^\dagger + a_p = -\I e^{\I \theta (a_p^\dagger + a_p)}.
\end{equation}
Similarly, we arrive at the following expressions for $U_2$ and $U_3$:
\begin{equation}
	U_2 = \frac{1}{\sqrt{2}} (-1)^{n_0}\left( (-1)^{n_1+1}\I e^{\I\frac{\pi}{2} (a_2^\dagger + a_2)} + (-1)^{n_2} \right)
\end{equation}
and
\begin{equation}
	U_3 = \frac{1}{\sqrt{2}} (-1)^{n_0}\left( (-1)^{n_1+n_2+1}\I e^{\I\frac{\pi}{2} (a_3^\dagger + a_3)} + (-1)^{n_3} \right),
\end{equation}
respectively.
The transformed Hamiltonian becomes
\begin{equation}
	\resizebox{\linewidth}{!}{$
		\begin{split}
			&\left(\frac{h^{0}_{0}}{2} + h^{2}_{2} + \frac{v_{02}^{02}}{2} - \frac{v_{02}^{20}}{4} + \frac{v_{23}^{23}}{4}\right) +\left(h^{0}_{0} + \frac{v_{01}^{01}}{2} + v_{02}^{02} - \frac{v_{02}^{20}}{2}\right) n_{0} +\left(- \frac{h^{0}_{0}}{2} - \frac{v_{02}^{02}}{2} + \frac{v_{02}^{20}}{4}\right) \left(a_{1} + {a^\dagger_{1}}\right)\\
			&+\left(- \frac{h^{2}_{2}}{2} - \frac{v_{02}^{02}}{4} - \frac{v_{23}^{23}}{4}\right) \left(a_{2} + {a^\dagger_{2}}\right) +\left(- \frac{h^{2}_{2}}{2} - \frac{v_{02}^{02}}{4} + \frac{v_{02}^{20}}{4} - \frac{v_{23}^{23}}{4}\right) \left(a_{3} + {a^\dagger_{3}}\right)\\
			&+\left(h^{2}_{2} + \frac{v_{02}^{02}}{2} + \frac{v_{23}^{23}}{2}\right) \left(a_{2} n_{1} + {a^\dagger_{2}} n_{1}\right) +\left(h^{2}_{2} + \frac{v_{02}^{02}}{2} - \frac{v_{02}^{20}}{2} + \frac{v_{23}^{23}}{2}\right) \left(a_{3} n_{1} + a_{3} n_{2} + {a^\dagger_{3}} n_{1} + {a^\dagger_{3}} n_{2}\right)\\
			&+\left(- \frac{v_{02}^{02}}{2} + \frac{v_{02}^{20}}{2}\right) \left(a_{2} n_{0} + a_{3} a_{1} n_{2} + {a^\dagger_{1}} a_{3} n_{2} + {a^\dagger_{1}} {a^\dagger_{3}} n_{2} + {a^\dagger_{2}} n_{0} + {a^\dagger_{3}} a_{1} n_{2}\right)\\
			&- \frac{v_{01}^{01}}{2} \left(a_{1} n_{0} + {a^\dagger_{1}} n_{0}\right) - \frac{v_{02}^{02}}{2} \left(a_{3} n_{0} + {a^\dagger_{3}} n_{0}\right)\\
			&- \frac{v_{01}^{23}}{4} \left(a_{2} a_{1} a_{0} + a_{3} a_{0} + {a^\dagger_{0}} a_{2} a_{1} + {a^\dagger_{0}} a_{3} + {a^\dagger_{0}} {a^\dagger_{1}} a_{2} + {a^\dagger_{0}} {a^\dagger_{1}} {a^\dagger_{2}} + {a^\dagger_{0}} {a^\dagger_{2}} a_{1} + {a^\dagger_{0}} {a^\dagger_{3}} + {a^\dagger_{1}} a_{2} a_{0} + {a^\dagger_{1}} {a^\dagger_{2}} a_{0} + {a^\dagger_{2}} a_{1} a_{0} + {a^\dagger_{3}} a_{0}\right)\\
			&+\left(\frac{v_{02}^{02}}{4} - \frac{v_{02}^{20}}{4}\right) \left(a_{3} a_{1} + {a^\dagger_{1}} a_{3} + {a^\dagger_{1}} {a^\dagger_{3}} + {a^\dagger_{3}} a_{1}\right)\\
			&+\frac{v_{01}^{23}}{4} \left(a_{1} a_{0} + a_{3} a_{2} a_{0} + {a^\dagger_{0}} a_{1} + {a^\dagger_{0}} a_{3} a_{2} + {a^\dagger_{0}} {a^\dagger_{1}} + {a^\dagger_{0}} {a^\dagger_{2}} a_{3} + {a^\dagger_{0}} {a^\dagger_{2}} {a^\dagger_{3}} + {a^\dagger_{0}} {a^\dagger_{3}} a_{2} + {a^\dagger_{1}} a_{0} + {a^\dagger_{2}} a_{3} a_{0} + {a^\dagger_{2}} {a^\dagger_{3}} a_{0} + {a^\dagger_{3}} a_{2} a_{0}\right)\\
			&+\frac{v_{02}^{02}}{4} \left(a_{2} a_{1} + {a^\dagger_{1}} a_{2} + {a^\dagger_{1}} {a^\dagger_{2}} + {a^\dagger_{2}} a_{1}\right) +\frac{v_{23}^{23}}{4} \left(a_{3} a_{2} + {a^\dagger_{2}} a_{3} + {a^\dagger_{2}} {a^\dagger_{3}} + {a^\dagger_{3}} a_{2}\right) +\left(v_{02}^{02} - v_{02}^{20}\right) \left(a_{2} n_{0} n_{1} + {a^\dagger_{2}} n_{0} n_{1}\right)\\
			&+v_{02}^{02} \left(a_{3} n_{0} n_{1} + a_{3} n_{0} n_{2} + {a^\dagger_{3}} n_{0} n_{1} + {a^\dagger_{3}} n_{0} n_{2}\right) +\left(- 2 h^{2}_{2} - v_{02}^{02} + v_{02}^{20} - v_{23}^{23}\right) \left(a_{3} n_{1} n_{2} + {a^\dagger_{3}} n_{1} n_{2}\right)\\
			&+\frac{v_{01}^{23}}{2} \left(a_{3} a_{0} n_{1} + a_{3} a_{0} n_{2} + {a^\dagger_{0}} a_{3} n_{1} + {a^\dagger_{0}} a_{3} n_{2} + {a^\dagger_{0}} {a^\dagger_{3}} n_{1} + {a^\dagger_{0}} {a^\dagger_{3}} n_{2} + {a^\dagger_{3}} a_{0} n_{1} + {a^\dagger_{3}} a_{0} n_{2}\right)\\
			&- v_{01}^{23} \left(a_{3} a_{0} n_{1} n_{2} + {a^\dagger_{0}} a_{3} n_{1} n_{2} + {a^\dagger_{0}} {a^\dagger_{3}} n_{1} n_{2} + {a^\dagger_{3}} a_{0} n_{1} n_{2}\right)- 2 v_{02}^{02} \left(a_{3} n_{0} n_{1} n_{2} + {a^\dagger_{3}} n_{0} n_{1}  n_{2}\right).
		\end{split}
		$}
\end{equation}
It is worth mentioning that, in second quantization, the transformed Hamiltonian does not conserve particle number and fermionic parity, a consequence of the fermionic transformation being non-Clifford.
Next, the tapered Hamiltonian in the $({+}{-}{-})$ symmetry sector, \cref{seq:st_H_+--}, reads in second quantization
\begin{equation}
	\mathscr{H}^{({+}{-}{-})} = \left[ h_0^0 + h_2^2 + \frac{1}{2}\left(v_{01}^{01} + v_{23}^{23}\right) + \left(h_2^2 - h_0^0 + \frac{1}{2} \left(v_{23}^{23} - v_{01}^{01}\right)\right)(I-2n) + v_{01}^{23} (a^\dagger + a) \right],
\end{equation}
while the UCCD unitary becomes
\begin{equation}
	e^{\I \theta Y} \xrightarrow{\text{JW}^{-1}} e^{-\theta(a^\dagger -a)}.
\end{equation}
Finally, the spinorbital tapered Schr\"{o}dinger equation for $\text{H}_2$/MBS reads:
\begin{equation}\label{seq:tapered_SE_sq}
	\resizebox{\linewidth}{!}{$
		\left[ h_0^0 + h_2^2 + \dfrac{1}{2}\left(v_{01}^{01} + v_{23}^{23}\right) + \left(h_2^2 - h_0^0 + \dfrac{1}{2} \left(v_{23}^{23} - v_{01}^{01}\right)\right)(I-2n) + v_{01}^{23} (a^\dagger + a) \right] e^{- \theta (a^\dagger - a)} \ket{1} = E_0 e^{- \theta (a^\dagger - a)} \ket{1}.
		$}
\end{equation}

\section{Classification of Lie Algebras generated by Fermionic anti-Hermitian Half-Body and Pair Operators}\label{ssec:Lie}

As reported in the main text, given a basis of $M$ single-particle states, the set of single excitation anti-Hermitian operators $\{A_q^p\}_{p<q}$ satisfies the commutation relations
\begin{equation}\label{seq:comm_singles}
	\left[A^p_q, A^r_s\right] = \delta_{qr} A^p_s - \delta_{qs} A^p_r - \delta_{pr} A^q_s +\delta_{ps} A^q_r,
\end{equation}
thus forming a basis of the $\frac{M(M-1)}{2}$-dimensional $\mathfrak{so}(M)$ Lie algebra.
When considering the set of pair anti-Hermitian operators $\{A_q^p, A^{pq}\}_{p<q}$, in addition to \cref{seq:comm_singles}, the following commutation relations arise:
\begin{equation}\label{seq:comm_pair}
	\left[A^{pq}, A^{rs}\right] = \delta_{qr} A^p_s - \delta_{qs} A^p_r - \delta_{pr} A^q_s +\delta_{ps} A^q_r
\end{equation}
and
\begin{equation}\label{seq:comm_singles_pair}
	\left[A^p_q, A^{rs}\right] = \delta_{qr} A^{ps} - \delta_{qs} A^{pr} - \delta_{pr} A^{qs} + \delta_{ps} A^{qr}.
\end{equation}
In what follows, we show that \cref{seq:comm_singles,seq:comm_pair,seq:comm_singles_pair} are the commutation relations defining the $\mathfrak{so}(M)\oplus\mathfrak{so}(M)$ Lie algebra.

Let $E_{pq}$ denote the matrix unit having a value of 1 in the $p$th row and $q$th column with all other elements being 0.
In the standard basis, the $\mathfrak{so}(M)$ Lie algebra is spanned by $\{L_{pq}\}_{p<q}$, with $L_{pq}\equiv E_{pq} - E_{qp}$, satisfying commutation relations analogous to \cref{seq:comm_singles}.
By taking the direct sum of $\mathfrak{so}(M)$ with itself, we generate the $M(M-1)$-dimensional $\mathfrak{so}(M)\oplus\mathfrak{so}(M)$ Lie algebra, a generic element of which is given by $(L_{pq},L_{rs})$.
Although the set $\{(L_{pq}, 0), (0, L_{pq})\}_{p<q}$ forms a natural basis of $\mathfrak{so}(M)\oplus \mathfrak{so}(M)$, we will use the alternative basis provided by the set $\{(L_{pq}, L_{pq}), (L_{pq}, -L_{pq})\}_{p<q}$.
These basis vectors give rise to the commutation relations
\begin{equation}\label{seq:som_som_singles}
	\begin{split}
		[(L_{pq}, L_{pq}), (L_{rs}, L_{rs})] &= ([L_{pq}, L_{rs}], [L_{pq}, L_{rs}])\\
		&= (\delta_{qr} L_{ps} - \delta_{qs} L_{pr} - \delta_{pr} L_{qs} +\delta_{ps} L_{qr}, \delta_{qr} L_{ps} - \delta_{qs} L_{pr} - \delta_{pr} L_{qs} +\delta_{ps} L_{qr})\\
		&= \delta_{qr}(L_{ps},L_{ps}) - \delta_{qs}(L_{pr},L_{pr})- \delta_{pr}(L_{qs},L_{qs}) + \delta_{ps}(L_{qr},L_{qr}),
	\end{split}
\end{equation} 
\begin{equation}\label{seq:som_som_pair}
	\begin{split}
		[(L_{pq}, -L_{pq}), (L_{rs}, -L_{rs})] &= ([L_{pq}, L_{rs}], [L_{pq}, L_{rs}])\\
		&= \delta_{qr}(L_{ps},L_{ps}) - \delta_{qs}(L_{pr},L_{pr})- \delta_{pr}(L_{qs},L_{qs}) + \delta_{ps}(L_{qr},L_{qr}),
	\end{split}	
\end{equation}
and
\begin{equation}\label{seq:som_som_singles_pair}
	\begin{split}
		[(L_{pq}, L_{pq}), (L_{rs}, -L_{rs})] &= ([L_{pq}, L_{rs}], -[L_{pq}, L_{rs}])\\
		&= (\delta_{qr} L_{ps} - \delta_{qs} L_{pr} - \delta_{pr} L_{qs} +\delta_{ps} L_{qr}, -\delta_{qr} L_{ps} + \delta_{qs} L_{pr} + \delta_{pr} L_{qs} -\delta_{ps} L_{qr})\\
		&= \delta_{qr}(L_{ps},-L_{ps}) - \delta_{qs}(L_{pr},-L_{pr})- \delta_{pr}(L_{qs},-L_{qs}) + \delta_{ps}(L_{qr},-L_{qr}),
	\end{split}
\end{equation}
which define the $\mathfrak{so}(M)\oplus\mathfrak{so}(M)$ Lie algebra.
After comparing \cref{seq:comm_singles,seq:comm_pair,seq:comm_singles_pair} to \cref{seq:som_som_singles,seq:som_som_pair,seq:som_som_singles_pair}, we realize that the Lie algebra spanned by $\{A^p_q, A^{pq}\}_{p<q}$ is isomorphic to $\mathfrak{so}(M)\oplus\mathfrak{so}(M)$, with $A^p_q \mapsto (L_{pq}, L_{pq})$ and $A^{pq} \mapsto (L_{pq}, -L_{pq})$.

When considering the Lie-algebraic structure of the set $\{A^p_q, A^{pq}, A^p\}$, in addition to \cref{seq:comm_singles,seq:comm_pair,seq:comm_singles_pair}, the following commutation relations emerge:
\begin{equation}\label{seq:comm_half}
	\left[A^p, A^q\right] = 2(A^{pq} - A^p_q),
\end{equation}
\begin{equation}\label{seq:comm_singles_half}
	\left[A^p_q, A^r\right] = \delta_{qr} A^p - \delta_{pr} A^q,
\end{equation}
and
\begin{equation}\label{seq:comm_pair_half}
	\left[A^{pq}, A^r\right] = \delta_{pr} A^q - \delta_{qr} A^p.
\end{equation}
Below we prove that this set spans a Lie algebra isomorphic to $\mathfrak{so}(M)\oplus\mathfrak{so}(M+1)$.

The direct sum $\mathfrak{so}(M)\oplus\mathfrak{so}(M+1)$ gives rise to an $M^2$-dimensional Lie algebra with generic elements $(L_{pq}, L_{rs})$, with $1 \le p < q \le M$ and $1 \le r < s \le M + 1$.
Instead of the standard basis $\{(L_{pq},0), (0, L_{rs})\}_{\substack{p<q\le M \\ r<s\le M+1}}$, we will work with the alternative basis $\{(L_{pq}, L_{pq}), (L_{pq}, -L_{pq}), (0, 2L_{r\,M+1})\}_{\substack{p<q\le M\\r < M+1}}$.
In addition to the commutation relations of \cref{seq:som_som_singles,seq:som_som_pair,seq:som_som_singles_pair}, we obtain
\begin{equation}\label{seq:som_som1_half}
	\begin{split}
		[(0, 2L_{p\,M+1}), (0, 2L_{q\,M+1})] & = 4(0, [L_{p\,M+1}, L_{q\,M+1}])\\
		&= 4(0, \cancelto{0}{\delta_{M+1\,q}} L_{p\,M+1} - \delta_{M+1\,M+1} L_{pq} - \delta_{pq} \cancelto{0}{L_{M+1\,M+1}} +\cancelto{0}{\delta_{p\,M+1}} L_{M+1\,q})\\
		&= 4(0, -L_{pq})\\
		&= 2\left[(L_{pq}, -L_{pq}) - (L_{pq}, L_{pq})\right],
	\end{split}
\end{equation}
\begin{equation}\label{seq:som_som1_singles_half}
	\begin{split}
		[(L_{pq}, L_{pq}), (0, 2L_{r\,M+1})] &= (0, [L_{pq}, 2L_{r\,M+1}])\\
		&= (0, \delta_{qr} 2L_{p\,M+1} - \cancelto{0}{\delta_{q\,M+1}} 2L_{pr} - \delta_{pr} 2L_{q\,M+1} + \cancelto{0}{\delta_{p\,M+1} 2L_{qr}})\\
		&= \delta_{qr}(0,2L_{p\,M+1}) - \delta_{pr} (0,2L_{q\,M+1}),
	\end{split}
\end{equation}
and
\begin{equation}\label{seq:som_som1_pair_half}
	\begin{split}
		[(L_{pq}, -L_{pq}), (0, 2L_{r\,M+1})] &= (0, -[L_{pq}, 2L_{r\,M+1}])\\
		&= \delta_{pr} (0,2L_{q\,M+1}) - \delta_{qr}(0,2L_{p\,M+1}).
	\end{split}
\end{equation}
After comparing \cref{seq:comm_singles,seq:comm_pair,seq:comm_singles_pair,seq:comm_half,seq:comm_singles_half,seq:comm_pair_half} to \cref{seq:som_som_singles,seq:som_som_pair,seq:som_som_singles_pair,seq:som_som1_half,seq:som_som1_singles_half,seq:som_som1_pair_half}, we realize that the Lie algebra spanned by $\{A^p_q, A^{pq}, A^p\}$ is isomorphic to $\mathfrak{so}(M)\oplus\mathfrak{so}(M+1)$, with $A^p_q \mapsto (L_{pq}, L_{pq})$, $A^{pq} \mapsto (L_{pq}, -L_{pq})$, and $A^p \mapsto (0, 2L_{p\,M+1})$.

\end{document}